%% file: snls3lc.tex
\documentclass{aa}

\usepackage{amsmath}
\usepackage{natbib}
\bibpunct{(}{)}{;}{a}{}{,} 
\usepackage{graphicx}
\usepackage{txfonts}
\usepackage{color}
\usepackage{longtable}
\usepackage{rotating}
\usepackage{ulem}
\usepackage{lscape}

\newcommand{\om}{\Omega_{\rm M}}
\newcommand{\ol}{\Omega_\Lambda}

\newcommand{\msb}{m_B^{*}}
\newcommand{\x}{X_{1}}
\newcommand{\col}{\mathcal{C}}

\newcommand{\phase}{\tau}
\newcommand{\bmv}{(B-V)}
\newcommand{\umb}{(U-B)}
\newcommand{\vmr}{(V-R)}
\newcommand{\utwomb}{(U_{02}-B)}
\newcommand{\umega}{$u_M$}
\newcommand{\gmega}{$g_M$}
\newcommand{\rmega}{$r_M$}
\newcommand{\imega}{$i_M$}
\newcommand{\zmega}{$z_M$}
\newcommand{\allmegasnfilts}{\gmega\rmega\imega\zmega}
\newcommand{\bdtruc}{BD$\,$+17$\,$4708}

\newcommand{\numberofspectroedIas}{285$\,$}
\newcommand{\numberoffailedphotom}{4$\,$}
\newcommand{\numberofbadsampling}{25$\,$}

\newcommand{\numberofgoodphotom}{252$\,$} 
\newcommand{\numberofredsne}{14$\,$} 
\newcommand{\numberofoutliers}{7$\,$} 
\newcommand{\numberofsiftofailed}{3$\,$} 
\newcommand{\numberinhubblediagram}{231$\,$}

\begin{document}

\title{The Supernova Legacy Survey 3-year sample: Type Ia Supernovae photometric distances and cosmological constraints
\thanks{
Based on observations obtained with MegaPrime/MegaCam, a joint project
of CFHT and CEA/DAPNIA, at the Canada-France-Hawaii Telescope (CFHT)
which is operated by the National Research Council (NRC) of Canada,
the Institut National des Sciences de l'Univers of the Centre National
de la Recherche Scientifique (CNRS) of France, and the University of
Hawaii. This work is based in part on data products produced at the
Canadian Astronomy Data Centre as part of the Canada-France-Hawaii
Telescope Legacy Survey, a collaborative project of NRC and CNRS.
Based on observations 
obtained at the European Southern Observatory using the Very Large Telescope
on the Cerro Paranal (ESO Large Programme 171.A-0486 \& 176.A-0589).
Based on observations (programs 
GS-2003B-Q-8,
GN-2003B-Q-9,
GS-2004A-Q-11,
GN-2004A-Q-19,
GS-2004B-Q-31,
GN-2004B-Q-16,
GS-2005A-Q-11,
GN-2005A-Q-11,
GS-2005B-Q-6,
GN-2005B-Q-7,
GN-2006A-Q-7,
GN-2006B-Q-10) 
obtained at the Gemini Observatory, which is
operated by the Association of Universities for Research in Astronomy,
Inc., under a cooperative agreement with the NSF on behalf of the
Gemini partnership: the National Science Foundation (United States),
the Particle Physics and Astronomy Research Council (United Kingdom),
the National Research Council (Canada), CONICYT (Chile), the
Australian Research Council (Australia), CNPq (Brazil) and CONICET
(Argentina).
Based on observations obtained at the W.M. Keck
Observatory, which is operated as a scientific partnership among the
California Institute of Technology, the University of California and the
National Aeronautics and Space Administration. The Observatory was made
possible by the generous financial support of the W.M. Keck Foundation.
Mark Sullivan acknowledges support from the Royal Society.
}
}

\include{authors}

\titlerunning{SNLS-3: SNe~Ia photometric distances and cosmological constraints}
\authorrunning{J. Guy et al., SNLS Collaboration}                         
\offprints{guy\@@in2p3.fr}

\date{Received Month DD, YYYY; accepted Month DD, YYYY}

\abstract
{} 
{We present photometric properties and distance measurements of \numberofgoodphotom high redshift Type
  Ia supernovae ($0.15 < z < 1.1$) discovered during the first three years  of the Supernova Legacy Survey (SNLS).
These events were detected and their
  multi-colour light curves measured using the MegaPrime/MegaCam instrument at
  the Canada-France-Hawaii Telescope (CFHT), by repeatedly imaging four one-square degree fields in 
  four bands. Follow-up spectroscopy was performed at
  the VLT, Gemini and Keck telescopes to confirm the nature of the
  supernovae and to measure their redshifts.} 
{Systematic uncertainties arising from light curve modeling are studied, 
making use of two techniques to derive the peak magnitude, shape and colour of the supernovae,
and taking advantage of a precise calibration of the SNLS fields.} 
{A flat $\Lambda$CDM cosmological
 fit to  \numberinhubblediagram SNLS  high redshift Type Ia supernovae alone gives 
$\om = 0.211 \pm 0.034\mathrm{(stat)} \pm 0.069\mathrm{(sys)}$. The dominant systematic uncertainty comes from uncertainties in the photometric calibration.
Systematic uncertainties from light curve fitters come next with a total contribution of $\pm 0.026$ on $\om$. No clear evidence is found for a possible evolution of the slope ($\beta$) of the colour-luminosity relation with redshift.
 } 
{} 

\maketitle

\section{Introduction}

Since 1998~\citep{Riess98b,Perlmutter99}, surveys of cosmologically distant Type Ia supernovae (SNe~Ia) 
have shown that the expansion of the Universe is accelerating,
distant SNe~Ia being fainter 
than expected in a decelerating Universe.
With the assumption that the Universe can be described on average 
as isotropic and homogeneous, this acceleration implies either the existence of a fluid with
negative pressure, usually called dark energy, a cosmological constant, or modifications of gravity on cosmological scales.
Several other cosmological probes such as  cosmic microwave background (CMB) anisotropies and baryon acoustic oscillations (BAO) confirm this result, but SNe~Ia observations are currently the most sensitive technique to study dark energy or its alternatives, since they can be used to directly measure the history of the expansion of the Universe.

Recent results from high redshift SNe~Ia surveys (\citealt{Astier06,Riess07,WoodVasey07,Freedman09}) and combinations of SNe~Ia samples (\citealt{Kowalski08,Hicken09b}) give consistent measurements of the effective equation of state parameter of dark energy ($w$, the ratio of pressure over density) in the redshift range accessible with SNe data. Values consistent with $w = -1$ are found, as expected for a cosmological constant, with uncertainties of order of 10\%, including systematics. 
In contrast with those analyses, \citet{Kessler09} report a much larger systematic uncertainty on $w$. They obtain a discrepancy $\delta w = 0.2$ when using two different techniques to estimate SNe distances. We discuss this issue in $\S$\ref{sec:fitter-selection}. 

This paper is the first of a set of three papers which present the cosmological analysis of the Supernova Legacy Survey (SNLS) after three years of operation. The SN~Ia sample is about four times as large as the first year sample presented in \citet[ hereafter A06]{Astier06}. Since systematic uncertainties are close to the statistical limits, large efforts on all aspects of the analysis have been made to identify, reduce and propagate them to the final cosmological result. A key feature of the work presented in this set of papers is that the full analysis has been performed twice with significantly different techniques, including for photometry, detection of candidates, spectroscopic identification, calibration, and distance estimate.
The step by step comparison of the results obtained with these different techniques made it possible to pin down systematics often ignored in previous works. This first paper describes the photometric data reduction and the estimation of light curve fit parameters for this sample. It also reports on cosmological constraints that can be obtained from SNLS supernovae alone. The second paper~(\citealt{Conley10}, hereafter C10) presents cosmological constraints obtained by combining the SNLS 3-yr sample with lower and higher redshift SN Ia data and the 3rd paper presents cosmological constraints when combining SN data with other cosmological probes~\citep{Sullivan10}.

After an overview of the survey in $\S$\ref{sec:snls-overview}, the photometric data reduction is described in $\S$\ref{sec:photometry} together with the photometric techniques used. 
The two light curve fitting techniques used in this analysis, SiFTO~\citep{Conley08} and SALT2~\citep{Guy07}, are presented in $\S$\ref{sec:lcfitting}. Their results are compared and combined in $\S$\ref{sec:hubble-diagram-om} to constrain $\om$ for a flat $\Lambda$CDM cosmological model. We conclude in $\S$\ref{sec:conclusion}.

\section{Overview of the Supernova Legacy Survey}
\label{sec:snls-overview}
The Supernova Legacy Survey (SNLS) uses data taken as part of the deep
component of the five-year Canada-France-Hawaii Telescope Legacy
Survey (CFHT-LS). CFHT-LS is an optical imaging survey using the
one square degree MegaCam camera \citep{MegacamPaper} on the
CFHT. The deep component conducted repeat imaging of 4 low
Galactic-extinction fields (see
A06 for the field coordinates). The survey ended in August 2008.
A detailed description of the telescope and camera 
is given in \citet{Regnault09} (hereafter R09).
  The data
are time-sequenced with observations conducted every 3--4 nights in
dark time. Four filters were used (denoted
\allmegasnfilts)\footnote{Supplementary \umega-band data was also
  acquired as part of CFHT-LS; these data were not time-sequenced and
  not used in the SN light curves.} similar to those of the Sloan
Digital Sky Survey (SDSS). This data allows us to obtain
high-quality multi-colour SN light curves. Since the first year
analysis in A06 the cadence and exposure time of the \zmega\
observations were increased to improve the signal-to-noise in the
light curves of the most distant SNe.

For the ``real-time'' SN searches (see \citealt{Perrett10} for details) the SNLS data was reduced by the
CFHT-developed Elixir data reduction system
\citep{Magnier04} and processed through two independent
search pipelines, from which combined candidate lists were
generated\footnote{Candidate lists are available from\\
  \texttt{http://legacy.astro.utoronto.ca/}}. 
Both pipelines generated
difference images, subtracting from the search images deep references
constructed from previous observations, leaving only sources which
had varied since the reference epoch.  The subtracted images were
searched using various automated techniques, with all likely
candidates visually inspected by human eye.  SN candidates were prioritised for
spectroscopic follow up using a photometric identification technique
\citep{Sullivan06} which identifies likely SNe~Ia from 2-3
epochs of real-time photometry, predicting redshifts, phases, and
temporal evolution in magnitude. New candidates were checked against
 a database of existing variable sources to exclude previously
known AGN or variable stars.

Spectroscopic follow-up was used to confirm SN types and measure
redshifts. 
The survey was allocated 60 hours per semester on
the Gemini North and South telescopes using the GMOS instruments
\citep{Hook04} and 120 hours per year at the European
Southern Observatory Very Large Telescopes (VLT) using the FORS-1/2
instruments \citep{FORS98}.  Spectroscopic time was also
obtained at the Keck telescopes using the LRIS~\citep{KECK-LRIS95} and DEIMOS~\citep{KECK-DEIMOS} instruments (PI: Perlmutter). 
Some additional types and
redshifts were obtained as part of a separate detailed studies program
also undertaken using Keck-I/LRIS~\citep{Ellis08}.

Data from the first
year of the Gemini program were presented in
\citet{Howell05}, from the second and third years in
\citet{Bronder08} and data from May 2006 to July 2006 in \citet{Walker10}.
Spectra obtained at VLT are published in
\citet{Balland09} (see \citealt{Baumont08} for the reduction technique used), and Keck spectra from the detailed studies program
are in \citet{Ellis08}. All spectra were analysed and
uniformly typed. Two sub-classes are considered in this work according to the confidence level of the spectroscopic identification: certain SN Ia are denoted ``SN~Ia'' (corresponding to the confidence indices CI 5 and 4 in the classification scheme of \citealt{Howell05}), and probable SN~Ia (CI 3) are labeled ``SN~Ia$\star$''. For most SNe, two identification techniques were used and their results cross-checked (based on the analysis presented in \citealt{Howell05} and \citealt{Balland09}).

\section{Photometry measurement}
\label{sec:photometry}

Elixir pre-processed images were retrieved from the
Canadian Astronomy Data Centre
(CADC)\footnote{http://www.cadc-ccda.hia-iha.nrc-cnrc.gc.ca/cadc/}. 
This process performs basic image ``de-trending'' (bias subtraction, flat-field correction and fringe removal in \imega\ and \zmega\ bands). The flat-fields and fringe maps are more precise than the ones used for the real-time SN search as they are constructed from an entire queue run of data (including non CFHT-LS data), from median stacks of twilight and science exposures for the flat-field and fringe maps respectively. 
The Elixir pipeline also attempts to
produce images with photometric uniformity (i.e. a constant zero point
across the mosaic) by constructing photometric correction 
 frames from dithered observations of dense stellar
fields. However, this process is not perfect and some radial trends in
the photometric zero point remain. This is shown in detail in R09, who found a centre to edge variation of $\simeq 0.02$ magnitude, and derived refined photometric correction frames.

The subsequent treatment of images including sky background subtraction, astrometry, and photometric correction has been performed in two independent pipelines.
 The first one follows A06.
SExtractor~\citep{Sex} is used to produce an image catalog and a sky background map that is subtracted from the image. 
Second order moments of the objects are derived (using an iterative Gaussian-weighted fit), and point-like sources are identified to estimate the image quality (hereafter IQ, given by the FWHM of the point spread function).  
A weight map is derived using the spatial sky variance, the bad pixel map provided by Elixir, and a map of cosmic rays hits and satellite trails identified with dedicated algorithms.
The astrometry of each CCD image is obtained using a match of the image catalog to an astrometric catalog derived from observations of \citet{Stone1999} astrometric
calibration regions and USNO-B \citep{USNO-B} or SDSS
\citep{SDSS5thRelease07} star catalogs. 
The photometric corrections (more precisely the ratio of the ones derived in R09 to those applied to Elixir images) are not applied to the images but to the fluxes of objects in the catalogs.
 The other pipeline performs the same  reduction steps but uses different software (for instance
   the Image Reduction and Analysis Facility package
\footnote{IRAF is distributed by the National Optical Astronomy
  Observatories, which are operated by the Association of Universities
  for Research in Astronomy, Inc., under cooperative agreement with
  the National Science Foundation.}). 

\subsection{Measurement of the supernova fluxes}

Two photometry techniques are considered. We present both and compare their outcome. The goal of this study is to
cross-check the methods, and evaluate photometric systematic uncertainties from
 the level of agreement reached in this comparison. This also allows us to 
 select the best technique for deriving light curve parameters of the SNe.

\subsubsection{Simultaneous fit of galaxy and supernova fluxes}
\label{sec:simultaneous-fit}

This first method (hereafter method A) was described in A06. It consists first in 
resampling all the images of a field in each pass-band to the pixel grid of the best IQ
image (hereafter called reference image) for which a PSF (point spread
function) model is derived.  Convolution kernels of the reference
image to each aligned image are then derived for subsequent use. The
photometric ratio of the two images is simply given by the integral of
the kernel.  The photometric fit then consists in fitting
simultaneously an image of the host galaxy (at the sampling and the
image quality of the reference image) together with the position of
the supernova and its fluxes in all images, convolving both the galaxy
and PSF models using the kernels obtained previously. The supernova
flux is forced to zero in images where its flux is negligible. When we
calculate the pixel uncertainties, we deliberately ignore the
contribution of the supernova and host galaxy fluxes, so that PSF
inaccuracies affect the flux of bright and faint sources in the same
way (this is further discussed in $\S$\ref{sec:calibration}).  Note
that the supernovae considered here are faint enough for the departure
from statistical optimality to be negligible.

  Whereas DAOPHOT~\citep{DAOPHOT} was used in A06 to model the PSF, we have since then developed an independent PSF modeling code for this analysis.
Also, some improvements have been made in the selection of images for alignment. The reference frame is larger than the actual reference CCD image to cover gaps between CCDs in the MegaCam focal plane. For large dithers, images from other CCDs overlapping the reference image are included so that different fractions of the same image can be aligned on different reference frames. This allows us to measure fluxes for all SNe in the field, including those that sometimes fall in the gaps between CCDs (depending on the actual pointing).

Some cuts are applied to the images:
 an IQ better than 1.5$\arcsec$ is required, the $\chi^2$ per degree of freedom of the convolution kernel fit has to be better than 1.5, and the  image correlation coefficient\footnote{The dimensionless image correlation coefficient at a lag $\Delta$ is given by $\sum \left(f_{i}-f_{i+\Delta}\right)^2 / \sum f_{i}^2$, where $f_{i}$ is the flux in the pixel $i$ (the average value of $f_i$ is close to zero as the images were previously background subtracted).}  at a 50 pixel lag has to be lower than 0.08. This latter cut allows us to get rid of images with large fringe residuals (in \imega\ and \zmega\ bands) or background subtraction problems.

The ratio of images that pass the cuts, the number of exposures and total exposure time for each field and band for this dataset is shown in Table~\ref{table:stat_exposure}, along with the average IQ of the selected images.
 The relatively low efficiencies of about 80\% in \imega\ band for the fields D2 and D3 are due to issues with  fringe subtractions for some MegaCam runs.

\begin{table}
\begin{center}
\caption{
Total exposure time and average IQ of the
SNLS 3 year data set. The selected images are those that pass the quality cuts
(see text for details). 
\label{table:stat_exposure}
}
\begin{tabular}{ccccc}
\hline
\hline
 & $\#$ exp. & exp. time & ratio of selected & average IQ\\
 &           &  (hours)  & images & (FWHM) \\
D1 \\
\hline
\gmega & 389  & 24  & 99.4\% & 0.92$\arcsec$ \\
\rmega & 559  & 49  & 98.6\% & 0.88$\arcsec$ \\
\imega & 769  & 104 & 95.6\% & 0.84$\arcsec$ \\
\zmega & 520  & 52  & 97.2\% & 0.83$\arcsec$ \\
\\
D2 \\
\hline
\gmega & 244  & 15 & 94.7\% & 0.97$\arcsec$ \\
\rmega & 379  & 32 & 91.5\% & 0.90$\arcsec$ \\
\imega & 509  & 68 & 78.7\% & 0.85$\arcsec$ \\
\zmega & 272  & 27 & 94.2\% & 0.80$\arcsec$ \\
\\
D3 \\
\hline
\gmega & 387  & 23 & 96.1\% & 0.99$\arcsec$ \\
\rmega & 496  & 40 & 95.9\% & 0.89$\arcsec$ \\
\imega & 745  & 88 & 80.6\% & 0.90$\arcsec$ \\
\zmega & 410  & 40 & 98.4\% & 0.82$\arcsec$ \\
\\
D4 \\
\hline
\gmega & 398  & 24 & 97.7\% & 0.97$\arcsec$ \\
\rmega & 551  & 48 & 88.3\% & 0.89$\arcsec$ \\
\imega & 713  & 94 & 88.7\% & 0.85$\arcsec$ \\
\zmega & 515  & 51 & 93.0\% & 0.82$\arcsec$ \\
\end{tabular}
\end{center}
\end{table}

For a known fixed position of the SN in the images, the photometry fit is linear and we do not expect any bias for a perfect PSF model.
However, when the position of the SN is fit simultaneously with the fluxes, we have (for a Gaussian PSF, see $\S$~\ref{section:psf-bias} for a proof) the following fractional bias in flux: 
\begin{equation}
\frac{\delta f}{f} \equiv \frac{E[\hat f] -f}{f} = - \frac{Var(f)}{f^2} 
\end{equation}
where $Var(f)$ is the variance of the light curve amplitude $f$, obtained from a fit combining observations at different epochs.

\begin{figure}
\centering
\includegraphics[width=\linewidth]{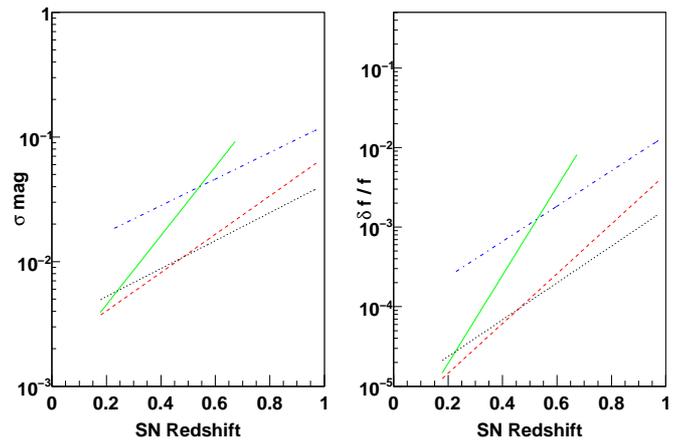}
\caption{Statistical uncertainty on magnitudes (left panel) and expected bias on PSF photometry with a simultaneous fit of the position (right panel) as a function of supernova redshift. The different \allmegasnfilts\ MegaCam bands are shown respectively with green solid, red dashed, black dotted, and blue dashed-dotted curves.
\label{fig:psfbias}
}
\end{figure}

Figure~\ref{fig:psfbias} shows the expected bias of the observed SNe as a function of redshift in \allmegasnfilts\ bands.
For the \rmega\ and \imega\ bands, the biases are respectively of 1~mmag and 0.6~mmag at $z=0.8$, and 5~mmag and 2~mmag at $z=1$.
The \gmega\  and \zmega\  light curves have a low signal to noise ratio ($S/N$) at high redshift so we do not fit the position for those bands and use instead the weighted average of the ones obtained in the \rmega\ and \imega\ band fits. With an astrometric precision better than 10~milliarcsec for the transformation of coordinates from one image to another (estimated from the RMS of the residual match of the catalogs), and an average IQ of $0.9''$, we do not expect a bias due to the transformation of coordinates larger than 0.3~mmag (see Eq.~\ref{eq:bias_var_pos}). As a consequence, this effect can be ignored, and the total bias is given by that obtained in \imega\ which has the highest $S/N$ (reaching a maximum value of  2~mmag at $z=1$). 
 Those biases are an order of magnitude smaller than calibration uncertainties for redshifts $z<0.8$, and other sources of uncertainties at higher $z$ (see $\S$\ref{section:systematic_uncertainties}), so we do not correct for them and ignore their negligible contribution to uncertainties in the following. The correlation introduced in this process between the magnitudes in different bands will also be neglected.

As in A06, the photometry fit provides us with flux estimates for all exposures in a given night. 
 Since we do not expect significant variation of the SN luminosity on this time scale, 
we average these measurements and use the additional
scatter between measurements to account for the imperfect kernel evaluation and under-estimated statistical uncertainties due to pixel correlations (introduced by the resampling of images).
A few outliers (about 1\% of the data) are rejected at this level, some of which may be due to unidentified cosmic ray hits. Since the measurements on the light curve are correlated since they share the uncertainty on the host galaxy flux and SN position (in \rmega\ and \imega\ bands), their covariance matrix is recorded for a subsequent usage in the light curve fits.  The quality of the photometry fit and the fraction of outliers rejected at this level are the same as in A06.

\subsubsection{PSF photometry on image subtractions}
\label{sec:photometry-on-subtractions}

We now describe the second photometric method (method B).
The PSF and IQ are measured for each image from a list
of isolated stellar objects with 35-50 objects used for each CCD.
Photometric alignment is performed using a multiplicative scaling
factor from a comparison to the set of tertiary calibrating stars,
discussed in $\S$\ref{sec:calibration}.  Aperture photometry with a
  radius of 4\arcsec\ is used for this alignment, 
and  the average offset to PSF photometry is recorded at the same time. 

SN flux measurement are performed on a CCD-by-CCD basis; even though
the telescope is dithered during observations on a given night, we
measure the flux on each observation independently. 
A series of deep reference images are constructed for each season by
combining the data from every other season in bins of IQ.
The calibrated images (and their weight-maps) entering a given reference frame 
are geometrically re-sampled to a common pixel coordinate system
using a kernel conserving flux. They are then combined using a
weighted mean with 5-$\sigma$ outlier rejection, generating a deep
reference and associated weight-map. Each field/filter/season
combination typically has five or more statistically independent
references each with a different mean IQ and each at least 6 times
deeper (or 36 times the integration) than the individual exposures
from which the SN measurements are made.

For each calibrated image containing SN light, the reference with the
closest (but superior) IQ is geometrically re-sampled to the same pixel
coordinate system, and the PSFs of the two images matched by degrading
the reference image to match the image containing the SN light. This
adjusted reference image is then subtracted from the science
image. This process avoids any geometric re-sampling of SN pixels and
introduces only a minimum correlated noise in the subtraction of the
deep reference. Flux measurement is then performed on the difference
images using a custom-written PSF-fitting program using PSFs measured
on the unsubtracted images, weighting the fit using the appropriate weight map.

\subsubsection{Flux measurement uncertainties}

The SNe data set considered consists of a sub-sample of all the SNe~Ia detected and spectroscopically confirmed by the SNLS up to July 2006. They are listed Table~\ref{table:spectro}.
For those labeled ``Ia$^*$'', one can not completely exclude Ib/c core collapse SNe.
Light curves from \numberoffailedphotom  SNe out of \numberofspectroedIas could not be obtained for the following
 reasons: lack of observations (beginning or end of an observing run, bad weather conditions, instrument failures), another variable object at the position of the SN (active host galaxy, or another SN in the same host), vicinity of a very bright star, or a specific location on the edge of the focal plane. 

The calibration of light curves on tertiary standard stars obtained with method A is detailed in $\S$\ref{sec:calibration}. Light curves obtained with method B were calibrated in two steps, first with a 4\arcsec radius aperture photometry, then with an aperture to PSF photometry correction. We checked that both methods were equivalent, so that we can safely assume for this comparison that both light curve sets have identical calibration.

In order to compare the flux scale of each individual light curve obtained with the two methods, 
both light curve sets were fitted with a fiducial light curve model and
the resulting light curve amplitudes were compared. In this process, only the photometric points present 
in both reductions were considered to minimise the statistical noise of the comparison 
(a residual statistical noise is expected since different cuts for the selection of images within a night were applied and references for image subtractions vary from night to night for method B).

From this comparison, a systematic over-subtraction of the galaxy was
identified in the data set obtained with PSF photometry on image
subtractions (method B; about 2\% of the galaxy flux at the SN
location, estimated with the PSF of the image with best IQ).
This bias was traced to a systematic normalisation offset of
the convolution kernel: its integral was forced to unity 
after the current and reference images were
photometrically aligned using 4$\arcsec$ aperture photometry of field
stars. This approach 
turned out to be incorrect. As the impact on SN fluxes is a linear
function of the host galaxy flux at the location of the
SN, we resorted to pursue the comparison, focusing on SNe in faint
hosts, and correcting approximately for the identified bias.
 We account for this effect (for comparison purposes only) by fitting a linear relation between the difference of peak magnitudes obtained with both pipelines and the galaxy to SN flux ratio ($f_{gal}$):
$m_A - m_B = \Delta m + \epsilon \times f_{gal}$.

We exclude from this fit noisy light curves ($S/N<50$ in \gmega\rmega\imega\ and 20 in \zmega, where the $S/N$ is given by the uncertainty on the fitted light curve amplitude) and assign a 20\% uncertainty on  $f_{gal}$  which is not well defined (in addition to the statistical uncertainty on the SN flux that enters in the ratio).  
The results are listed in Table~\ref{table:photom_comparison}. 
For \gmega\rmega\imega\ bands, the residual dispersion of order of 0.025 mag can be attributed to imperfect PSF/kernel modeling in the pipelines. The larger dispersion in \zmega-band of 0.05 mag is due to a much lower signal to noise ratio and fringe residuals in the images (note that we do not use exactly the same images in both pipelines so that the Poisson photon noise impacts the comparison). 
The average offsets from 0.002 or 0.009 mag depending on the pass-band are marginally significant, the correction for the host galaxy over-subtraction of the order of 0.01 magnitude on average for all bands being only approximate.

\begin{table}
\begin{center}
\begin{tabular}{ccccc}
\hline
\hline
band & $\Delta m$ & $\epsilon$ & RMS for $f_{gal}<0.3$\\
\hline
\gmega & $-0.003 \pm 0.005$ & $-0.046 \pm 0.007$ & 0.028 \\ 
\rmega & $-0.006 \pm 0.004$ & $-0.025 \pm 0.003$ & 0.023 \\ 
\imega & $-0.009 \pm 0.004$ & $-0.010 \pm 0.002$ & 0.025 \\ 
\zmega & $-0.002 \pm 0.010$ & $-0.028 \pm 0.005$ & 0.054
\end{tabular}
\caption{Average difference of the fitted peak magnitudes obtained with methods A and B after the correction  for the host galaxy over-subtraction ($\Delta$), slope of the correlation with $f_{gal}$ ($\epsilon$), and RMS for high S/N SNe in faint galaxies (see text for details).
\label{table:photom_comparison}}
\end{center}
\end{table}

\begin{table}
\begin{center}
\begin{tabular}{cc}
\hline
\hline
band & $<f_\mathrm{zero}/f_\mathrm{max}>$\\
\hline
\gmega & $-0.0007 \pm 0.0008$ \\
\rmega & $-0.0003 \pm 0.0008$ \\
\imega & $-0.0014 \pm 0.0012$ \\
\zmega & $\phantom{-}0.0066 \pm 0.0037$
\end{tabular}
\caption{Ratio of measured flux before 22 restframe days (before maximum) to the
fitted lightcurve amplitude, averaged over events, for method A.
\label{table:average_zero_flux}}
\end{center}
\end{table}

Light curves obtained with the simultaneous fit (method A) are respectively  13\%, 13\%, 3\% and 16\% less scattered in \gmega\rmega\imega\ and \zmega\ bands than those obtained with photometry on subtractions (method B) with an uncertainty of 2\% for all bands. Those numbers are based on the standard deviation of residuals to a light curve fit per pass-band with a free amplitude, width and date of maximum light.
About half of this difference can be attributed to the relative shallowness of the reference images used for the subtractions.

In order to assess the accuracy of the host subtraction technique of method A, 
we compute the average measured flux before 22 rest frame days
before maximum light ($f_\mathrm{zero}$) for each supernova and band, when available. 
Table \ref{table:average_zero_flux} reports 
the average ratio of $f_\mathrm{zero}$ to the fitted 
light curve amplitude for method A, and does not display any significant bias.

To summarise the comparison of photometric methods:
Method B over-subtracts host galaxies by $\sim$2\%, which we later
traced to the photometric alignment method prior to subtraction.
Both methods agree on the amplitude of light curves (including 
the calibration transfer from field stars to supernovae)
if one concentrates on supernovae hosted in faint galaxies.
Method A is free from host subtraction biases at 1 to 3 mmag accuracy, 
depending on band.

We therefore choose to use the light
curve photometry from the simultaneous fit method (method A) in the subsequent
analysis.

\subsection{Photometric calibration uncertainties}
\label{sec:calibration}

A catalog of calibrated tertiary stars has been obtained for each 
of the four CFHT-LS deep fields by R09 
 (those stars were calibrated against the  \citealt{Landolt92} catalog of secondary $UBVRI$ standard stars).

It consists of a set of \allmegasnfilts\  magnitudes for each
 star in a {\it local natural system} (average airmass of the observations, transmission function at the average position of the star in the focal plane).
The photometric calibration  consists in transfering the raw SN light curves to this system.
For this purpose, a zero point ($ZP \equiv mag+2.5 \log_{10}(flux)$) has to be associated with each light curve.
It is simply obtained by the flux measurements of tertiary stars using the same photometry technique on the same set of images as the ones used to derive the SN light curve.

The light curves obtained with the simultaneous photometric fit
($\S$\ref{sec:simultaneous-fit}) are calibrated as follows. The fluxes
of tertiary stars on the same CCD as the SN are measured with the same
PSF photometry technique as the one used for SNe, the only difference
being that a galaxy model is not considered. Least squares optimality
dictates that the pixel uncertainties used for the fit include the
contribution of the star itself, but we do not do so: 
we do not want the relative weights of pixels to change between supernovae (faint objects) 
and tertiary stars (mostly bright objects) such that PSF innaccuracies affect 
both in the same way. 
So, the pixel weighting
we adopt for bright stars is the one that is optimal for faint objects,
and this choice does not adversly affect the statistical uncertainty of the 
flux ratio of SN to tertiary stars, since the later are
brighter and more numerous.

Since we obtain a flux
measurement for each observation, this allows us to check for the
photometric alignment, and discard variable stars.
Before assigning a zero point to the SN using the average fluxes of stars and their associated magnitudes, the photometric corrections for the reference image, provided by R09, are applied to the stars and SN measurements, depending on their location on the focal plane. At this level, the zero point uniformity within each reference CCD frame is checked. 
 
\begin{figure}
\centering
\includegraphics[width=\linewidth]{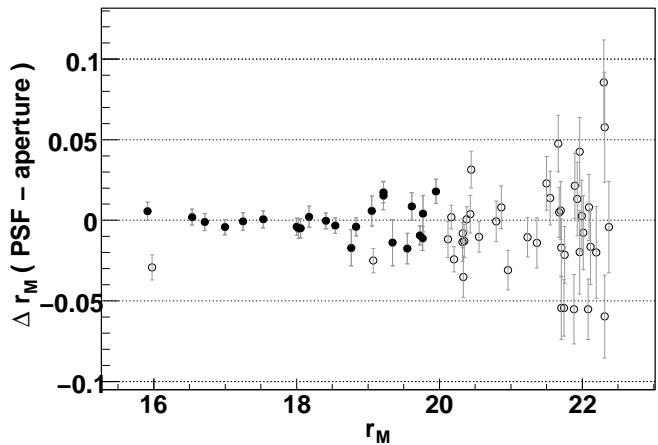}
\caption{Residuals from the calibration fit ( PSF magnitude minus aperture magnitude of the catalog) for CCD \#13, band \rmega\,  in field D1. Each point is a tertiary star. Those marked with open symbols were excluded from the  fit based on cuts on RMS, magnitude, and number of observations. The fit is performed iteratively with a 2.5 $\sigma$ outlier rejection.
\label{fig:residuals-of-psf-calibration}
}
\end{figure}

\begin{figure}
\centering
\includegraphics[width=\linewidth]{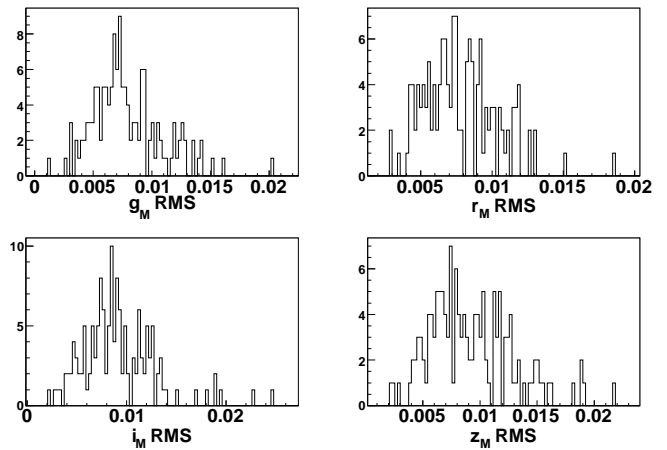}
\caption{RMS of the differences between PSF and aperture magnitude of the tertiary stars for the \allmegasnfilts\ bands. 
In each histogram, there are $36 \times 4$ entries, each corresponding to a CCD/field combination, for which a zero point is determined.\label{fig:rms-psf-calibration}
}
\end{figure}

As an  example, residuals from the calibration  of a central CCD in field D1 for \rmega\ band are shown in Fig.~\ref{fig:residuals-of-psf-calibration}. The RMS of the differences between the PSF magnitudes and the aperture magnitudes (used in R09) for the tertiary stars are shown in Fig.~\ref{fig:rms-psf-calibration}, they are typically of 0.008 mag. With  about 30 stars per CCD on average, the typical statistical uncertainty on the zero point determination of a light curve is of order of 0.002 mag. Since this number is averaged out when combining SNe observed on different fields on different CCDs, it adds a negligible uncertainty to the cosmological analysis.    

\begin{figure}[!h]
\centering
\includegraphics[width=0.8\linewidth]{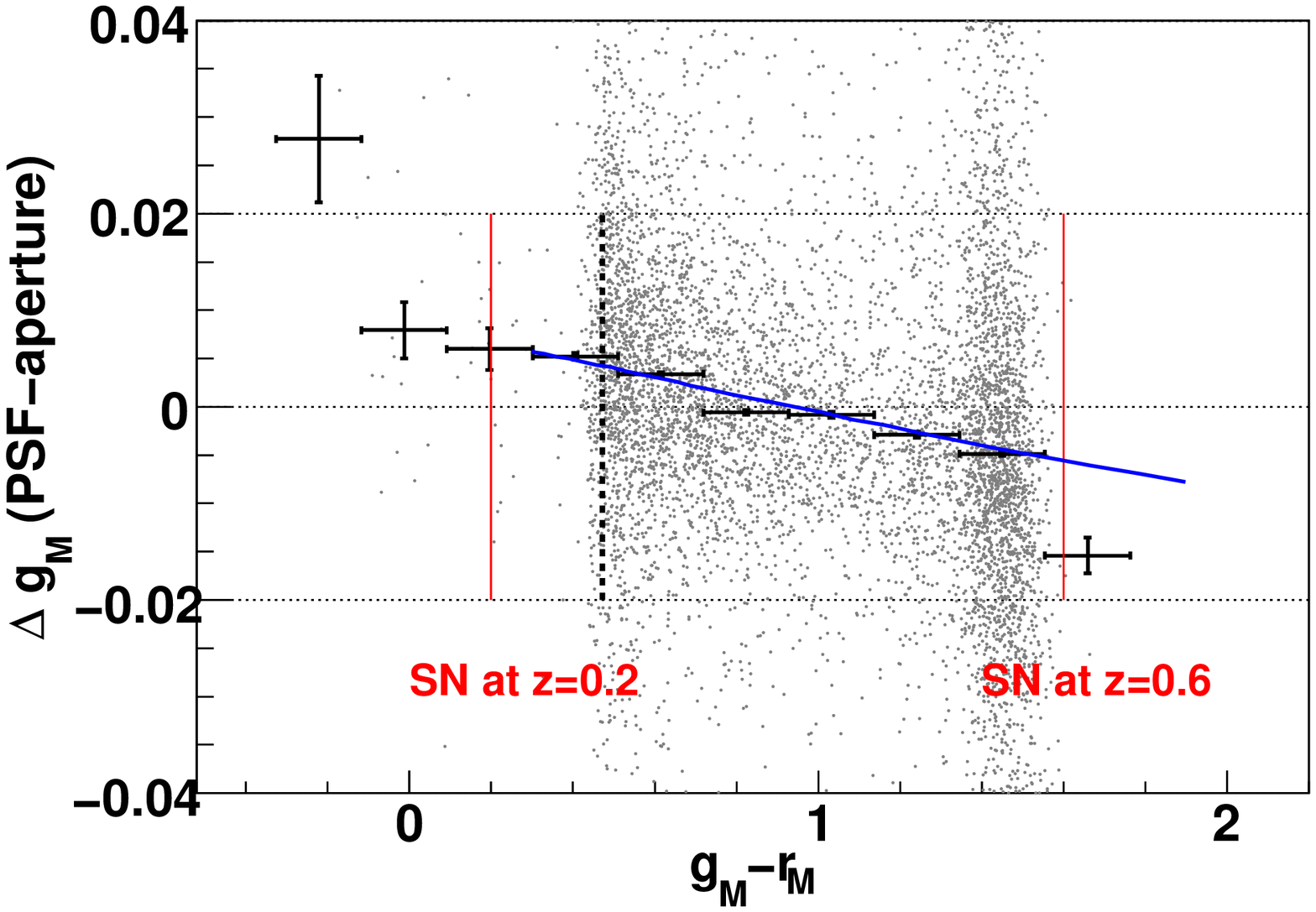}
\includegraphics[width=0.8\linewidth]{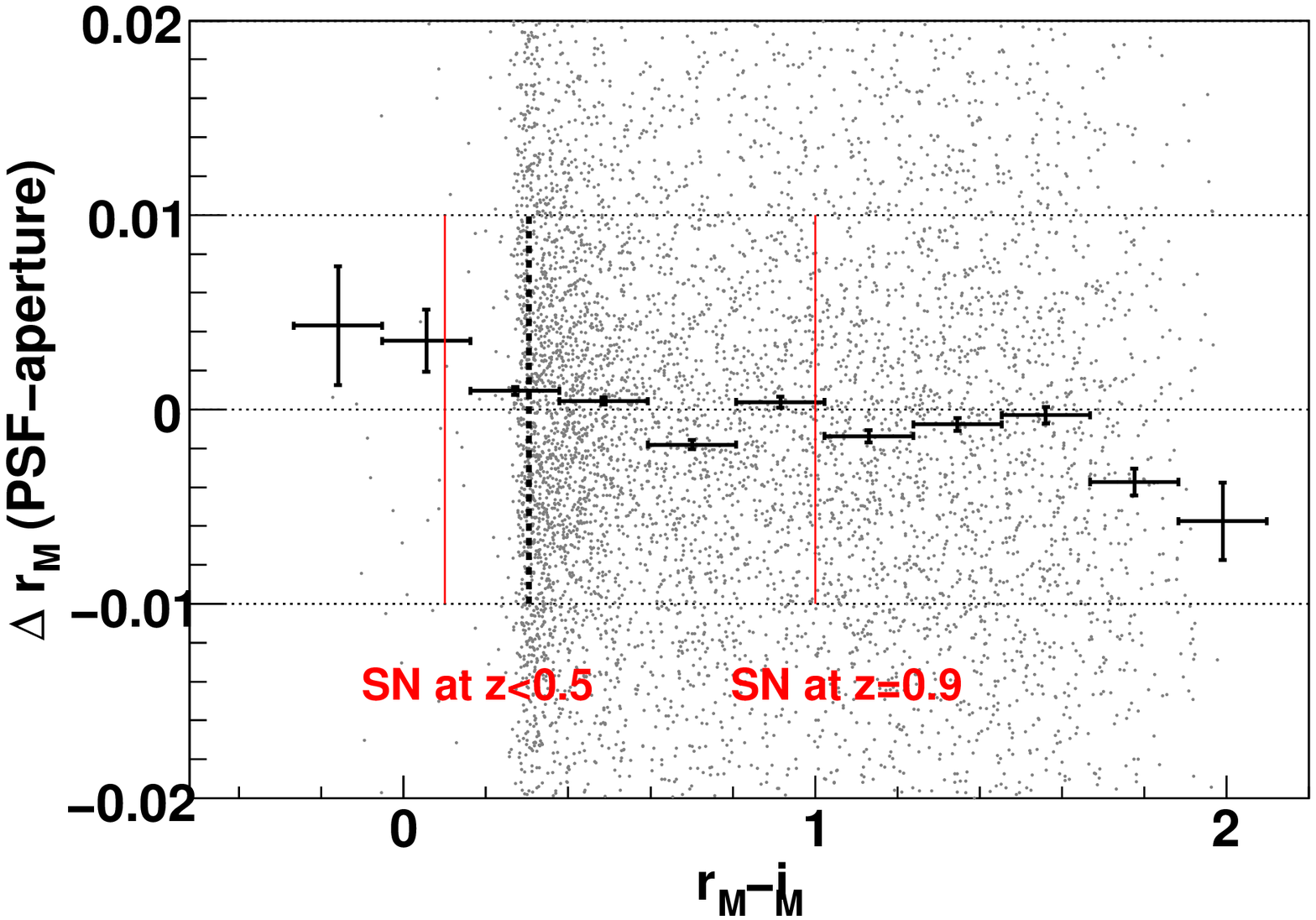}
\includegraphics[width=0.8\linewidth]{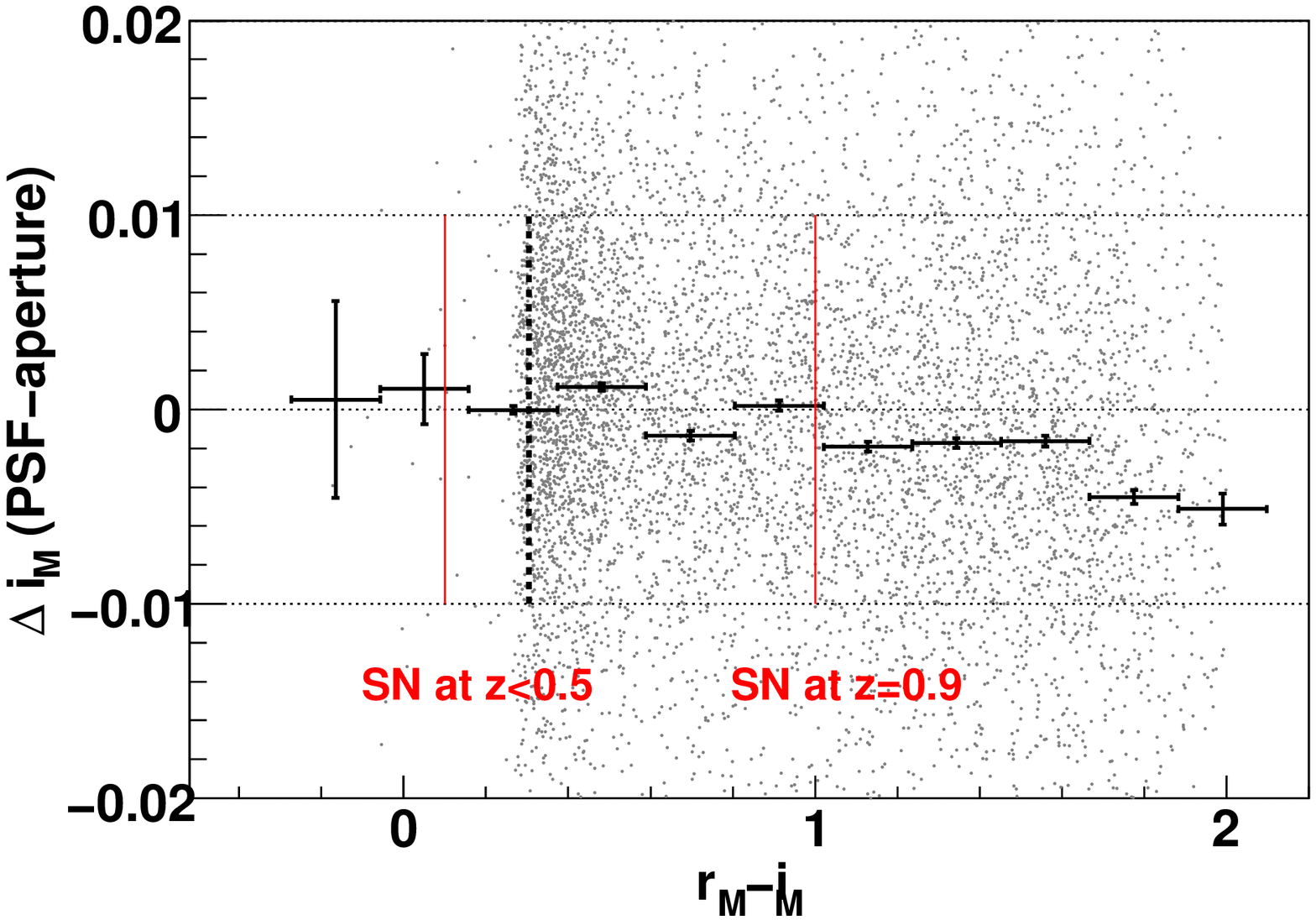}
\includegraphics[width=0.8\linewidth]{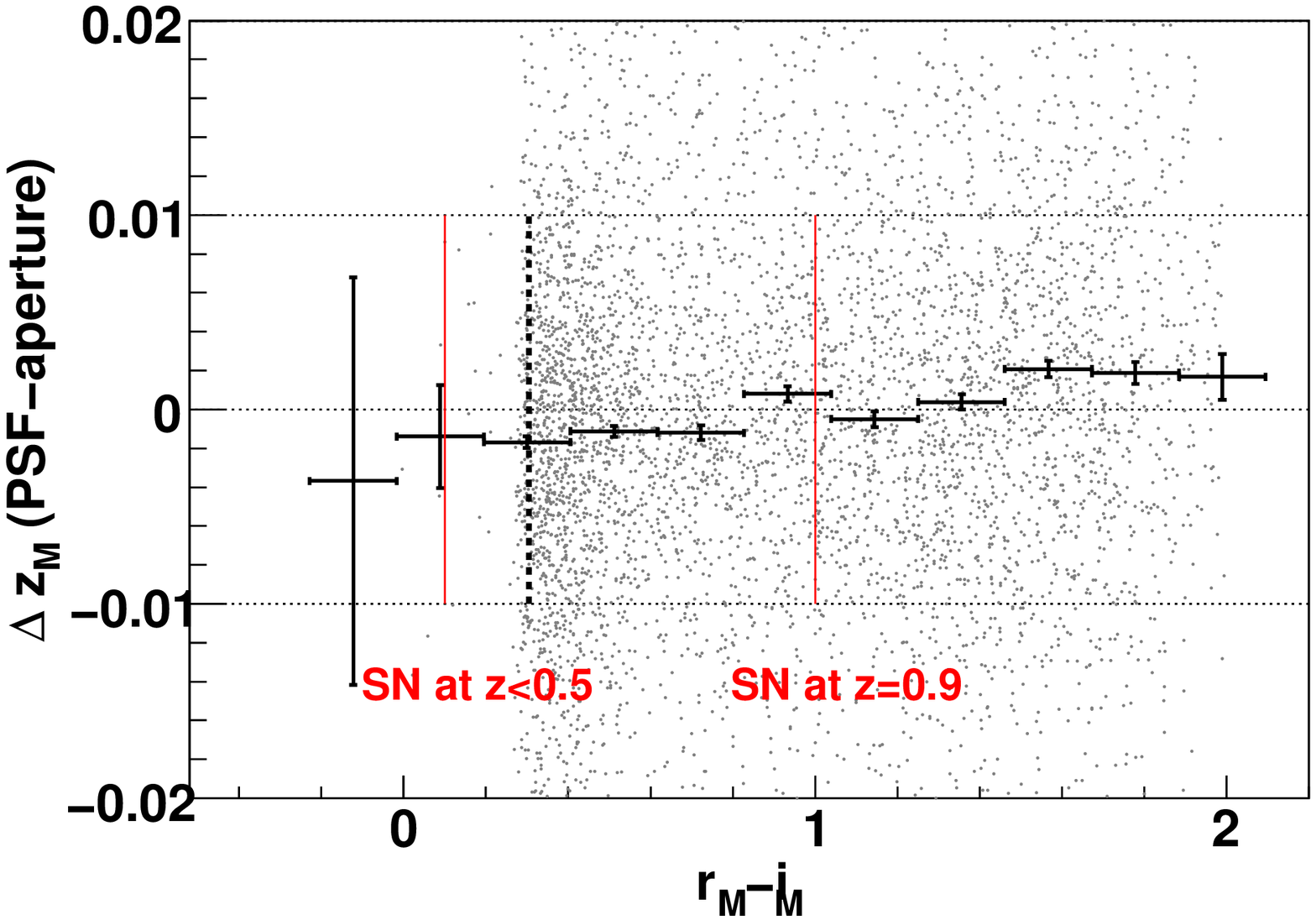}
\caption{Differences between PSF and aperture magnitudes of the tertiary stars in the \allmegasnfilts\ bands as a function of the colour of stars (gray dots). The black points with error bars represent the average deviation and its uncertainty in bins of colour. Typical SNe colours at maximum light are marked with red dotted vertical lines. The blue curve on the top panel shows the effect of the PSF wavelength dependent correction on synthetic magnitudes obtained with PHOENIX stellar models. The colours of BD+17~4708 are marked by the black vertical lines.\label{fig:chromatic-effect}
}
\end{figure}

Chromatic systematic effects were identified in the residual differences between PSF and aperture magnitude as shown in Fig.~\ref{fig:chromatic-effect}. 
These are expected because of the PSF variation with wavelength which induces a colour term in the comparison of PSF and aperture magnitudes (for an achromatic PSF model per pass-band).
At level of a few millimagnitudes, one cannot exclude some chromatic effects in the aperture photometry, which is contaminated by light reflected in the MegaPrime optical system, forming haloes around stars in the images.

In the \rmega\imega\zmega\ bands, the colour dependent offset is of order of 0.002 mag. For such a small effect, we cannot separate it from the residual offset as a function of magnitude (see below) since there is a correlation between the magnitudes and colours of the tertiary stars. We will hence treat this offset as an additional source of systematic uncertainty, rather than try and correct for it.
In \gmega\ band, the effect is however much larger. It is clearly related to the colours of stars. It induces a relative change of calibration of 0.015 magnitude for SNe at redshifts 0.2 and 0.6 (at higher redshifts, \gmega-band light curves are not used to derive distances, since they correspond to observations at rest-frame wavelengths shorter than 3000~$\AA$, see $\S$\ref{sec:lcfitting}). 

In order to account for this PSF variation with wavelength, we have modified the effective transmission of the instrument in \gmega-band with a multiplicative correction depending linearly with wavelength: $corr(\lambda) = 1 + 0.048 (\lambda- 4979~\AA )/(1000~\AA)$. 
 This correction was adjusted on the residuals of Fig.~\ref{fig:chromatic-effect} using synthetic magnitudes of stellar spectra obtained with the Phoenix model \citep[][ and references therein]{Hauschildt97, Baron98, Hauschildt99}\footnote{The study presented here relies on version 2.6.1 of the Phoenix / GAIA spectral library, that can be retrieved from the Phoenix ftp server: {\tt ftp://ftp.hs.uni-hamburg.de/pub/outgoing/phoenix/GAIA}}. The resulting colour dependent effect is shown as a blue curve in Fig.~\ref{fig:chromatic-effect} (for \gmega-band only).

\begin{figure}[!h]
\centering
\includegraphics[width=0.8\linewidth]{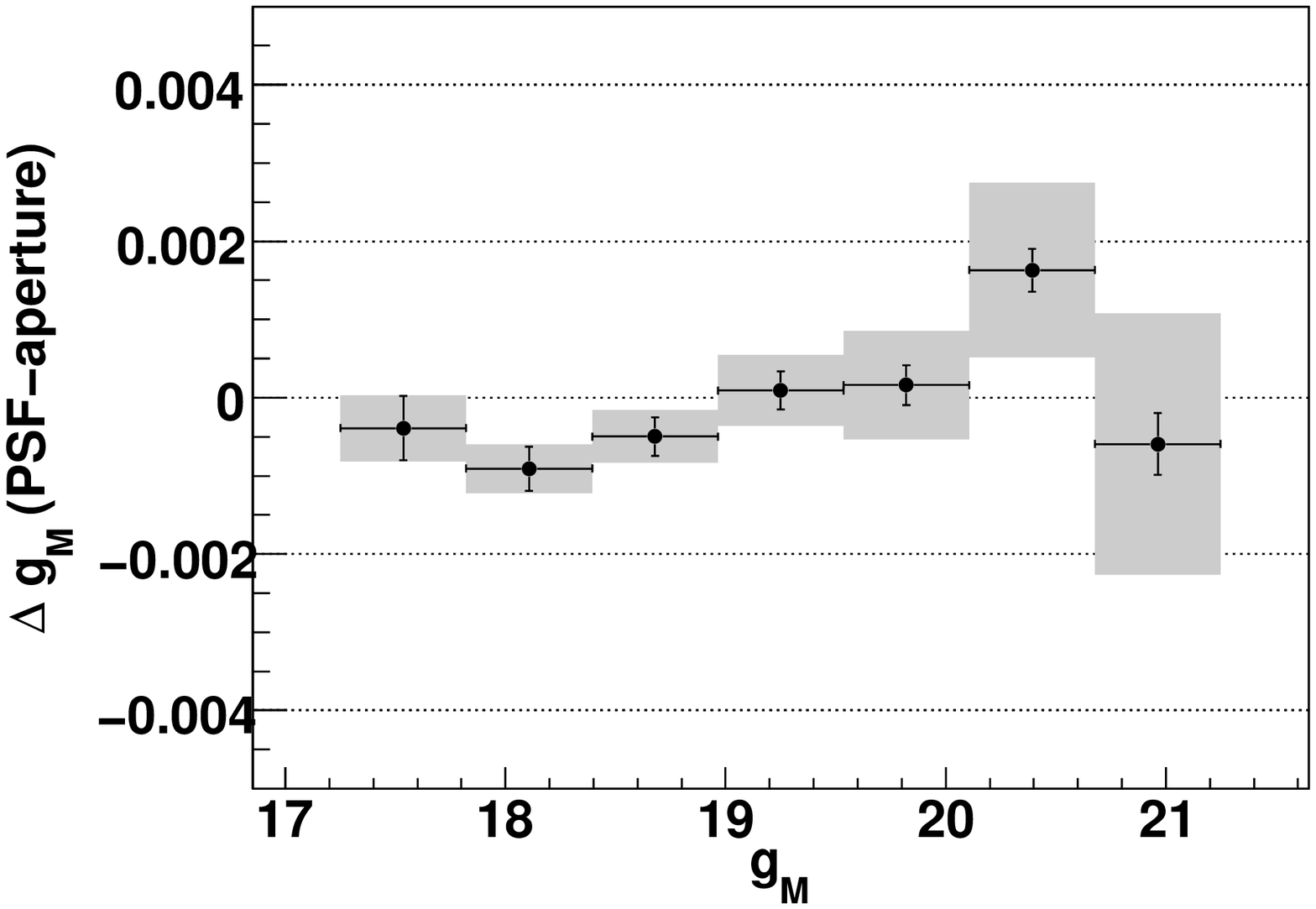}
\includegraphics[width=0.8\linewidth]{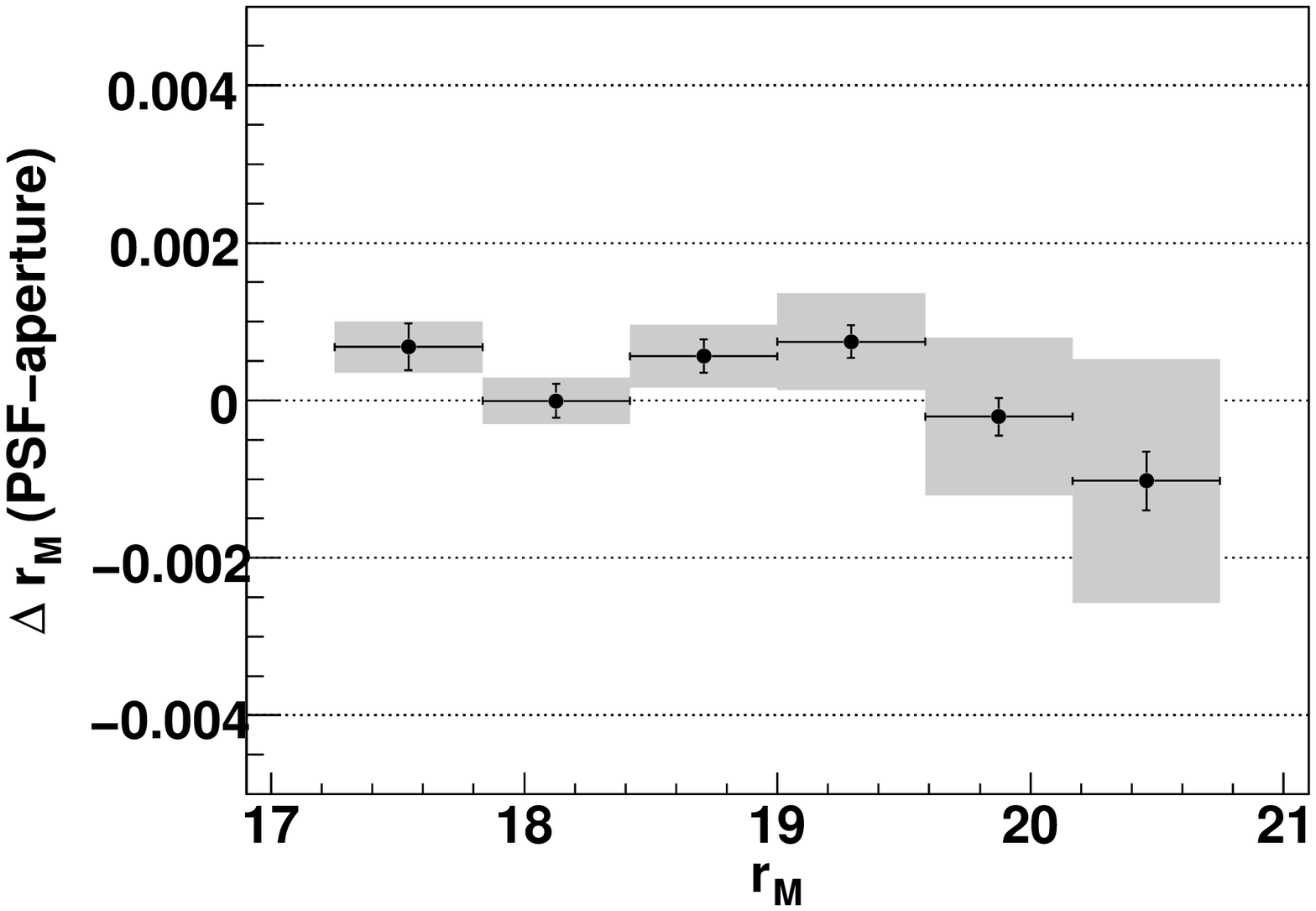}
\includegraphics[width=0.8\linewidth]{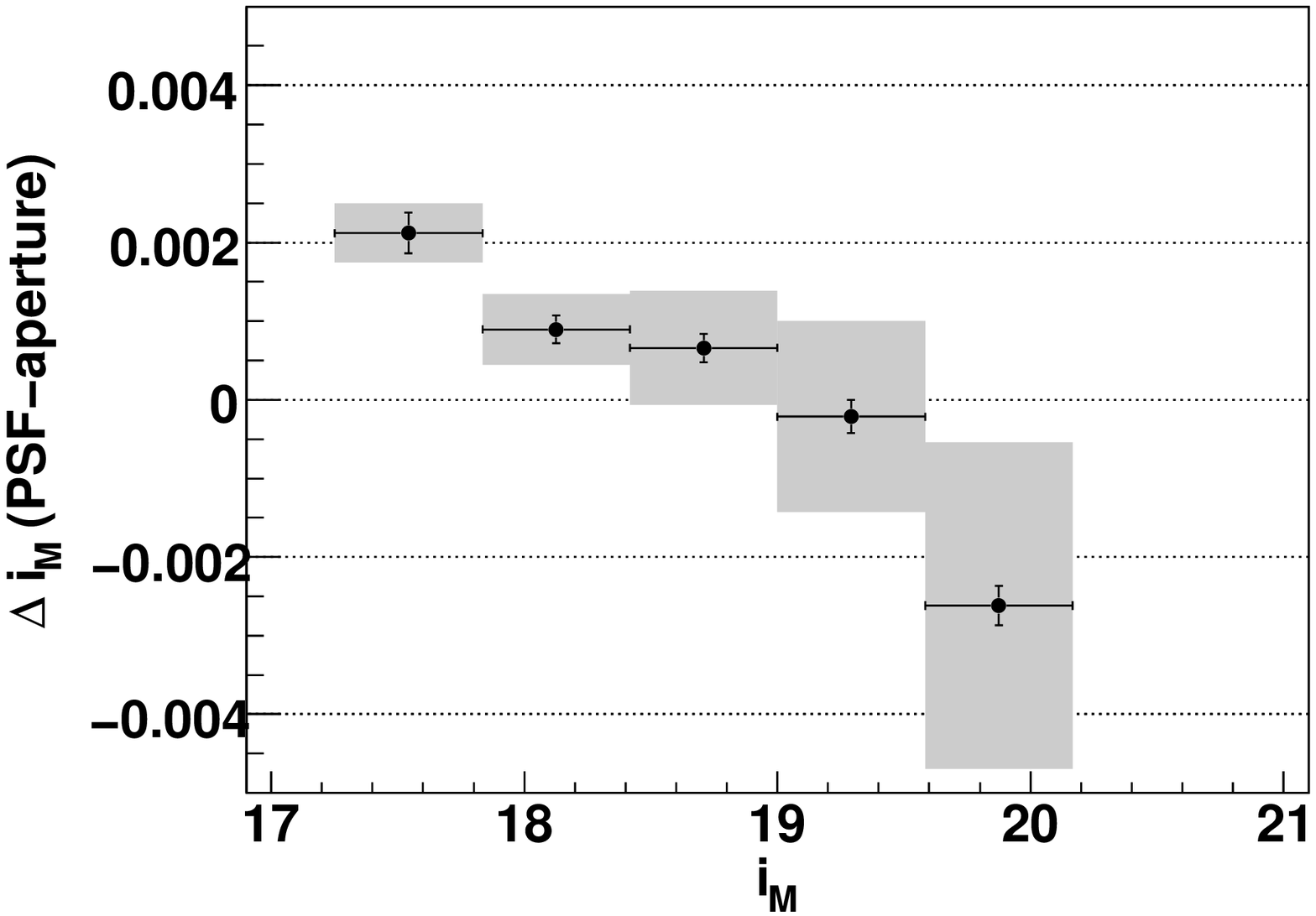}
\includegraphics[width=0.8\linewidth]{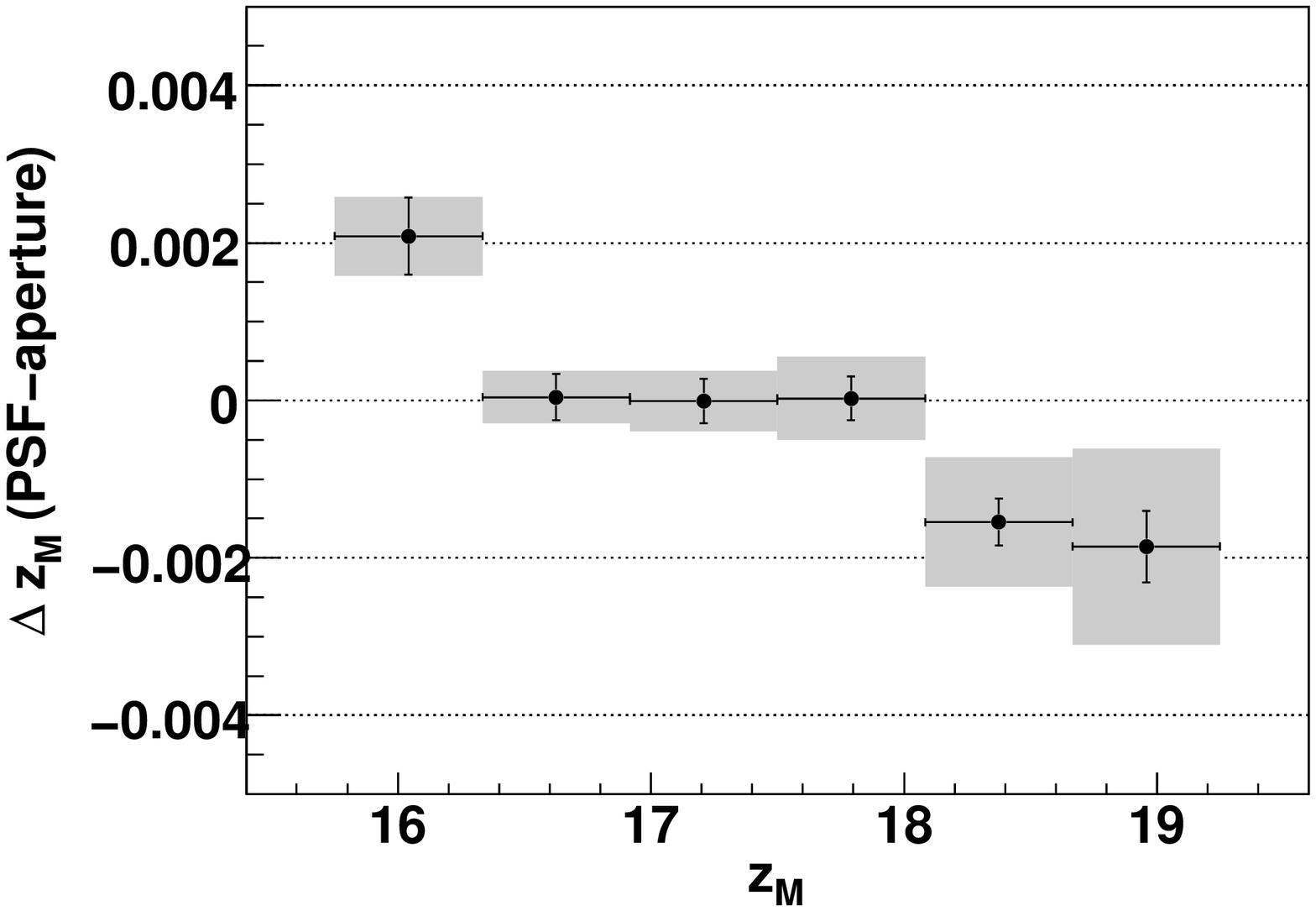}
\caption{Average differences between PSF and aperture magnitude of the tertiary stars in \allmegasnfilts\ bands as a function of the star magnitude (once corrected for the colour dependent terms). The back error bars represent the statistical uncertainty on the average. The shaded areas represent the uncertainty on aperture magnitudes due to a systematic uncertainty on the residual background in the images used for calibration ($0.1$~ADU per pixel for \gmega\rmega\zmega, $0.2$~ADU for the \imega-band)\label{fig:calib-vs-mag}
}
\end{figure}

As a last check of the calibration to tertiary stars, Figure~\ref{fig:calib-vs-mag} presents the average differences between PSF and aperture magnitudes as a function of the magnitudes of the tertiary stars, where PSF magnitudes were corrected for the colour dependent terms. The brightest stars which are saturated on a fraction of the images (depending on image IQ, exposure time, sky transparency and Poisson fluctuation) were excluded from this analysis because of the potential biases introduced by the selection of unsaturated observations (excluding positive statistical fluctuations reaching the saturation level results in a bias on the average). The dimmest stars were also discarded to avoid biases due to the preferential detection of positive Poisson fluctuations of the signal. The range of magnitude considered was obtained by selecting stars on the plateau of Fig. 16 in R09 which presents the number of observations of the tertiary stars as a function of magnitude.
Discrepancies for the dimmest stars are visible: those are consistent
with a residual background in the aperture of the tertiary stars. In 
R09~$\S$4.2, a residual background of order of $+0.06$, $-0.03$, $-0.23$, and
$-0.04$~Analog Digital Units (hereafter ADU) per pixel is found for the \allmegasnfilts\ bands. Assigning 
 a systematic uncertainty of $0.1$~ADU per pixel on this background for \gmega\rmega\zmega, and
 $0.2$~ADU for the \imega-band, leads to the shaded error bars shown in Fig.~\ref{fig:calib-vs-mag}. 
Accounting for this, we assign a 0.002 mag systematic uncertainty on the photometric calibration transfer.

In this section, we have presented the photometry and calibration of the SNLS third year SN sample.
One of the two techniques developed in the collaboration has been selected based on a comparison 
 of their performances. We use this light curve data set in $\S$\ref{sec:results} to determine for each supernova
 the parameters needed to estimate its distance. 

\subsection{Interpretation of the photometric calibration}
\label{sec:using-calibration}

The calibrated data are formally the ratio of the target flux $F_{SN}$ as measured by the detector to that of a reference star $F_{ref}$ that would have been observed in the exact same conditions. 
In practice, we have determined a flux $F_{meas}$ in ADU and a zero point $ZP_{meas}$ which provides a normalisation coefficient.

As we will see in the next section, the supernova light curves are
fit with a spectral sequence model $M_{SN}(\lambda,t,...)$ in order to extract from the light curves
a flux intensity and additional parameters that characterise the diversity of SNe~Ia.
 For this purpose, one has to be equipped with a model of the instrument response as a function of 
wavelength $T(\lambda)$ for each pass-band in order to compare the model to the observations.

The light curve fit consists in comparing the measured 
quantity 
\begin{equation}
F_{meas} \times 10^{-0.4 (ZP_{meas}-m_{ref})} \label{eq:meas_data}
\end{equation}
to the model
\begin{equation}
D_{model} =  F_{SN} /F_{ref} = \frac{\int T(\lambda) M_{SN}(\lambda,t,...) d \lambda}{\int T(\lambda) M_{ref}(\lambda) d \lambda}\label{eq:meas_model}
\end{equation}
where $M_{ref}(\lambda)$ is the spectrum of the reference star and $m_{ref}$ its magnitude. The integrals in Eq.\ref{eq:meas_model} have the dimension of counts in the detector per unit time (i.e. in units of ADU$\,$s$^{-1}$), as they model the actual observations\footnote{As a consequence, if $M_{SN}$ and $M_{ref}$ are spectral energy densities (in erg$\,$cm$^{-2}\,$s$^{-1}\,\AA^{-1}$ for instance), then $T(\lambda)$  is the product of an effective collection area (accounting for all the transmission and reflection efficiencies in the optical path), a dimensionless quantum efficiency, the CCD gain, and the inverse of the photon energy (in $\AA \, $erg$^{-1}$).}. However, Eq.~\ref{eq:meas_model} tells us we do not need to know the normalisation of the instrumental response function, even from pass-band to pass-band, as it cancels in the ratio.

For SNLS, the magnitude system defined in R09 has been anchored to the star \bdtruc\ with a prescription for its magnitudes in this system. As we have photometrically aligned the SN light curves to the catalog of stars defining this system, we use \bdtruc\ as our reference star and consider for its magnitudes those provided by R09: $g_M = 9.6906$, $r_M = 9.2183$, $i_M = 8.9142$, and $z_M = 8.7736$. Note that those magnitudes are conventional, one could have changed them in the definition of the magnitude system as long as the magnitudes of the tertiary stars and hence the values of $ZP_{meas}$ in Eq.~\ref{eq:meas_data} are modified accordingly.
We use for $M_{ref}(\lambda)$ the spectrum of \bdtruc\ as measured on HST by \citet{Bohlin2004b}; we retrieved the latest version of this spectrum on the CALSPEC database\footnote{\tt ftp://ftp.stsci.edu/cdbs/current\_calspec/ \\ bd\_17d4708\_stisnic\_002.ascii}. 

Concerning the instrumental model ($T(\lambda)$), a complication arises from the fact that the transmission of MegaCam filters varies across the focal plane, following a radial pattern, with a typical variation of the central wavelength up to 5~nm from the centre to the edge. 
The magnitude system has been designed so that no colour transformations are required to translate our SN measurements to this system;  
the small residual airmass-dependent colour corrections cancel out on average and do not contribute significantly to
 the uncertainty of the measurements, and the  variation of the effective pass-bands within the dithering pattern of the DEEP field observations are negligible. 
But the  magnitude system has to be interpreted using a model of the response function of the instrument that follows this radial pattern. We use the transmissions provided by R09 to integrate the spectral model of supernovae, using the average position of the SNe observations on the focal plane.
Note also that the magnitudes of \bdtruc\ have been chosen to be the same everywhere in the MegaPrime focal plane despite the fact that the filter transmissions vary.

The SNe Ia light curves of the SNLS 3-year sample are available on-line at the {\it Centre de Donn\'ees astronomiques de Strasbourg} (CDS). An example of the published data is given in Table~\ref{table::photometry}.

\section{Modeling the supernova light curves}
\label{sec:lcfitting}

\subsection{General considerations}
\label{sec:lcfitter_considerations}

The only two light curve parameters that have been found so far to correlate with the luminosity of SNe~Ia are the width of their light curves (first measured in Johnson B-band, see e.g.~\citealt{Phillips93}, and then extended to other bands) and their colour as measured, e.g., by the difference of magnitudes (or ratio of fluxes) in rest-frame $B$ and $V$ bands.
Once corrected for those, the absolute maximum luminosity has a dispersion of the order of 15\%.
All recent high-redshift cosmological analyses use those parameters (luminosity, light curve shape and colour) in a more or less obvious way to derive distances~\citep{Riess04,Riess07,Astier06,WoodVasey07} with the notable exception of the CMAGIC technique which relies on a colour-magnitude diagram of SNe~Ia~\citep{Wang03,Conley06}. 
The challenge of a cosmological application is to derive those parameters with a minimal redshift dependent bias.

\subsubsection{An empirical modeling}

The goal of the light curve fitting is then to evaluate for each SN those parameters from observations performed with a limited set of observer-frame filters and a limited cadence of observations.
This requires a model of the spectral sequence of the SN in order to interpolate among observations.
Despite the fact that 
there is a broad consensus as to the basic physical picture of the explosion mechanism, 
it is extremely difficult currently to make quantitative predictions for the observed signal based on a physical model. Indeed extremely precise 3D modeling is required in order to simulate the flame propagation in the SN progenitor.
As a consequence, an empirical modeling of the observables is needed.
Historically, light curve templates were built in a limited set of filters from a sample of nearby SNe (see for instance \citealt{Goldhaber01}). This required a correction of the observations for redshifted supernovae, usually called $K$-correction~\citep{Nugent02}.
Those were performed using an average spectral sequence based on a set of spectra obtained at different phases (days after maximum light) of the SN. This method is applied for the MLCS2k2 light curve fitter~\citep{Jha07}, with tabulated $K$-corrections as a function of phase, redshift and colour. 
More recently, techniques based on an explicit modeling of the spectral sequence have been developed. The data are not corrected to rest-frame filters but directly compared to the integral of the spectra in a model of the instrumental response (SALT(2):~\citealt{Guy05,Guy07}, hereafter G07, SiFTO:~\citealt{Conley08}, hereafter C08). The advantage of this approach is to keep track of the correlations between the light curve shape, colours and the spectral properties in the fitting process. 

\subsubsection{Impact of a limited training sample: using high-z SNe}

This light curve fitting technique is a fundamental ingredient of the cosmological analysis.
Especially, the assumed broad-band colour relations (i.e. the relative amplitude of the SN spectral model at a wavelength scale of order of $1000\AA$, beyond a simple colour tilt, in other words the curvature of the spectra) in the wavelength range of validity of the model have a direct impact on the derived distances. In order to illustrate this, let us consider two SNe observed in \rmega\ and \imega\ band at redshifts of 
 0.5 and 0.8. Since those \rmega\ and \imega\ observations correspond to the rest-frame B and V,  and U and B bands respectively, the ratio of distances derived for those two SNe is directly a function of the $(U-B)-(B-V)$ colour\footnote{The SNe colours mentioned in this section are always considered at maximum light. Note however that the estimates of the colours resulting from a light curve fit are actually an average of the difference between the data and the model with a weight that varies with the phase of the observations.} difference of the model. Since all light curve models are empirically derived from a limited training set, this latter colour 
 has an uncertainty which introduces a redshift-dependent correlation among the derived supernova distances (see for instance~\citealt{Knop03}).
Since many more SNe have been observed at high-redshift than at low redshift, high-z SNe must be considered in the training of the light curve models in order to overcome the statistical limitations of the nearby sample. This has been done with the SALT2 and SiFTO models; it was possible since both techniques do not make use of distances in their training process.
 
\subsubsection{Modeling of the near UV emission}

High-z SNe allow the observation of the rest-frame near UV emission from the ground 
without the need of space telescopes.
The near UV is modelled in SALT2 and SiFTO using SNLS photometric (in \gmega\ and \rmega\ bands up to a redshift of 1) and spectroscopic observations (see references in Table~\ref{table:spectro}). Using near UV data allows for a drastic improvement of the colour and hence distance estimate for SNe at redshifts of order of unity, where the sensitivity of the rest-frame B and V is limited by the quantum efficiency drop of MegaCam CCDs in the \zmega\ band. However, we still lack spectroscopic observations at early and late phases (the primary goal of the SNLS spectroscopic program was to provide an identification of the SNe which is easier at maximum brightness).

\subsubsection{Diversity of SNe~Ia colours: intrinsic variation or absorption by dust}

There is still much debate about the treatment of the SN colour parameter (generally anchored to $B-V$
at maximum light). 
Whereas all cosmological analyses based on SNe perform a linear correction of distance moduli (i.e. logarithm of distances) with the measured colour,
the value of the coefficient used and its interpretation differ significantly from one analysis to another.
In A06, this coefficient $\beta$ is marginalised over in the cosmological fit, without any
 attempt to separate the reddening effect of dust absorption or a potential intrinsic variation.
On the contrary, the MLCS2k2 technique used in ESSENCE~\citep{WoodVasey07}, GOODS~\citep{Riess04,Riess07} and SDSS surveys ~\citep{Kessler09}, assumes that the derived $(B-V)$ colour offset  primarily comes from extinction by dust, and therefore that the $\beta$ parameter should be identified with the $R_B$ value of the \citet{Cardelli89} extinction law. When $\beta$ is fit at the same time as cosmology, values ranging from about $2$ to $3$ are found depending on the technique used to derive it. Those values are systematically smaller than the value of $4.1$ in the \citet{Cardelli89} extinction law.

The large range of values obtained for $\beta$ is likely to be a
consequence of different assumptions on the uncertainties of the
$(B-V)$ colour estimates (and to a lesser extent the colour range of
the SNe sample considered).  For a given data set, the larger the
assumed uncertainties on $(B-V)$, the larger the fitted $\beta$ value.
This issue is raised by~\citet{Freedman09}; we come back to it in
$\S$\ref{sec:beta-evolution}.  Whereas in previous papers (including
papers from the SNLS collaboration: A06, G07, C08), low values of $\beta
\simeq 2$ were found, we get larger values when accounting for an
intrinsic scatter in SNe colour relations in this paper (see
$\S$\ref{sec:residual-scatter} and $\S$\ref{sec:beta-evolution}).
Fitting for $\beta$ or not has some consequences.  For instance,
\citet{Conley07} have shown that either we live at the centre of an
under-dense region of the Universe as proposed by \citet{Jha07}, or
the relation between SN colours and luminosity does not follow the one
expected for the Galactic extinction and $\beta < R_B$.

This low value of $\beta$ points to either an unusual extinction
law in host galaxies of SNe~Ia or an intrinsic colour variation that
dominates the effect of extinction.  One hint is that the colour
variation law (which describes how the SN flux varies with colour as a
function of wavelength) can be derived from the SN data themselves,
and differs significantly from the \citet{Cardelli89} extinction law
in the near UV and U-band, even for extreme values of
$R_B$~(see~\citealt{Guy05,Guy07}, and
Fig.~\ref{fig:colorcorrection}). In SiFTO, the derived relation
between the $(U-B)$ and $(B-V)$ colours of SNe can not be explained
with an extinction law either.  While there is not yet a definitive
proof that the colour variation we observe is intrinsic to the SN, we
still have to relax the assumption that it is purely due to dust
extinction as modelled by \citet{Cardelli89}. This has some
consequences for the cosmological analysis. Indeed, applying an
incorrect correction to luminosity introduces a redshift dependent
bias since the average colour of SNe varies with redshift because of
Malmquist bias (bluer SNe are brighter and hence dominant near the
detection limit of a survey). This occurs at the highest redshifts of
all surveys but also for nearby SNe that were observed by other means
(see e.g. \citealt{Conley07}).

In the MLCS2k2 approach, a colour excess $E(B-V)$ is measured as the difference between the observed colour and that of the model. In this model, the intrinsic variability of SNe is addressed with a single parameter ($\Delta$), and any possible additional intrinsic variation is unaccounted for. As a consequence, one expects that, at some level, the $E(B-V)$ estimate resulting from the light-curve fit combines both dust reddening and a possible residual intrinsic colour variation. 
There are several examples of SNe~Ia being clearly extinguished by dust. In contrast, there is no proof that part of the $(B-V)$ colour variation is driven by intrinsic SNe properties. The only hints come from observations which point to a value of $\beta < R_B$ and a colour variation law incompatible with a standard dust extinction law. Nevertheless, as the physical mechanism responsible for the SN~Ia explosions obviously involves more than a single parameter (composition of the progenitor, ignition conditions, ...), it is very possible that some properties of the SN impact on the observed $(B-V)$ colour beyond that addressed by $\Delta$. 

As an example, \citet{Kasen2009} present a class of supernovae simulations with various ignition conditions, deflagration to detonation transition (DDT), and viewing angles, which exhibit intrinsic variations of colour. The models with the same DDT criterion present a range of $(B-V)$ colours (at maximum light) that are not fully correlated with the decline rate\footnote{We selected models with the DDT criterion 3 in Table 2 of \citet{Kasen2009} (supplementary material) and found  a residual $(B-V)$ dispersion of 0.03 after correcting for a marginal correlation with decline rate. The model spectra were kindly provided to us by the author.}. Such an effect, which is present in those simulations, has not been ruled out on real data.

It is further possible that the slope of the colour-luminosity relation evolves 
as a function of redshift. A larger range of dust opacities is expected
 at higher redshifts where galaxies host younger stellar populations, so that the
 relative weight of dust extinction and intrinsic variation in the colour-luminosity relation 
should evolve with redshift.
 \citet{Kessler09} found an evolution of the $\beta$ parameter with redshift
 using SALT2, but with an opposite trend with respect to the expectations (they found a lower $\beta$
 at higher $z$). We study this issue in $\S$\ref{sec:beta-evolution}.

\subsection{Comparison of light curve fitters and distance estimate}
\label{sec:fitter-selection}
Using different light curve fitters to estimate distances allows us to
quantify the systematic uncertainties associated with this step of the
analysis. As shown in C10 and to a lesser extent in
$\S$\ref{sec:om-measurement} of this paper, the combined uncertainties
on cosmological parameters are dominated by systematics, so the
statistical precision of light curve fitters do not play a major role
in their selection. For a cosmology application, the most crucial aspects are
those that may lead to a redshift-dependent bias: i)
$K$-corrections uncertainties, ii) any bias that could arise when
fitting low signal to noise light curves, iii) biases associated with
selection effects.

Several authors have already performed this comparison of light curve
fitters. The largest systematic effects were found by
\citet{Kessler09} (hereafter K09), who applied MLCS2k2 and SALT2 to a
sample combining low redshift SNe, HST data and the first release of SDSS-II,
ESSENCE and SNLS data. They found a difference on $w$ of
0.2 when using MLCS2k2 or SALT2. This difference is larger that any
other source of uncertainty. The authors have looked for the sources of the observed discrepancy (see $\S$11 in K09 for a detailed discussion). For this purpose, light curves were fit with modified versions of MLCS2k2 designed to replicate some elements of the SALT2 model. 
 They explain the discrepancies by two main differences between the light curve fitters: differences in the light curve templates, and the use of colour priors. We now review these two effects, and 
discuss afterwards the differences in the treatment of colour. We eventually
summarise this comparison section.

\subsubsection{Systematic differences in the light curve templates}
\citet{Kessler09} provide evidence that SALT2 and MLCS2k2 
predict differently the rest-frame $U$-band. 
Excluding rest-frame $U$-band light curves from the fit of SDSS SNe at $z>0.21$ changes the MLCS2k2 and SALT2 distance moduli by $\simeq 0.13$ and $\simeq 0.07$ magnitude respectively\footnote{We consider here the relative change of distance moduli with respect to SNe at $z<0.21$}. Note that the offset found for SALT2 is marginally consistent (at the $2\sigma$ level) with the calibration systematic uncertainties of order of 0.01 magnitude for all bands reported both for the SDSS-II and SNLS first year releases (the SALT2 version used in K09 was trained on SNLS first year data).

The SALT2 and SiFTO light curve fitters are more reliable than MLCS2k2 in rest-frame $U$-band because they were calibrated using high $z$ supernovae. They benefit from the SNLS precise calibration and are not sensitive to the systematic uncertainties that are known to affect the observer-frame $U$-band calibration of low redshift SNe. In particular, the effective $U$-band response function is poorly known for many SNe observations because of important variations of the atmospheric transmission at wavelengths shorter than 350~nm. Also, when fitting with SALT2 or SiFTO (see $\S$\ref{sec:residual-scatter} and $\S$\ref{appendix:salt2-errors}), we find a larger dispersion of residuals in the $U$-band for low redshift SNe (0.1 mag, RMS) than in  the \gmega\ band at $z \simeq 0.4$ where \gmega\ roughly corresponds to rest-frame $U$ (0.05 mag). The systematic offset found by K09 could possibly be affected by some evolution of SNe properties in the UV. There is no evidence for such an effect, but even in this case, using SNe at $z \simeq 0.4$ to calibrate the model rest-frame $U$-band is more reliable than using SNe at lower $z$. Indeed, the model rest-frame $U$-band is primarily used to estimate distances of SNe at high redshift, and since a monotonous change of the average SNe UV emission with redshift is expected due to an evolution of metallicity or age of the progenitor population, $U$-band measurements done at the highest possible redshift should therefore be used.

\subsubsection{Using priors on SNe colours.}

In the MLCS2k2 approach, a prior is used that forces the extinction $A_V$ to be positive.
This is equivalent to applying a prior on colour excess.
However, since there is no evidence that the observed variation of SNe~Ia colours is entirely and solely due to extinction by dust, applying such a prior is not justified. 

Furthermore, even if the model was qualitatively correct, applying a prior is non optimal since any bias on this prior will produce a bias on cosmology and hence artificially large systematic uncertainties. For instance, more than half of the difference on $w$ between the analysis of \citealt{WoodVasey07} and that of K09 is due to the choice of prior (see $\S$10.1.4 in K09). Alternatively, not using a prior on colours does not bias the estimate; one only obtains larger and reliable statistical uncertainties.

\subsubsection{Estimating distances}

SALT2 and SiFTO use the colour in their distance modulus whereas
MLCS2k2 uses a colour excess. This might be regarded as conceptually
different, but, as shown below, the distance moduli used in the two approaches are mathematically equivalent, as long as the slopes of the brighter-slower and colour-luminosity relations are fit to the Hubble diagram.

MLCS2k2 defines the colour excess as $E(B-V) = (B-V)_{obs} - (B-V)_{model}$. At first order in $\Delta$, the shape parameter, 
$(B-V)_{model} = a + b \times \Delta$ and the distance modulus can then be written (at first order in shape and colour parameters): 
\begin{eqnarray}
\mu_{MLCS}  &=& m_B - M_B + \alpha \, \Delta - R_B \, E(B-V) \nonumber \\
&=& m_B - M_B + (\alpha + R_B \, b ) \, \Delta - R_B \, (B-V)_{obs} + R_B \, a \nonumber \\
&=& m_B - M_B' + \alpha' \, shape - \beta \, \col \nonumber
\end{eqnarray}
where $shape$ is any shape parameter and $\col=(B-V)_{obs}$ or any affine function of it.
This last expression is exactly the one used in A06 and in this paper (Eq.~\ref{eq:distance-modulus}). 
As a consequence, choosing to use a colour excess or directly the observed colour to parametrise the distance modulus only modifies the meaning and 
values of $M_B$ and $\alpha$ coefficients, and does not change distances.

So at first order, when no prior is used on colour and $R_B$ is treated as a free parameter ($\beta$), the MLCS2k2 approach is contained in the SALT2/SiFTO as both the colour variation law (or colour relations, see $\S$\ref{sec:sifto-description}) and the slope of the colour-luminosity relation of SALT2/SiFTO are parametrised in a way that allows the model to match a standard dust extinction law.
 Disentangling the contribution of intrinsic colour variation and dust extinction 
requires adding parameters beyond the simple one-parameter description of the SN intrinsic variation implemented in MLCS2k2, SiFTO and SALT2 ($\Delta$, stretch, $X_1$). 
As a consequence, when using these fitters, 
the observed colour luminosity relation could evolve with redshift  as it combines two effects whose relative weight is expected to change with redshift. We investigate this in $\S$\ref{sec:beta-evolution}. 

\subsubsection{Summary of the light curve fitter comparison}

We have shown above that using a colour or colour excess in the distance estimator is not the source of the differences between SALT2/SiFTO and the MLCS2k2-like approaches. MLCS2k2-like parametrisations assume a Cardelli-like colour variation law and in some cases a colour-luminosity relation, while the SALT2 or SiFTO approach allows one to derive these quantities from the data itself. The values obtained for these quantities differ significantly from the assumptions made in MLCS2k2. In addition, colour priors are used in the MLCS2k2 approach, which we think are unnecessary and subject to additional systematics. 

To estimate systematic uncertainties arising for the light curve fitting process, we therefore chose to fit all SN light curves using both SALT2 and SiFTO models as described in the following section. 
     
\subsection{Fitting with SALT and SIFTO models}

We use SALT2 and SiFTO to fit light curves and derive three parameters for each SN: an amplitude conventionally described by the peak rest-frame $B$-band magnitude $m^*_B$, a shape parameter, and a colour labeled ``$\col$'' that roughly corresponds to the rest-frame $B-V$ colour at maximum light. The two models that were fit on low-z and SNLS data minimise the modeling statistical uncertainty and cover a large wavelength range (300--700~nm) well suited for the SNLS data set. For both fitters, the light curves are obtained by integrating a spectral energy density (SED) varying with phase in a model of the instrumental response function. 
The effective colour parameter that is derived is due to a combination of extinction and intrinsic variation, and this has to be kept in mind in the subsequent usage of those parameters to derive distances. 
Those two techniques differ substantially in their detailed parametrisation of observables and in the procedures considered for training and light curve fitting. As a consequence, in the following, we use both in order to estimate the impact on cosmology of the modeling choices that were made. Before comparing the result of the fits obtained on the SNLS sample, we present both approaches, along with the minor modifications that were applied since the publication of the methods in G07 and C08. The two approaches are compared and combined in $\S$\ref{sec:results}.

\subsubsection{SALT2}

\begin{figure}
\centering
\includegraphics[width=\linewidth]{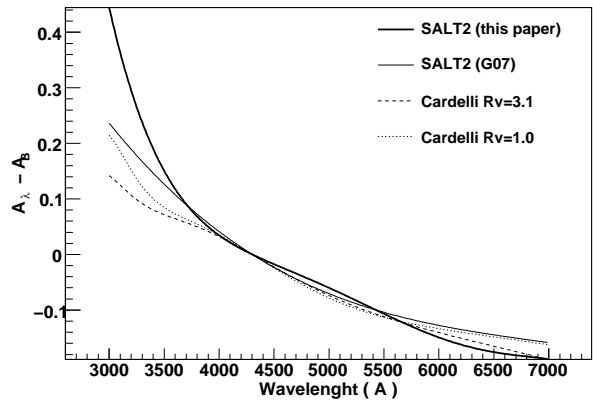}
\caption{SALT2 colour variation law for $\col=0.1$ (thick solid curve),  along with the SALT2 colour variation law of G07 and \citet{Cardelli89} extinction law for $E(B-V)=0.1$ and two different values of $R_V$ (1.0 and 3.1 for the dotted and dashed blue curves).
\label{fig:colorcorrection}
}
\end{figure}

The SALT2 method consists in modeling the SNe~Ia spectral energy density (SED) variation with time and its diversity using a linear combination of several principal components multiplied by the exponential of a colour dependent function of wavelength (which we will call ``colour variation law'', although it can model a pure extinction law).
This model is trained on a large sample of nearby SNe listed in Table~\ref{table:salt2_nearby_training_sample} and a sub-samble of the SNLS SNe~Ia listed in Table~\ref{table:spectro}. We use only SNLS SNe with redshifts $z\le 0.7$, with an unambiguous spectroscopic identification (excluding the $Ia\star$) and a good light curve sampling (see $\S$\ref{sec:sampling-requirements}).
Both light curves and spectra are used in the fit.
Since significant calibration uncertainties are expected for most spectra (due for instance to slit losses), wavelength dependent corrections for each spectrum are included in the model.
The model parameters, the calibration correction coefficients of the spectra, and the SNe parameters are fitted simultaneously. Only two principal components have been considered; one that represents the average SN~Ia, and one which can be identified as the shape variation of light curves of SiFTO. 
This training sample is larger than the one considered in~G07. 
A few technical modifications, detailed in Appendix~\ref{appendix:salt2}, have been applied to the training procedure: higher resolution for the components and the colour variation law, a new regularisation scheme, improved handling of the residual scatter about the model, and propagation of the model statistical uncertainties.
The colour variation law is shown in Fig.~\ref{fig:colorcorrection} for $\col=0.1$ (i.e. for a $B-V$ colour excess of 0.1). It differs significantly from the \citet{Cardelli89} extinction law for wavelengths $\lambda < 370$~nm even when extreme values of $R_V$ are considered. Adding more parameters to the colour law with respect to G07 has resulted in a steeper variation at short wavelength.

For the fit of SNe entering the Hubble diagram, all light curves for which the effective wavelength of the associated response function lies in the model validity range (300--700~nm rest-frame) are fit simultaneously. 
 The actual fit is a simple least square minimisation taking into account both the  covariance matrix of the flux measurements and a model uncertainty (described in $\S$\ref{appendix:salt2-errors}). Since the uncertainty on the model depends on the parameters that are fitted, 
 the fit is performed iteratively with updates of the model uncertainties at each step until convergence is reached.
The model parameters and the code to use them are public; see $\S$\ref{sec:salt2-release} for details.

\subsubsection{SiFTO}
\label{sec:sifto-description}
\begin{figure}
\centering
\includegraphics[width=\linewidth]{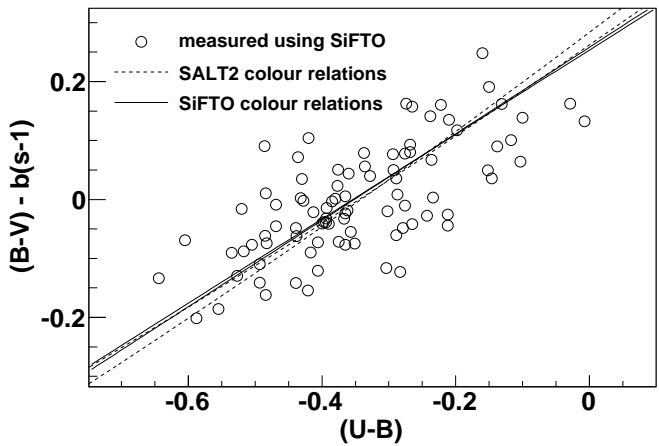}
\caption{$\bmv$ corrected for stretch as a function of $\umb$ 
for a selected sample of SNLS SNe as measured by SiFTO (open circles) along with the linear relations of
Eq.~\ref{eq:colrelation} and Table~\ref{table:sifto-color-relations} ($a$ and $c$ terms, solid lines) and the synthetised ones of SALT2 (dashed line) for the two colour scatter hypothesis. SALT2 does not contain explicitly colour relations as SiFTO, but the $a,b$ and $c$ coefficients can still be derived from the model.\label{fig:colrelations}
}
\end{figure}

The SiFTO model consists of a SED sequence whose time evolution has been
 calibrated on a large SNe~Ia sample combining low-$z$ and SNLS data, starting from the spectral sequence derived by~\citet{Hsiao07}. 
The light curve shape variability is modelled with a time-stretching of the SED sequence about the date of maximum light in rest-frame $B$-band, with a wavelength-dependent stretch factor ($s$) indexed by its value in $B$-band.
The broad-band wavelength dependent calibration of the SED sequence is not performed at this stage of the model building.
Indeed, the derivation of the amplitude and colour parameters is obtained in two steps: 
i) Contrary to SALT2, the amplitude of each light curve is a fit parameter, leading to an observer-frame peak magnitude for each pass band. 
ii) These peak magnitudes are then used to adjust the SED at maximum-light
with a smooth multiplicative function of wavelength so that the flux integrated in any rest-frame filter can be evaluated. 
Up to 5 rest-frame filters are used --
$U_{02}UBVR$\footnote{$U_{02}$ is an artificial filter defined as $U$
  blueshifted by $z=-0.2$} -- resulting in 4 colour combinations of
$\utwomb$, $\umb$, $\bmv$ and $\vmr$, with the limitation that only the rest-frame filters within 450\AA\ of an observer-frame filter are considered. 
The second step consists in converting these colours into an estimate of 
$\bmv$, denoted $\bmv_{\mathrm{pred}}$, using equations of the form
\begin{equation}
\label{eq:colrelation}
\bmv_{\mathrm{UBpred}}=a\left[\umb + 0.5\right] + b\left(s-1\right) + c
\end{equation}
with  similar equations relating
$\bmv$ with $\utwomb$ and $\vmr$. 
The data used to derive the linear relation of the $\bmv$ vs $\umb$ relation is shown in Fig.~\ref{fig:colrelations}. 
The estimation of the $a,b,c$ parameters entering in these relations depends on the assumed intrinsic scatter of the SNe colours, this is discussed in $\S$\ref{sec:residual-scatter}.

The final SiFTO colour parameter $\col$ is determined by a weighted
combination of each $\bmv_{\mathrm{pred}}$ together with any actual
measurement of $\bmv$, degrading each $\bmv_{\mathrm{pred}}$ by the dispersion measured for each relation. The uncertainties in the derived colour
relations introduce correlations between SNe that are propagated in the analysis.

Since C08, SiFTO has been retrained with a much larger sample of SNe including a sub-sample of the SNLS light curves presented in this paper (with stringent requirements on the time sampling) and many more at low redshift. It also uses an updated version of the SED sequence from \citet{Hsiao07} that has been extended in the near infra-red.

As a conclusion to the description of the techniques, SiFTO does not
contain an explicit colour variation law as a function of wavelength, 
nor a broad-band colour
calibration of the SED sequence as in SALT2.  These two pieces of
information, which are essential for distance estimate, are coded into
the linear colour relations of Eq.~\ref{eq:colrelation}.  Also, the
light curve stretching behaviour is a prior of the SiFTO
parametrisation, whereas it is only approximately realised as an
outcome of the SALT2 training.  Indeed the second principal component
of SALT2 turns out to be close 
to a derivative of the first one with respect to a stretch factor, hence
mimicking a time stretching at the first order.
As a consequence, based on these differences, we do not expect to get
exactly the same light curve shapes so that the peak magnitudes and
colours obtained with both fitters may have small systematic offsets
and a stretch dependence.  Some additional dispersion due to the differences 
in the data weighting in the least
square minimisation is also anticipated.

\subsection{Residual scatter}
\label{sec:residual-scatter}
The SALT2 and SiFTO empirical models do not fully account for the diversity of SNe~Ia. As a consequence,
 some intrinsic scatter about the best fit model is expected on top of the measurement uncertainties.
 The wavelength dependence of this residual scatter has to be evaluated as it affects the determination
of some of the model parameters. As an example, for SiFTO, the fitted slope of the relation between $\umb$ and $\bmv$
depends on whether the scatter about the linear relation is attributed to the estimate of $\umb$ or $\bmv$. The same effect applies to the determination of the SALT2 colour variation law.
Since a bias on this slope (for SALT2, this slope is encoded in the second derivative of the colour variation law in the $UBV$ wavelength range) leads to bias on the average colour of SNe at high redshift where blue SNe are preferentially selected, we have to determine the amplitude of this scatter as a function of wavelength.

\begin{figure}[h]
\centering
\includegraphics[width=\linewidth]{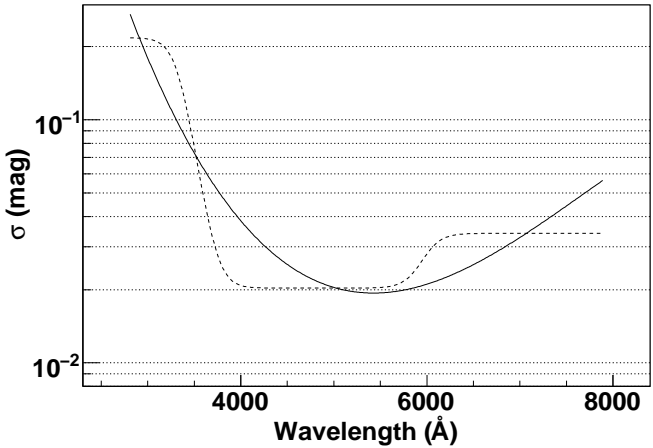}
\caption{SALT2 training sample broad-band colour dispersion as a function of wavelength for two different functional forms (solid curve:~EXPPOL, dashed curve: SIGMOID).
\label{fig:salt2_broadband_colour_dispersion}
}
\end{figure}

This cannot be obtained in the SiFTO framework where only two colours are compared at a time to derive the model parameters. On the contrary, it is possible with SALT2 since it assumes a single relation between all the peak magnitudes of a given supernova (up to five for low redshift SNe: $UBVRI$). Note however that with the SiFTO method, this unique relation between all observations is restored in the final colour estimate when the various $\bmv_{\mathrm{pred}}$ are combined. The technical details of the determination of the residual scatter are given in $\S$\ref{appendix:salt2-errors}. Using two parametrisations of the scatter as a function of wavelength (the exponential of a polynomial, or a combination of sigmoids, labeled  respectively hereafter ``EXPPOL'' and ``SIGMOID'', see Eq~\ref{eq:colorscatterfunction}), one obtains the two estimates of the residual scatter shown in Fig.~\ref{fig:salt2_broadband_colour_dispersion}. This scatter model is translated into scatter in the reference bands used for SiFTO colour relations in Table~\ref{tab:colour_relation_scatter}. The values of the SALT2 residual scatter at the central wavelength of the $U_{02}BVR$ filters is used. The assumption that this scatter on magnitudes is uncorrelated between bands translates into correlated uncertainties in the colours used for the SiFTO colour relations.  
When accounted for in the determination of the colour relations, this leads to two sets of $a,b,c$ parameters for each colour relation that are listed in Table~\ref{table:sifto-color-relations}.

In this determination of the residual scatter, we  identified an additional scatter of 0.1 mag in observer $U$-band for low redshift SNe (see also C10 for a discussion about this issue). This is significantly larger than the dispersion obtained in the same rest-frame wavelength range with SNLS data (given by \gmega\ band observations at $z \simeq 0.35$, see Table~\ref{tab:colour_relation_scatter}). We hence attribute this additional dispersion to calibration uncertainties or varying effective filter transmissions from one $U$-band light curve to another. We de-weighted those light curves accordingly in the training of SALT2, and subsequently in the light curve fits. As for SiFTO, they were discarded for the fit of the $\umb$ $\bmv$ colour relation and were not used for the determination of the colour parameters.

\begin{table}
\begin{tabular}{l|ccccc}
 & $U_{02}$ & $U$ & $B$ & $V$ & $R$ \\ 
\hline
EXPPOL & 0.179 & 0.063 & 0.029 & 0.019 & 0.025 \\
SIGMOID & 0.215 & 0.048 & 0.020 & 0.020 & 0.034 
\end{tabular}
\caption{
Residual scatter in SiFTO reference pass-bands for the two parametrisations of the scatter of Eq~\ref{eq:colorscatterfunction}.\label{tab:colour_relation_scatter}}
\end{table}

\begin{table}
\begin{minipage}[c]{0.49\linewidth}
\begin{center}
EXPPOL 
\begin{tabular}{ll}
\multicolumn{2}{c}{$\bmv$ vs $\utwomb$} \\
\hline
$a$ & $= +0.186 \pm 0.032$ \\
$b$ & $= +0.371 \pm 0.134$ \\
$c$ & $= -0.112 \pm 0.015$
\\
\multicolumn{2}{c}{$\bmv$ vs $\umb$}  \\ 
\hline
$a$ & $= +0.734 \pm 0.040$ \\
$b$ & $= +0.323 \pm 0.059$ \\
$c$ & $= -0.109 \pm 0.009$
\\
\multicolumn{2}{c}{$\bmv$ vs $\vmr$} \\
\hline
$a$ & $=  +1.755 \pm 0.114 $ \\
$b$ & $=  +0.411 \pm 0.061 $ \\
$c$ & $=  -0.885 \pm 0.060 $
\\
\end{tabular}
\end{center}
\end{minipage}
\begin{minipage}[c]{0.49\linewidth}
\begin{center}
SIGMOID
\begin{tabular}{ll}
\multicolumn{2}{c}{$\bmv$ vs $\utwomb$} \\
\hline
$a$ & $= +0.225 \pm 0.032$ \\
$b$ & $= +0.443 \pm 0.134$ \\
$c$ & $= -0.130 \pm 0.015$
\\
\multicolumn{2}{c}{$\bmv$ vs $\umb$}  \\ 
\hline
$a$ & $= +0.716 \pm 0.040$ \\
$b$ & $= +0.333 \pm 0.059$ \\
$c$ & $= -0.104 \pm 0.009$
\\
\multicolumn{2}{c}{$\bmv$ vs $\vmr$} \\
\hline
$a$ & $=  +2.011 \pm 0.114 $ \\
$b$ & $=  +0.473 \pm 0.061 $ \\
$c$ & $=  -1.018 \pm 0.060 $
\\
\end{tabular}
\end{center}
\end{minipage}
\caption{Parameters of the SiFTO colour relations on $(B-V)$ obtained with the two residual scatter estimates of Table~\ref{tab:colour_relation_scatter}.\label{table:sifto-color-relations}}
\end{table}

Figure~\ref{fig:colrelations} presents the difference between SiFTO and SALT2 colour relation between $\umb$ and $\bmv$. Part of the differences of slopes can be explained by the use of different samples to derive those relations, and part can be attributed to the intrinsic differences in the methods used to derive those relations. Still, one can see that they have the same {\it average} estimate of $(B-V)$ given $(U-B)$, so we do not expect this to have a significant impact on cosmology, as we will see in the next section.

\subsection{Light curve sampling}
\label{sec:sampling-requirements}
We investigate in this section the reliability of the light curve parameters
 determined on poorly sampled light curves, in particular those which lack photometric measurements 
 before the date of maximum light (hereafter $T_{max}$). 
Any bias on $T_{max}$ induces a bias on the parameters needed to estimate a distance (magnitude, shape and colour), and such a bias may arise if the modeling  
 of the decline of the light curves is not sufficiently accurate. 
This is not really an issue for the SNLS sample presented in this paper as only  2\% of the SNe lack photometry in the rising part of the light curves (between  -10 and -1 days with respect to $T_{max}$, rest-frame), but it has to be studied for the selection of external samples (see C10) for which this fraction is much larger (especially at low redshift because of a follow-up triggered by the spectroscopic identification).

For the purpose of this study we select well sampled SNe data with high $S/N$ from SNLS (using only those at $z<0.4$) and from external samples: SNe at $z<0.1$ from various surveys (see references in C10) and SNe at $0.06<z<0.3$ from the SDSS~\citep{Holtzman08}. We compare the light curve parameters $\theta \in \{m^*_B, X_1, \col, \mu \}$  derived from the full light curve fits -- hereafter $\theta_F$ -- to those resulting from fits ignoring a fraction of the data points. For each SN, we derive several estimates of $\theta$ removing sequencially an increasing number of early points, and index each of those estimates by  $\tau_f = (T_{first}-T_{max})/(1+z)$ the phase of the first remaining data point.
Of course for real data $\tau_f$ is not directly observable and we have to rely on $\widetilde{\tau_f}$, an estimate of $\tau_f$  which results from the fitted value of $T_{max}$. We study the average offsets $\Delta \theta = \theta(\tau_f) - \theta_F$ as a function of $\widetilde{\tau_f}$. 
One first has to correct for a trivial bias (showing up even for an unbiased estimator) that is due to the finite width of the $\tau_f$ distribution. For instance, if we consider an input distribution with $\tau_f>0$, we will find biased $\Delta \theta$ for negative values of $\widetilde{\tau_f}$ simply because we are looking at events for which $\widetilde{\tau_f}$ is systematically underestimated (and hence $T_{max}$ over-estimated).
Within each bin around a given value of $\widetilde{\tau_f}$, we have an average bias on a parameter $\theta$ given by 

\begin{eqnarray}
 \Delta_{t} \theta (\widetilde{\tau_f}) &=& \frac{ \int_{u=\widetilde{\tau_f}-\epsilon}^{\widetilde{\tau_f}+\epsilon} \frac{1}{n} \sum_{i=1}^n   \left[ u-\tau_f^i \right ] \, \partial_\tau \theta \, p(u|\tau_f^i) d u }{\int_{u=\widetilde{\tau_f}-\epsilon}^{\widetilde{\tau_f}+\epsilon} \frac{1}{n} \sum_{i=1}^n  p(u|\tau_f^i) d u } \nonumber \\  
&\simeq& \frac{ \sum_{i}  \left[ \widetilde{\tau_f}-\tau_f^i \right]  \, \partial_\tau \theta  \, p(\widetilde{\tau_f}|\tau_f^i) } { \sum_{i}  p(\widetilde{\tau_f}|\tau_f^i) } \nonumber
\end{eqnarray}
where the sum is on the number of tests performed (indexed by $\tau_f^i$), and $p(\widetilde{\tau_f}|\tau_f)$ the likelihood of the estimator $\widetilde{\tau_f}$ knowing the true value $\tau_f$ ($\epsilon$ is half the bin size).
$\partial_\tau \theta = -(1+z) \, \partial \theta / \partial T_{max}$ is the derivative of $\theta$ with respect to $\tau_f$. 
Such a bias can be estimated for each value of $\widetilde{\tau_f}$ and corrected for assuming a Gaussian likelihood (with a $\sigma$ given by the redshift corrected measurement uncertainty on $T_{max}$).
We will consider in the following $\Delta' \theta = \Delta \theta - \Delta_{t} \theta$  to study potential biases.

\begin{figure}
\includegraphics[width=0.49\linewidth]{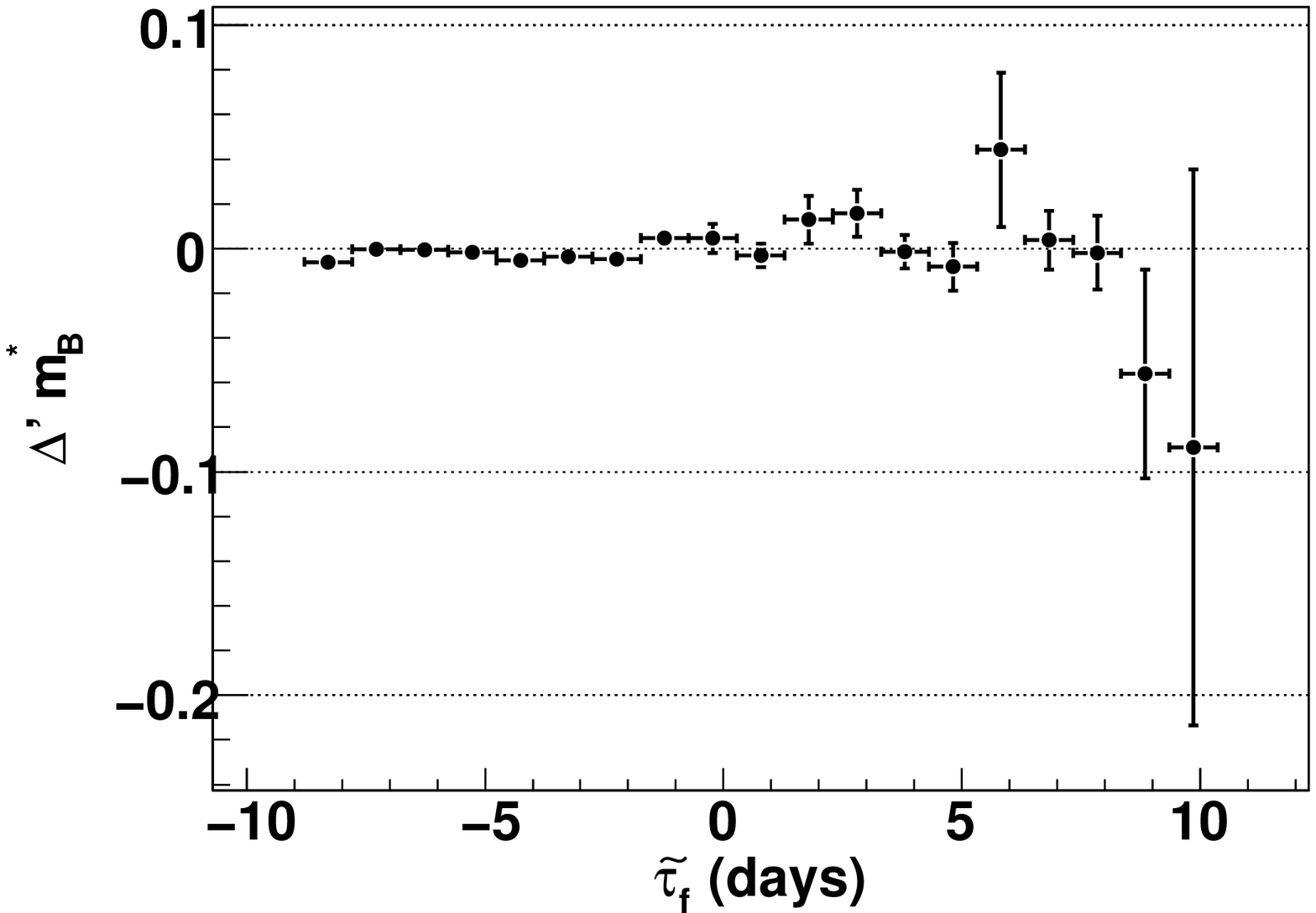}
\includegraphics[width=0.49\linewidth]{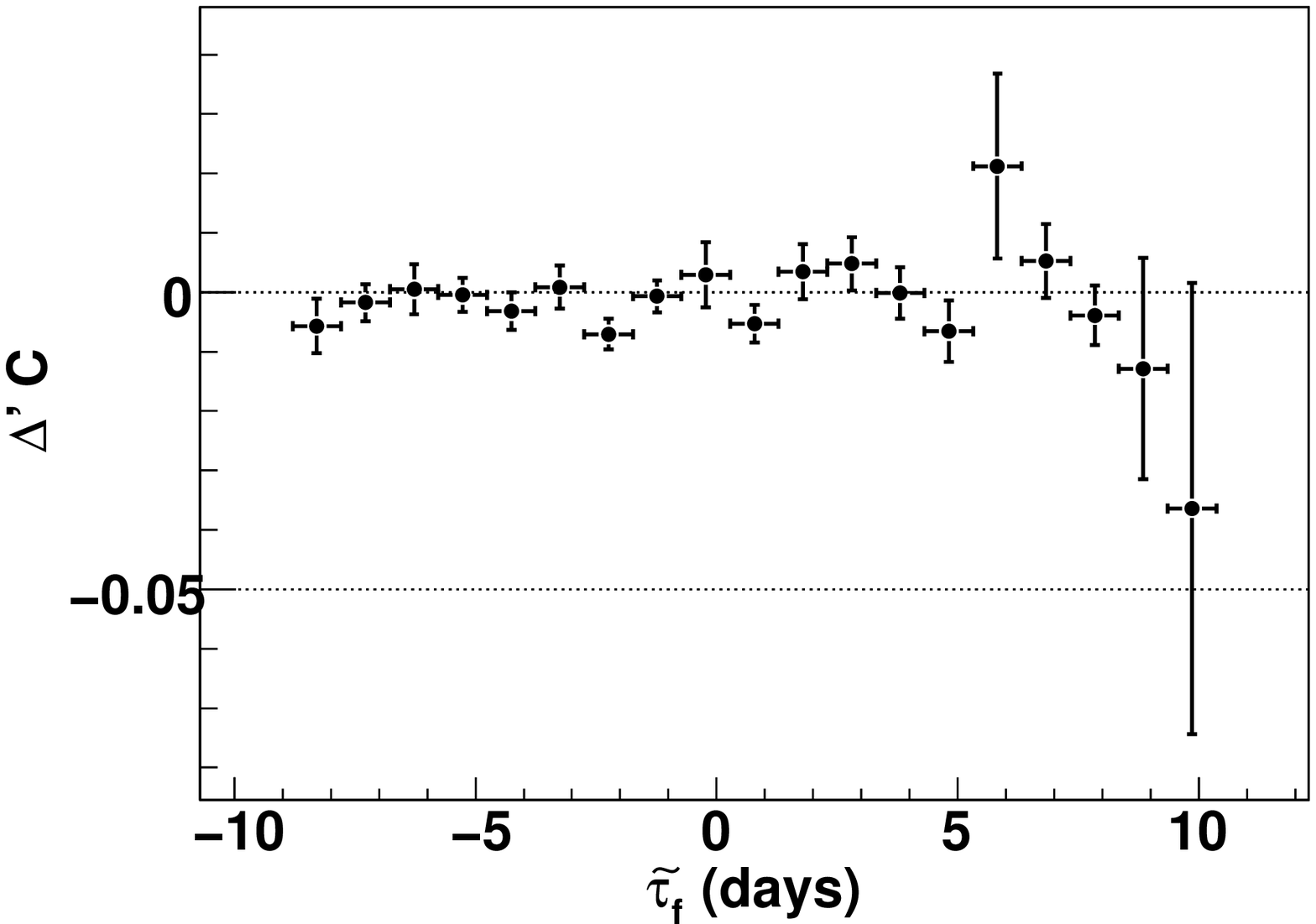}
\includegraphics[width=0.49\linewidth]{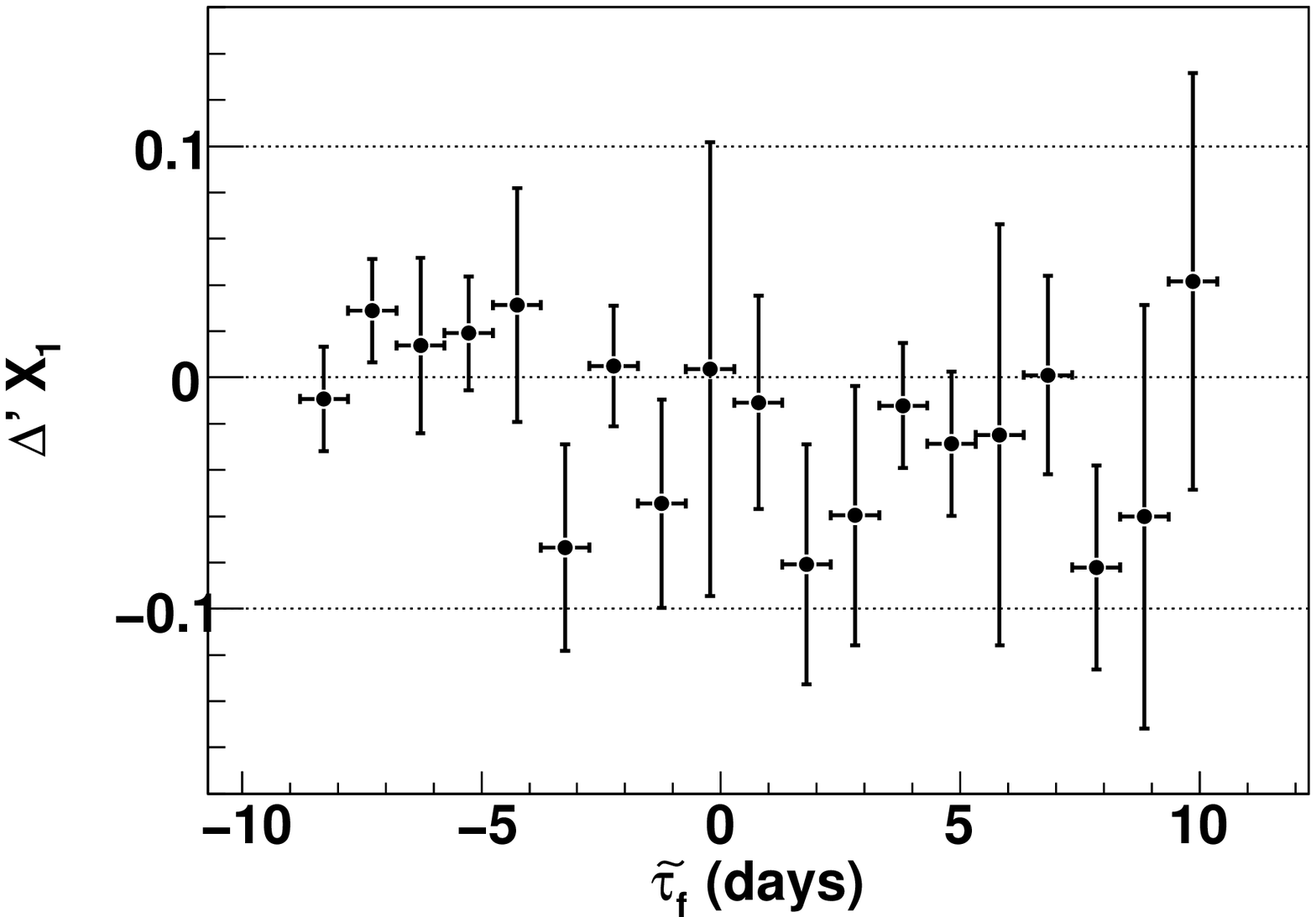}
\includegraphics[width=0.49\linewidth]{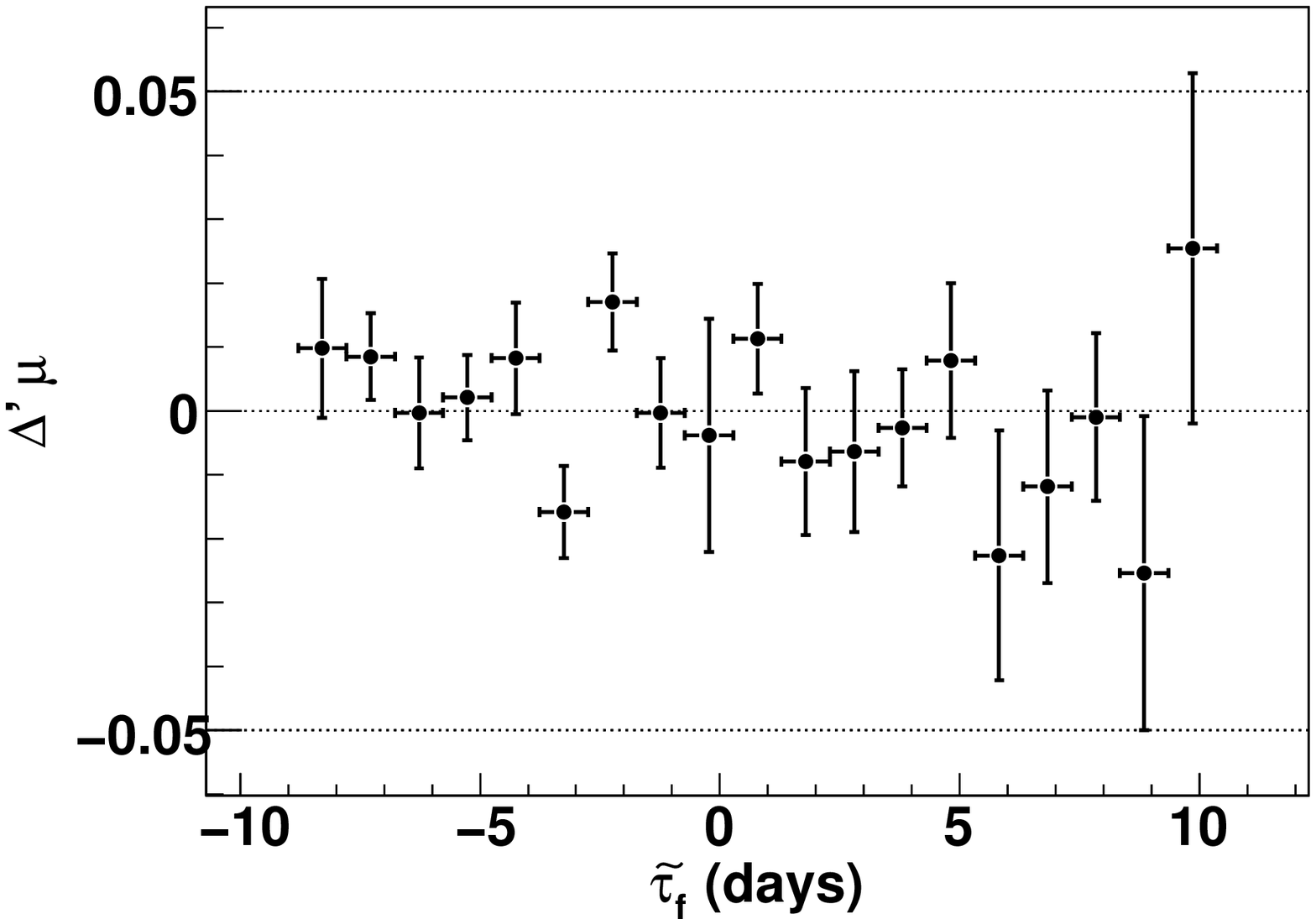}
\caption{From top left to bottom right, $\Delta' m^*_B$, $\Delta' \col$, $\Delta' X_1$ and $\Delta' \mu$ for SALT2 as a function of $\widetilde{\tau_f}$ the estimated phase of the first measurement.\label{fig:sampling-bias}}
\end{figure}

The results for the rest-frame peak magnitude in $B$-band, the peak $(B-V)$ colour and shape parameter are shown in Fig.~\ref{fig:sampling-bias}. The potential bias on the distance modulus $\mu$ is also shown (using best fit values for $\alpha$ and $\beta$, see Eq.~\ref{eq:distance-modulus} and $\S$\ref{sec:hubble-diagram}). For light curves without premax data but with $\widetilde{\tau_f}<5$~days, we do not find significant biases on $m^*_B$ and $(B-V)$ but a small bias on $\x$. The uncertainties on those biases are difficult to estimate, as the de-biasing is only approximate.

For subsequent analyses, we will select SNe according to the following minimum requirements
(based on the available phases $\phase = (T_{obs}-T_{max})/(1+z)$ of photometric observations):\\
(i) Measurements at four different epochs or more in the range $-10 < \phase < +35$~days. Three are mandatory to estimate both the date of maximum light and the shape parameter, one more is needed to get at least one degree of freedom. \\
(ii) At least one measurement in the range $-10 < \phase < +5$~days (equivalent to $\tau_f<5$~days).\\
(ii) At least one measurement in the range $+5 < \phase < +20$~days for a reasonable evaluation of the shape.\\
(iii) At least two bands with one measurement or more in the range $-8 < \phase < +10$~days in order to evaluate the SN peak luminosity and colour with confidence.

With this selection applied to the test data set described above, the estimated bias on distance moduli for the selected SNe without premax data is compatible with zero ($\Delta' \mu = -0.004 \pm 0.004$). However an additional uncertainty due to the de-biasing of order of 0.01 cannot be excluded. We use this latter number as a systematic uncertainty on the estimated distance moduli of SNe without premax data.

\section{Light curve parameters of SNLS SNe Ia and measurement of $\om$}
\label{sec:hubble-diagram-om}
We present in this section a fit of the SNLS Hubble diagram based on the results obtained with the two light curve fitters described above. 
The goals of this study are: i) to compare the SALT2 and SiFTO light curve parameters, ii) see how their
differences impact distance estimates and cosmological fits, iii) propagate the uncertainties of the whole analysis chain to the
 cosmological results, iv) present the constraints on cosmology provided by a single survey of high redshift supernovae.

\subsection{Light curve parameters of SNLS SNe Ia}
\label{sec:results}

We apply in this section the two techniques described above to determine the peak magnitudes, colours and light curve shapes of the SNLS SNe~Ia. Those are the basic ingredients to determine distances, they are used in~C10, along with external SNe data samples, to build a Hubble diagram and constrain cosmological models. Here we focus on consistency tests and comparisons of the outcome of the two light curve fitters, any difference being considered as a systematic uncertainty that will add to the photometry and calibration uncertainties discussed in $\S$\ref{sec:photometry}.

In order to get reliable estimates of magnitude, colour and light curve shape for each SN,
 we apply the selection described in $\S$\ref{sec:sampling-requirements}. 
This set of sampling cuts discards \numberofbadsampling SNe so that we are left with \numberofgoodphotom SNe~Ia/Ia$\star$, excluding the peculiar SNe 03D1cm, 03D3bb, 05D1by and 05D3gy. The discarded SNe are identified by the label ``(s)'' in Table~\ref{table:spectro}.

\subsubsection{Internal consistency check of the light curve fitters}
\label{sec:consistency-test}

In this section we try to quantify one of the primary virtues of the light curve fitters, which consists of deriving redshift-independent parameters (magnitude, shape and colour), from observations in a limited set of filters. 
This is possible in SNLS thanks to the high quality light curves obtained in \gmega\rmega\ and \imega\ bands. For supernovae in the redshift range $[0.2,0.7]$, this corresponds to observations spanning the rest-frame wavelength range 2900--6400~$\AA$.

\begin{figure}
\centering
\includegraphics[width=\linewidth]{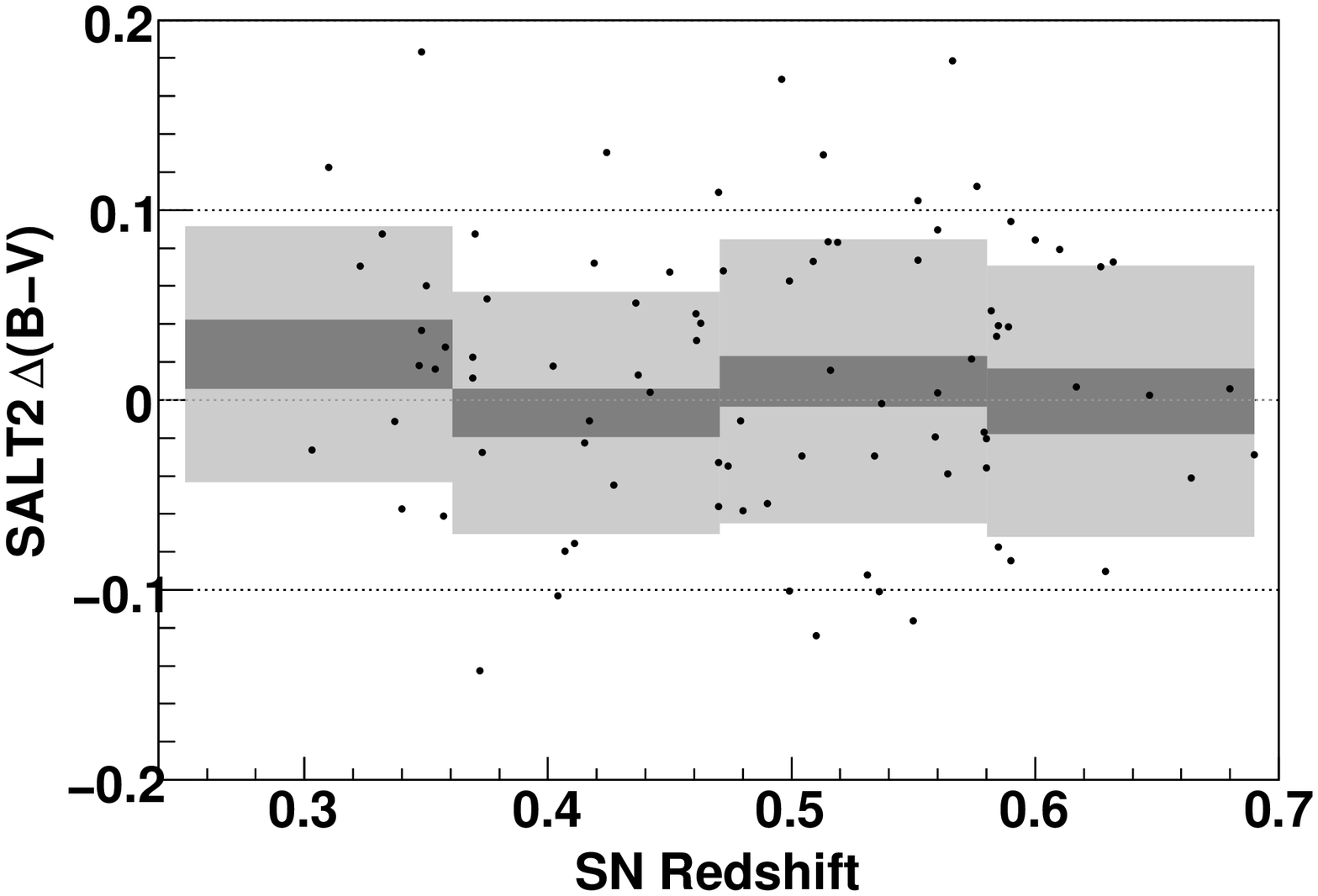}
\includegraphics[width=\linewidth]{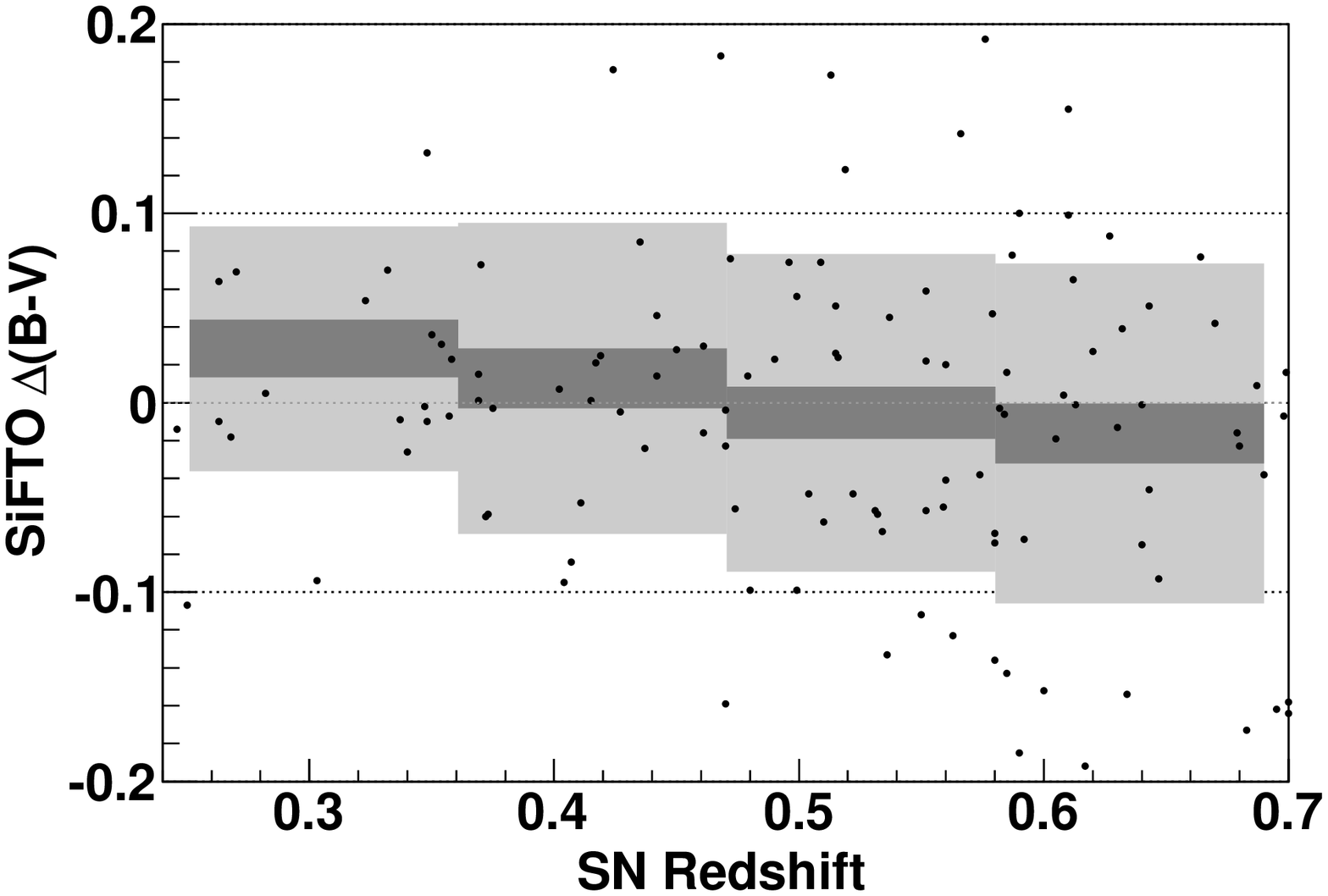}
\caption{Differences of estimated $(B-V)$ colour at maximum from the fit of \rmega\imega\ or \gmega\rmega\ light curves for both fitters (top:~SALT2, bottom:~SiFTO).
\label{fig:consistencycheck}
}
\end{figure}

The fitters should on average give the same estimates of the parameters for fits based on either \gmega$+$\rmega\ or \rmega$+$\imega\ light curves. The results of such a test are shown in Fig.~\ref{fig:consistencycheck} for the difference $\Delta \bmv$ of the estimates of rest-frame $\bmv$ colour, dispersed as a function of redshift. The average values of $\Delta \bmv$ in redshift bins are compatible with zero (the maximum deviations are respectively of $0.024 \pm 0.02$ and $0.020 \pm 0.012$ for SALT2 and SiFTO). This tells us that we can reliably use rest-frame $U$ and $B$ observations to estimate the $\bmv$ colour at high-redshift where we do not have rest-frame $V$-band observations. 
Note however that the statistical uncertainties on the colour relations are accounted for: uncertainty on the $c$ parameter of the colour relation for SiFTO (of order of $0.01$ for $\umb$), and covariance matrix of the training for SALT2 propagated to the distance moduli as detailed in $\S$\ref{sec:error_model_propagation}.

\subsubsection{Comparison of the SALT2 and SiFTO light curve parameters}

We compare in this section the output of the two light curve fitters.
 The results obtained with SALT2 and SiFTO are listed in Table~\ref{table:lcfit}.
Since those empirical models were primarily designed for distance estimates, it is important to briefly review
  how the three resulting parameters (rest-frame magnitude, shape and colour) are combined 
for this purpose. As in A06, a distance modulus is defined by the following linear combination of the
parameters:
\begin{equation}
\mu = m^*_B - M + \alpha \times shape - \beta \times  \col \label{eq:distance-modulus}
\end{equation}
where $m^*_B$ is a rest-frame $B$ magnitude, the $shape$ parameter is $(s-1)$ for SiFTO (where $s$ is the stretch factor), $X_1$ for SALT2, and $\col$ is the colour parameter. The absolute magnitude $M$, and the linear coefficients $\alpha$ and $\beta$ are fitted simultaneously with the cosmological parameters (and also marginalised over for the cosmological constraints).
This approach will be applied and discussed in detail in C10.
We simply note here that any constant systematic offset between SiFTO and SALT2 parameters will have no effect on cosmology as they will be absorbed in the absolute magnitude, that any multiplicative factor between the SiFTO and SALT2 estimates of the $shape$ and $C$ parameters will be compensated for by the $\alpha$ and $\beta$ coefficients, and finally that a $shape$ dependence of the differences of $m^*_B$ and $C$ estimates with SiFTO and SALT2 are also accounted for by the $\alpha$ term. 

As a consequence, in order to compare of the outcome of the two fitters, we will first transform the SALT2 parameters to match those of SiFTO with the following equations:
\begin{eqnarray}
m'_B &=& m^*_B + a_B^X \,  X_1 + a_B^0 \nonumber \\
\col'   &=& a_C^C \, \col + a_C^X \, X_1 + a_C^0  \nonumber \\
s'   &=& a_s^X \, X_1 + a_s^0 \label{eq:salt2-transfo}
\end{eqnarray}
$a_s^X$ and $a_s^0$ are inevitable because of the fundamentally different modeling of the variability of light curve shapes. $a_C^0$ and $a_C^C$ are due to the fact that for SALT2, $\col$ is an internal parameter of the model that only approximately correspond to a $(B-V)$ colour at maximum, with a constant term so that the average value of $\col$ on the training sample is zero, whereas for SiFTO $C$ is exactly the $(B-V)$ colour at maximum. $a_B^0$, $a_B^X$ and $a_C^X$ can be different from zero since the shape of the $B$ and $V$ bands light curves  are not exactly the same for SALT2 and SiFTO. The values of the coefficients obtained by the comparison of the data of Table~\ref{table:lcfit} are listed in Table~\ref{table:salt2-transfo-coefs}.
We apply this transformation of SALT2 parameters to focus on differences between the two light curve fitters that lead to biases on distance moduli (Eq.~\ref{eq:distance-modulus}). The values of the transformation coefficients themselves are simply related to differences in the definition of the parameters in both models and do not contain any meaningful information.

\begin{figure} 
\centering
\includegraphics[width=\linewidth]{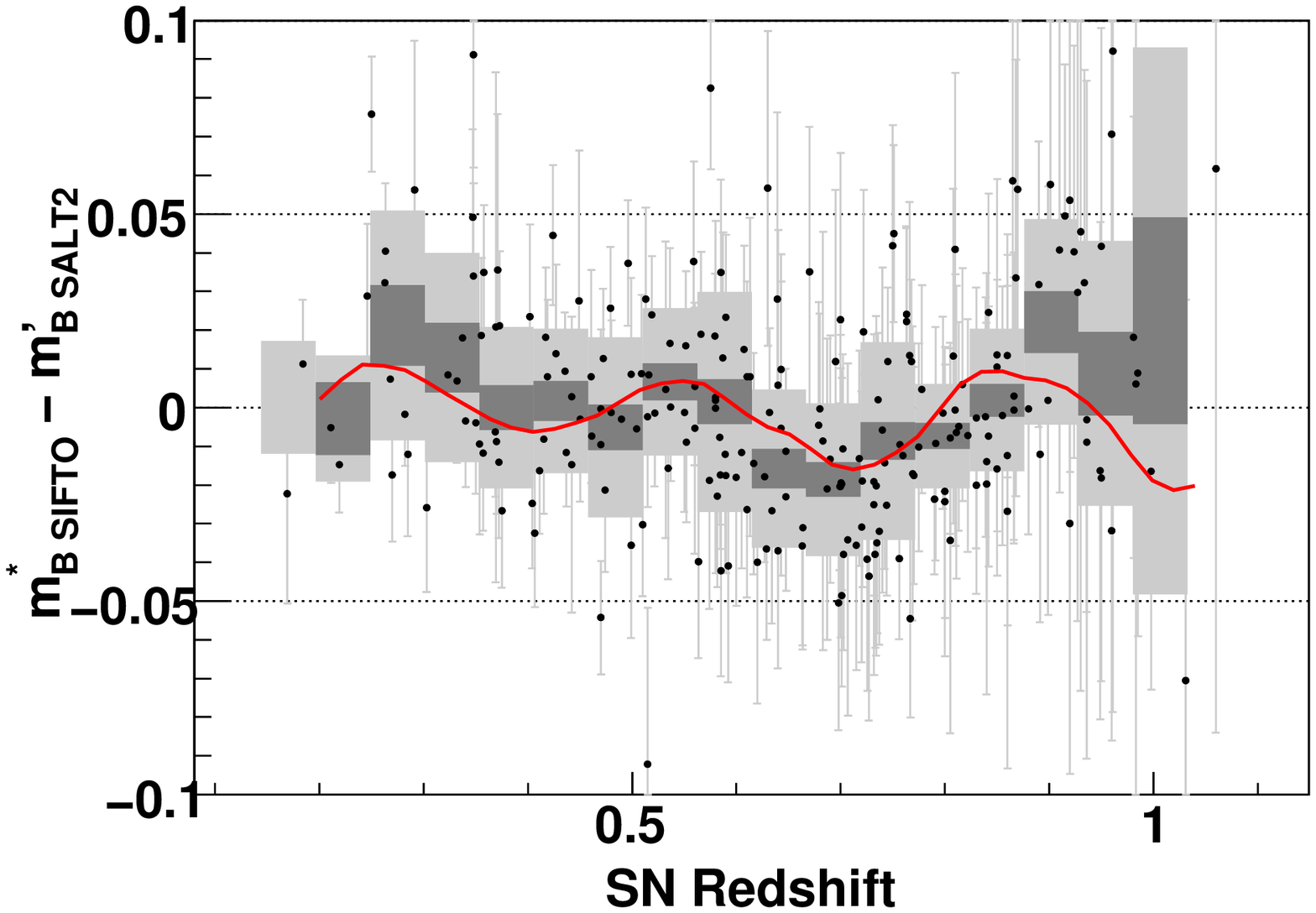}
\includegraphics[width=\linewidth]{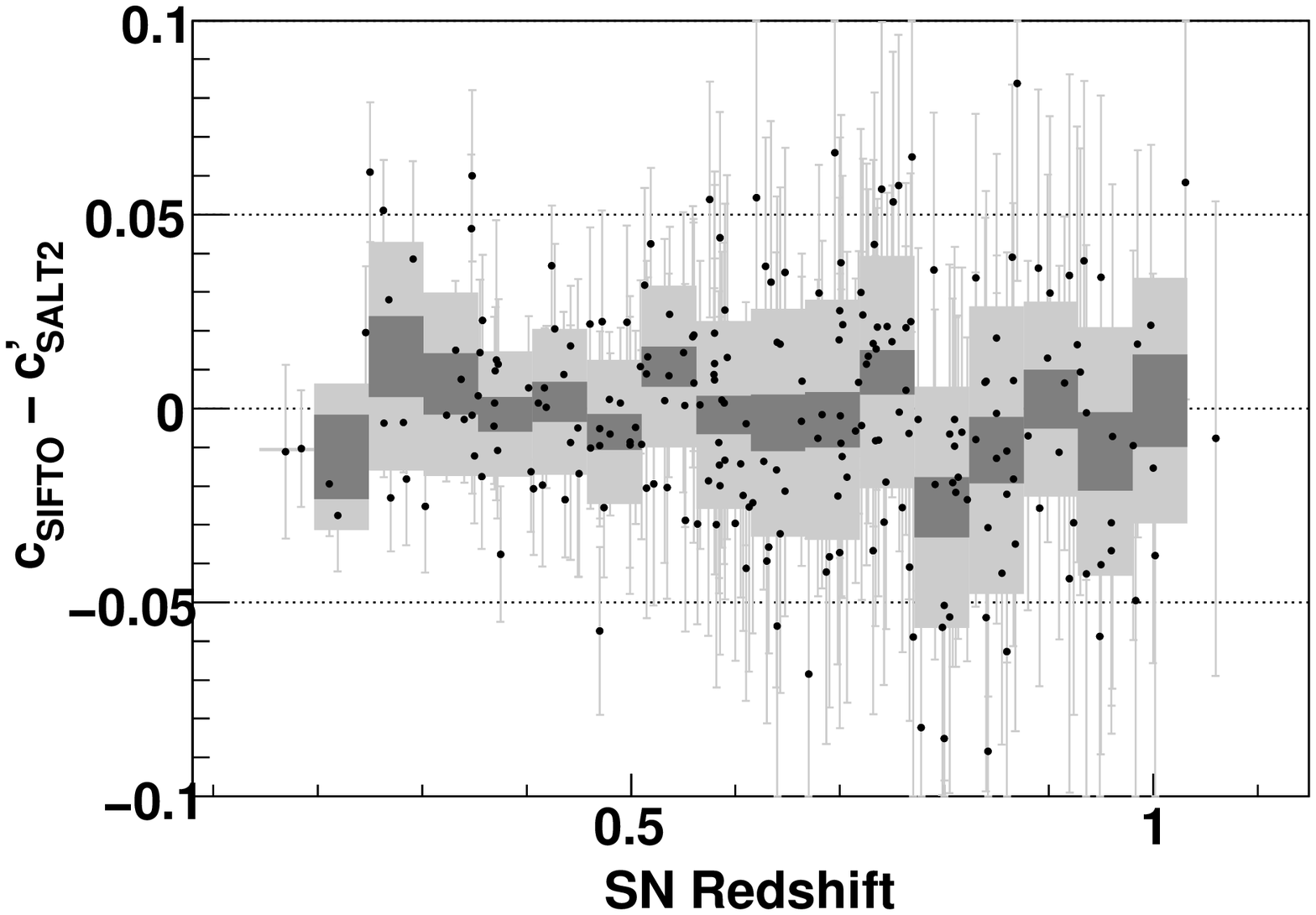}
\includegraphics[width=\linewidth]{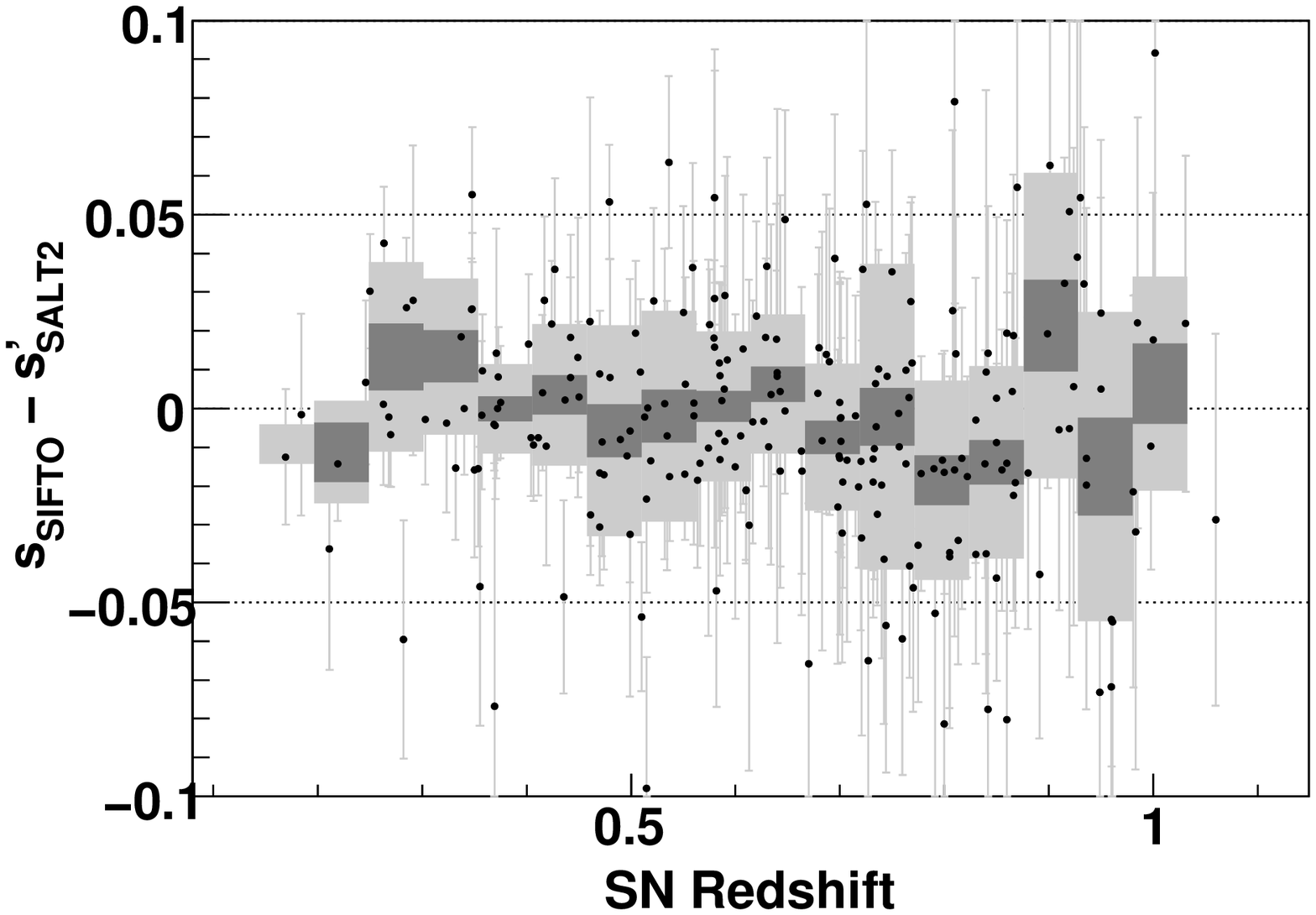}
\caption{Comparison of SALT2 and SiFTO parameters for a selected sub-sample of SNe of Table~\ref{table:lcfit}. Linear transformations of Eq.~\ref{eq:salt2-transfo} and Table~\ref{table:salt2-transfo-coefs} were applied to SALT2 parameters. Each dot is a SN, the light grey areas represent the RMS of the distribution in redshift bins, and the dark areas the uncertainty on the average values.
The red curve on the top panel for the rest-frame $B$ band peak magnitude is the expected discrepancy due to the differences of the SALT2 and SiFTO spectral sequences. 
\label{fig:fitter-comparison}
}
\end{figure}

The differences of those transformed magnitude, shape and colour parameters of SALT2 with those obtained with SiFTO are dispersed as a function of redshift in Fig.~\ref{fig:fitter-comparison}.

Whereas the shape and colour parameters do not show any significant deviation from zero when averaged in redshift bins, there is a systematic difference on $m_B$ for some redshifts ($-0.02 \pm 0.005$ at $z \simeq 0.7$ and $+0.02 \pm 0.005$ at $z \simeq 0.9$). This points to different effective $K$-corrections between SiFTO and SALT2. Those differences can be attributed to some discrepancies in the spectral sequences used (that of SALT2 and the one derived by~\citealt{Hsiao07}), as the spectral distortions performed in SiFTO do not compensate for differences of fluxes on a wavelength range much shorter than the gaps between pass-bands (see the red curve on Fig.~\ref{fig:fitter-comparison} top panel). However the choice of the empirical modeling of the SN diversity (global stretching or linear additive components) can also play a role, as the differences in the training sample and technical details of the training procedures.
 We will hence treat those differences as an additional source of systematic uncertainty.

\begin{table}
\begin{center}
\begin{tabular}{ll}
$a_B^X$ & $= -0.008 \pm 0.005$ \\ 
$a_B^0$ & $= 0.013 \pm 0.004$ \\ 
$a_C^C$ & $= 0.997 \pm 0.097$ \\ 
$a_C^X$ & $= 0.002 \pm 0.009$ \\ 
$a_C^0$ & $= 0.035 \pm 0.008$ \\ 
$a_s^X$ & $= 0.107 \pm 0.006$ \\ 
$a_s^0$ & $= 0.991 \pm 0.006 $ 
\end{tabular}
\caption{Coefficients of the linear transformation of SALT2 parameters to match those of SiFTO as defined by Eq.~\ref{eq:salt2-transfo}.\label{table:salt2-transfo-coefs}}
\end{center} 
\end{table}
 
\subsection{SALT2 and SIFTO based Hubble diagrams}
\label{sec:hubble-diagram}

We present in this section fits of the Hubble diagram based on SALT2 and SiFTO parameters. 
SALT2 parameters are transformed to match those of SiFTO (see Eq.~\ref{eq:salt2-transfo} and Table~\ref{table:salt2-transfo-coefs}). We use the distance modulus $\mu$ given in Eq.~\ref{eq:distance-modulus} with the $shape$ and $C$ parameters matching the SiFTO stretch parameter and the $(B-V)$ colour at maximum.
We  consider here a $\Lambda$CDM model, composed of unrelativistic matter and a cosmological constant, and further assume a spatial flatness, so that we are left with a single cosmological parameter to fit: $\om$ (or alternatively $\ol \equiv 1 - \om$). We apply the same procedure as the one described in A06. The estimation consists of minimising   
\begin{equation}
\chi^2 = \sum_{SN} \frac{\left ( \mu - \mu_{\mathrm{cosmo}} \right)^2}{\sigma^2(\mu)+\sigma^2_{int}} \label{eq:cosmo-chi2-0}
\end{equation}
where $\mu_{\mathrm{cosmo}} = 5 \log_{10} \left[ d_L(\om,z)/ 10 \, \mathrm {pc}  \right]$, $d_L$ being the luminosity distance, and $\sigma_{int}$ is an additional dispersion which value
is chosen in order to obtain a reduced $\chi^2=1$. 
This additional dispersion accounts for all the sources of variability of SNe luminosities beyond their first-order correlation to the shape of light curves and the colours of SNe. The measurement uncertainties $\sigma(\mu)$ are a function of the parameters $\alpha$ and $\beta$.
We use  $H_0 = 70$~km\,s$^{-1}$\,Mpc$^{-1}$. This is purely conventional as it only alters the value of the absolute magnitude $M$ in Eq.~\ref{eq:distance-modulus} through the normalisation of $d_L$, and $M$ is fitted simultaneously with the cosmological parameters. 

In addition to the sampling cuts listed in $\S$\ref{sec:sampling-requirements}, we discard SNe with a peak rest-frame $(B-V)>0.2$. Those red SNe are  found only at $z<0.6$ in our sample because they are fainter than the average, and hence undetected (or unidentified spectroscopically) at higher redshifts. Discarding them minimises the potential bias on distance moduli due to an inadequate colour correction. Indeed the average colour correction coefficient ($\beta$) we derive from the bulk of the SNe may not apply to those red SNe that are more likely to be extinguished by dust in their host galaxy than the bluer ones. This cut, applied to both SALT2 and SiFTO samples, discards \numberofredsne SNe\footnote{The  \numberofredsne  SNe discarded because of their colour are at $z<0.71$. For those redshifts, the average uncertainty on the colour is 0.03.}. We mention in $\S$\ref{sec:uncertainties-on-distances} the impact of a change of this colour cut on the cosmological fit.
In addition to this, the peak magnitudes of \numberofsiftofailed SNe could not be obtained with SiFTO because of a lack of observations in \gmega\ and \zmega\ bands.
Finally, a $3\sigma$ clipping to both SALT2 and SiFTO  Hubble diagram residuals was applied, which removes  \numberofoutliers  more SNe (4 for SiFTO and 4 for SALT2 with only one in common, the others being consistent with a statistical fluctuation).

A Malmquist bias correction is applied to the distance moduli. This correction is based on a detailed study of the survey detection and spectroscopic identification efficiency using fake SNe included in the images entering the detection pipeline. This analysis is described in \citet{Perrett10}. The correction is equal to zero for $z<0.4$ and rises at higher $z$ reaching a value of $\simeq 0.025$ mag at $z=1$.

\begin{figure}[h]
\centering
\includegraphics[width=\linewidth]{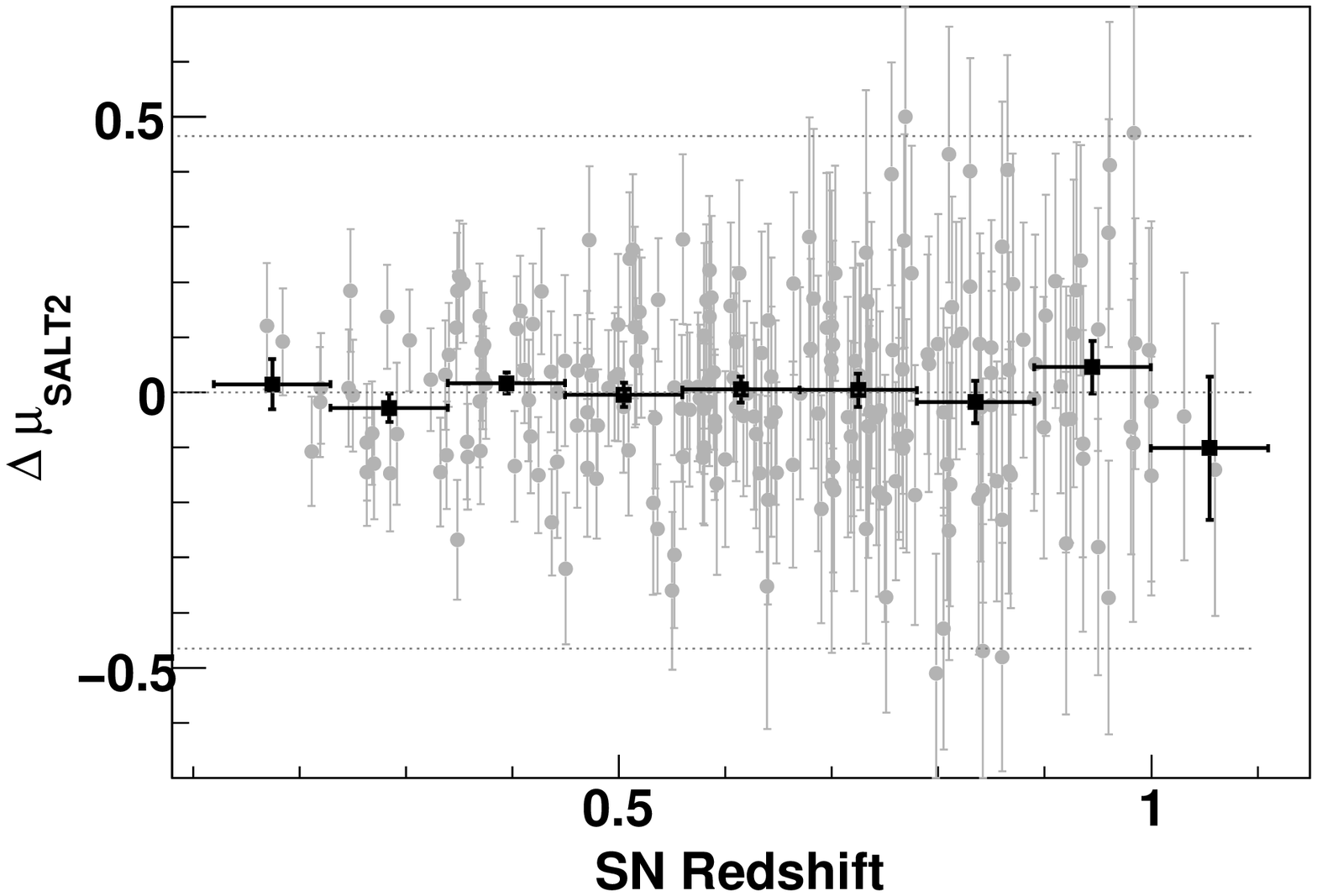}
\includegraphics[width=\linewidth]{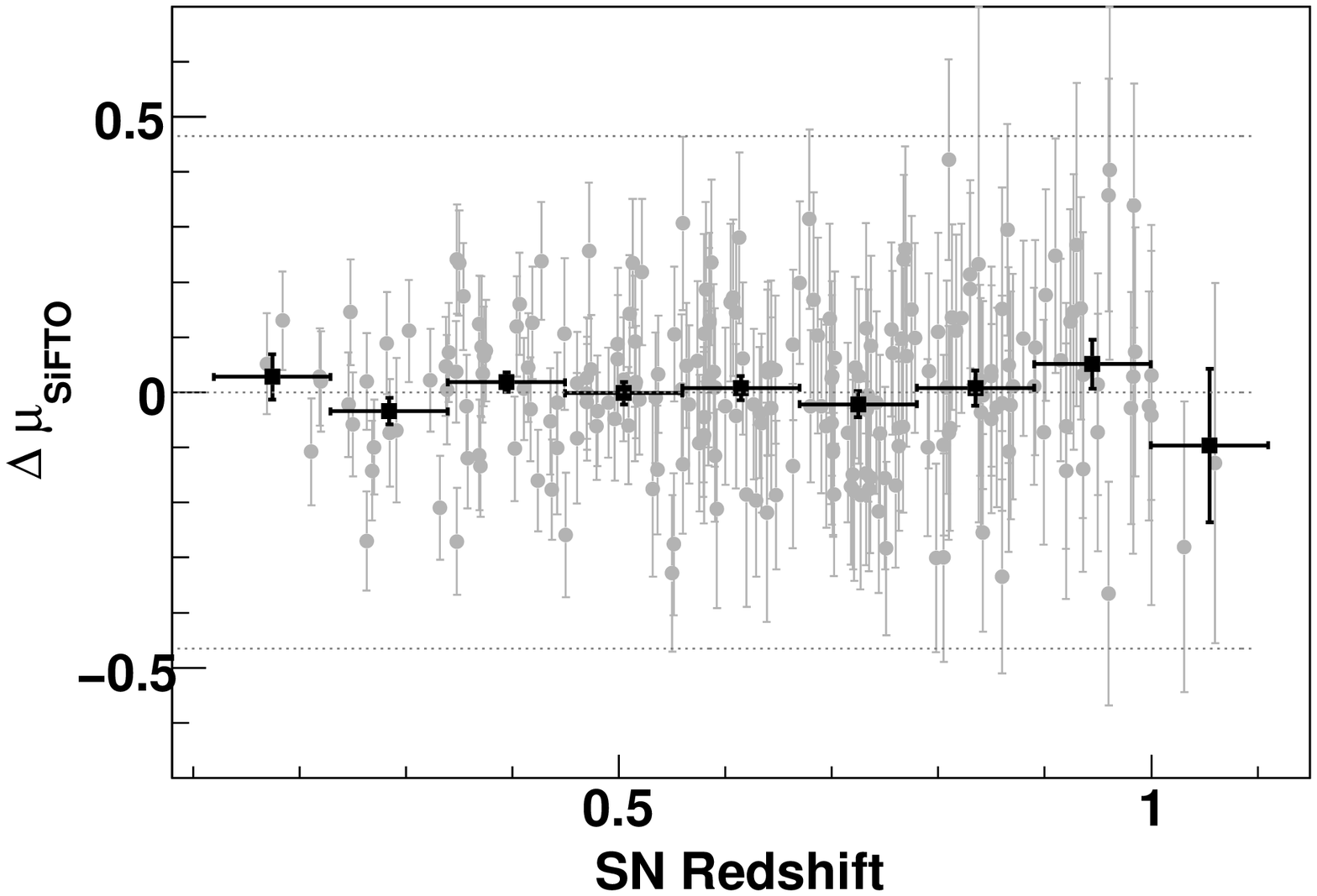}
\caption{Residuals from the Hubble diagram of the best fit $\Lambda$CDM cosmological model obtained with the two light curve fitters SALT2 (top) and SiFTO (bottom).
\label{fig:hubble-residuals}
}
\end{figure}

\begin{figure}[h]
\centering
\includegraphics[width=0.8\linewidth]{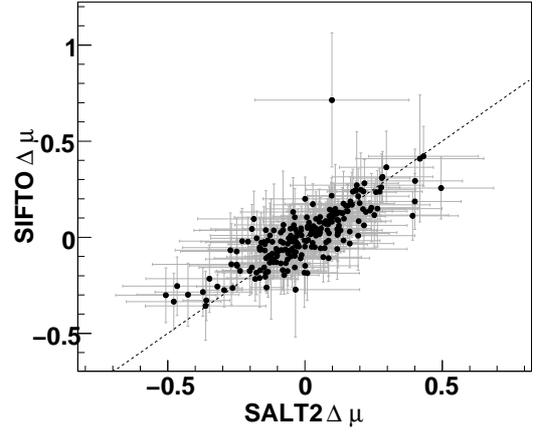}
\caption{Correlations of residuals from the Hubble diagram obtained with SiFTO and SALT2. The outlier is 04D2cc for which data in the rising part of light curves is missing; the statistical uncertainty on its distance modulus $\mu$ is large with both light curve fitters (0.47 with SiFTO and 0.28 with SALT2, the difference being due to differences in the model uncertainties, or ``error snake'').
\label{fig:correlation-of-residuals}
}
\end{figure}

The results obtained with both light curve fitters are listed in Table~\ref{table:salt2-sifto-omegam-stat}. The differences in the light curve fitters lead to a systematic difference of $0.020$ on $\om$ that corresponds to about two thirds of the statistical uncertainty. The RMS of the Hubble residuals are similar, SiFTO residuals being slightly less scattered than the SALT2 ones. The Hubble diagram residuals from the corresponding best fit models are shown in Fig.~\ref{fig:hubble-residuals}.
The correlation of the residuals from the Hubble diagram obtained with both fitters is shown in Fig.~\ref{fig:correlation-of-residuals}. The value of the correlation coefficient of 0.82 shows that most of the residual scatter is due to variability beyond that addressed by a single shape coefficient and a colour, irrespective of the way those parameters were derived.

\begin{table}
\begin{center}
\begin{tabular}{ccc}
 & SALT2 & SiFTO \\
\hline
$\om$    & $0.196\pm 0.035$ & $ 0.215\pm 0.033$\\
$M$     & $-19.218 \pm 0.032$  & $ -19.210 \pm 0.030$ \\
$\alpha$ & $1.295 \pm 0.112$  & $ 1.487 \pm 0.097$ \\
$\beta$  & $3.181 \pm 0.131$  & $ 3.212 \pm 0.128$ \\
RMS  & $0.173 \pm 0.008$  & $ 0.150 \pm 0.007$ \\
$\sigma_{int}$  & 0.087  & 0.087 
\end{tabular}
\caption{$\om$ and the nuisance parameters with their statistical uncertainties derived from the Hubble diagram fits using either SALT2 or SiFTO. SALT2 parameters were linearly transformed (see Eq.~\ref{eq:salt2-transfo} and Table~\ref{table:salt2-transfo-coefs}); in particular, the $\alpha$ coefficient applies to stretch and not the original SALT2 $X_1$ parameter.\label{table:salt2-sifto-omegam-stat}}
\end{center} 
\end{table}

\subsection{Combination of  SALT2 and SiFTO results}
\label{sec:salt2-sifto-combination}
The aim of using two light curve fitters is to evaluate the uncertainties
 on distances associated with the choices made in the development of those empirical models.
In the following, as we do not have any physical justification for choosing one or the other, 
 we  use as a central value the average of the two, and propagate the systematic differences of the two fitters to the uncertainties on cosmological parameters. 

We use for this purpose the method described in C10.
In a few words, all systematic uncertainties are accounted for in the covariance of the distance moduli.
The distance modulus of Equation~\ref{eq:distance-modulus} can be rewritten as  
$\mu = \theta^T \eta -M$ where $\theta = \{1,\alpha,-\beta\}$ is the vector of the  nuisance parameters, and
$\eta = \{m^*_B,s-1,\col\}$ is the vector of the SNe parameters.
With $H_i$ the Jacobian matrix of the derivative of $\eta_i$ with respect to the systematic uncertainties,
and $C_{sys}$ the covariance matrix of those uncertainties, the covariance of the distance moduli
of two supernovae is given by 
\begin{equation}
Cov(\mu_i,\mu_j) = \theta^T \left( \delta_{ij}  C_{\eta_i} + H_i^T C_{sys} H_j \right) \theta \label{eq:mu_covariance}
\end{equation}
where $C_{\eta_i}$ is the covariance matrix of the measurements $\eta_i$ (average of SALT2 and SiFTO results, including the intrinsic $\sigma_{int}$ obtained in $\S$\ref{sec:hubble-diagram} with each fitter independently) 
 assuming that the SALT2 and SiFTO results are fully correlated.
 
For the systematic uncertainty addressed here, we assign to $H$ the difference of SALT2 and SiFTO parameters, averaged over a redshift range about each SN\footnote{We use for the averaging about each SN redshift a Gaussian filter with $\sigma_z=0.05$.}, and consider a variance of one for this uncertainty in the covariance matrix $C_{sys}$.

\begin{figure}
\centering
\includegraphics[width=\linewidth]{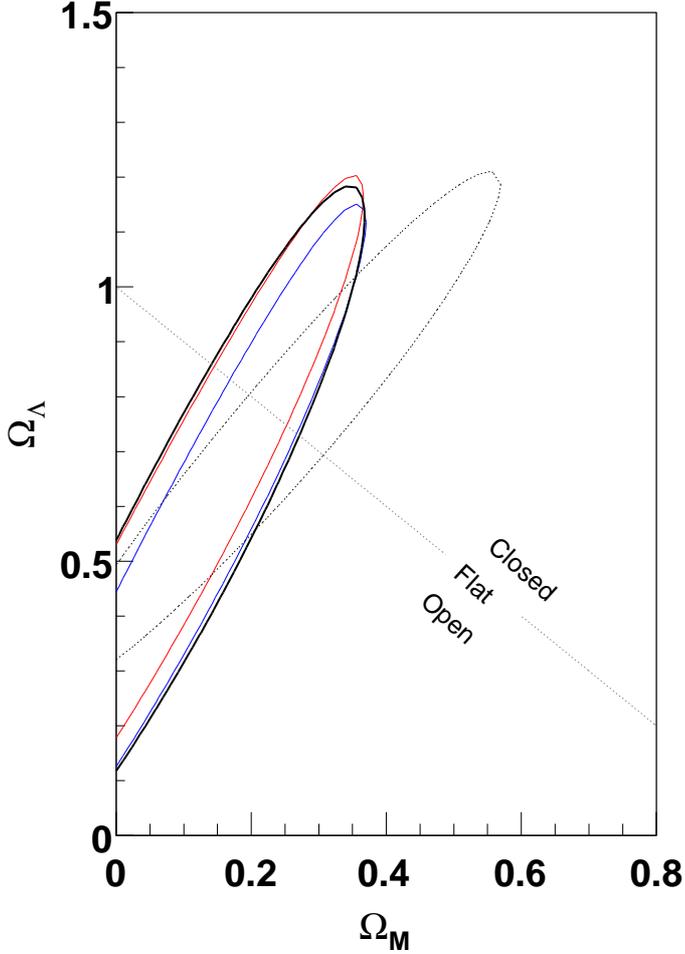}
\caption{Contours at 68.3\% confidence level (accounting only for
statistical uncertainties) for the fit of $\Lambda$CDM cosmology to 
the SNLS third year sample (without additional SNe samples) for 
SALT2 (thin red curve), SiFTO (thin blue) and the combination of the 
two (thick black), on the light curve parameters. The dotted inclined ellipse represents the 68.3\% confidence level from~\citet{Astier06} which includes 44 low redshift SNe along with the 71 SNe from the SNLS first year sample.\label{fig:omol-salt2-sifto}
}
\end{figure}

When fitting for a flat $\Lambda$CDM cosmological model, we obtain with this procedure $\om= 0.201 \pm 0.043$, where the uncertainty includes a statistical uncertainty of $\simeq 0.034$ and an uncertainty due to the difference of the fitters of $\simeq 0.026$.
If we now release the flatness prior, we obtain the confidence contours in the $\om$ $\ol$ plane shown in Fig.~\ref{fig:omol-salt2-sifto}. The combined confidence contour contains the ones obtained with SiFTO or SALT2 only.
Those results do not include the full sources of systematic uncertainties, which will be detailed in Sects.~\ref{section:systematic_uncertainties} and~\ref{sec:beta-evolution} and final results will be presented in Sect.~\ref{sec:om-measurement}.

\subsection{Systematic uncertainties from light curve fits}
\label{section:systematic_uncertainties}

We review here the list of systematic uncertainties that affect the determination of the light curve parameters and enter the covariance matrix $C_{sys}$.

\subsubsection{Calibration systematic uncertainties}

The calibration systematics have been identified and evaluated in R09, 
 they are listed in Table~\ref{table:calib_systematics}. 
All of them affect the light curve parameters. 
They can be divided in two sets: covariant magnitude uncertainties for each of the \allmegasnfilts\ bands and uncertainties on the modeling of filters.
The former magnitude uncertainties comprise  uncertainties 
on the calibration transfer from Landolt stars to tertiary stars, from tertiary stars to supernova light curves, and uncertainties on the magnitude to flux transformation given by the uncertainties of our flux standard BD+17~4708 magnitudes and its spectrum.
Additional uncertainties come from the calibration of the PSF photometry as discussed in $\S$\ref{sec:calibration}.

The uncertainties on the transmission function of the instrument are approximated by uncertainties on the central wavelength of each effective filter (an effective filter combines quantum efficiency of the CCDs, transmission of the MegaPrime optics and filters, reflectivity of the primary mirror, and a model of the atmospheric transmission at the average airmass of the observations, see R09); any higher order uncertainty on the effective transmission has a negligible impact on the total uncertainty budget.

\begin{table*}
\caption[]{Calibration systematic uncertainties\label{table:calib_systematics}}
\begin{tabular}{lcccc}
\hline
\hline
& \gmega & \rmega & \imega & \zmega \\
\hline
MegaCam magnitude system$^\dagger$ & $\pm 0.005 $ & $\pm 0.005 $ & $\pm 0.008 $ & $\pm 0.019 $ \\
Low S/N PSF photometry bias & $< 0.001^{*}$ & $< 0.002$ & $< 0.001$ & $< 0.001^{*}$ \\
Calibration of the PSF photometry & $\pm 0.002 $ & $\pm 0.002 $ & $\pm 0.002 $ & $\pm 0.002 $ \\
\hline
Total & $\pm 0.006 $ & $\pm 0.006 $ & $\pm 0.008 $ & $\pm 0.019 $ \\
\hline
Central wavelength of effective filters & $\pm 7 \AA$ & $\pm 7 \AA$ & $\pm 7 \AA$ & $\pm 25 \AA$ \\
\hline
\end{tabular}
\begin{list}{}{}
\item[$^\dagger$] including the uncertainties on the conversion of magnitudes to fluxes obtained with BD+17~4708 spectrum, see Table 12 of R09
\item[$^{*}$] potential bias due to the uncertainty on the weighted average position from \rmega\ and \imega\ fit.  
\end{list}
\end{table*}

In order to propagate these uncertainties, the derivative
 of the peak magnitude, colour and stretch (or $\x$ for SALT2) with respect to
each of these systematics ($H$ matrices in Eq.~\ref{eq:mu_covariance}) have to be evaluated for each SN.
For SALT2, this requires a full retraining of the model and new light curve fits for each kind of uncertainty. For the SiFTO model, calibration uncertainties only impact the linear colour relations.

\begin{figure}
\centering
\includegraphics[width=\linewidth]{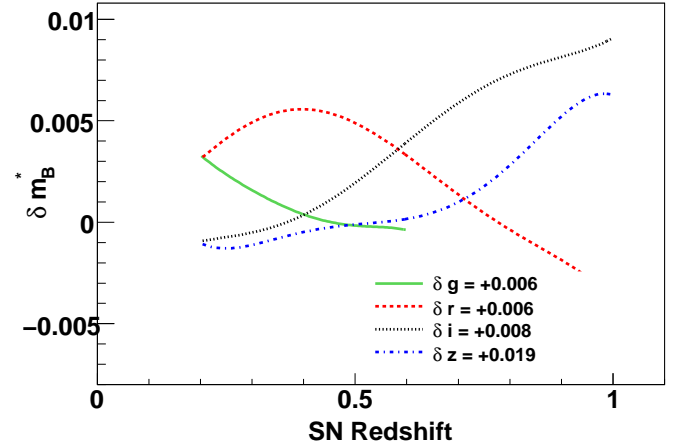}
\caption{Impact of the calibration uncertainties of Table~\ref{table:calib_systematics} on the SALT2 values of  $m_B^{*}$ as a function of redshift (\gmega\ and \rmega-light curves are not used on the whole redshift range because of the limited rest-frame wavelength range of SALT2).
\label{fig:calibration-systematics}
}
\end{figure}

As an example, the impact of the calibration uncertainties on the values of  $m_B^{*}$ are shown in Fig.~\ref{fig:calibration-systematics} for SALT2; they are similar for SiFTO. The sensitivity of $m_B^{*}$ to calibration offsets depend on the weight of each band in the fit which in turn depends on the assumed model uncertainties.

\subsubsection{Light curve fitter systematic differences}
\label{sec:lcfitter-uncertainties}
There are three sources of uncertainties associated with the light curve fitters. 
First, there are the intrinsic differences of modeling given by the differences of
 SALT2 and SiFTO outcomes as discussed previously.

Second, due to finite training samples, there is some statistical uncertainty associated with
 the determination of the parameters of the light curve fitters. 
For SALT2, those uncertainties are provided by the covariance matrix of the model that results
from the training fit (taking into account the residual scatter). The transformation to a covariance matrix of distance moduli is detailed in $\S$\ref{sec:error_model_propagation}, as is the computation of the uncertainty on the average distance modulus in a given
redshift bin.
As for SiFTO, only the uncertainties on the $a,b,c$ parameters of the
colour relations are considered. The uncertainties of
the spectral sequence (and especially its evolution with time) are ignored, 
because they induce negligible correlations of parameters derived for
different supernovae: an ``error snake'' is used to fit light curves, but the
covariance between different SNe is ignored.

Finally, as discussed in $\S$\ref{sec:residual-scatter}, there is some uncertainty on 
 the wavelength dependence of the residual scatter about the best fit model. We have considered two 
 functional forms for this dependence which provide us with an estimate of the associated 
 systematic uncertainty.

An additional  systematic uncertainty specific to the SALT2 technique is the regularisation weight.
 Regularisation is required  to train the model because of the lack of spectra in the near UV for some SN phases. Its weight should be minimal to prevent biases, but a weight that is too low generates unphysical high frequency wiggles in the resulting spectral sequence (this is a typical feature of all deconvolution processes). So the choice of the regularisation scale is partly arbitrary. Its impact on distance moduli is evaluated by the  difference of results obtained with two sets  of SALT2 trainings, one with the nominal weight and the other one with a weight five times smaller. However as shown in Fig.~\ref{fig:combined_uncertainties}, the differences obtained are typically of order of 0.005 and hence can be neglected.

\subsubsection{Resulting uncertainties on distance moduli}
\label{sec:uncertainties-on-distances}
\begin{figure}
\centering
\includegraphics[width=\linewidth]{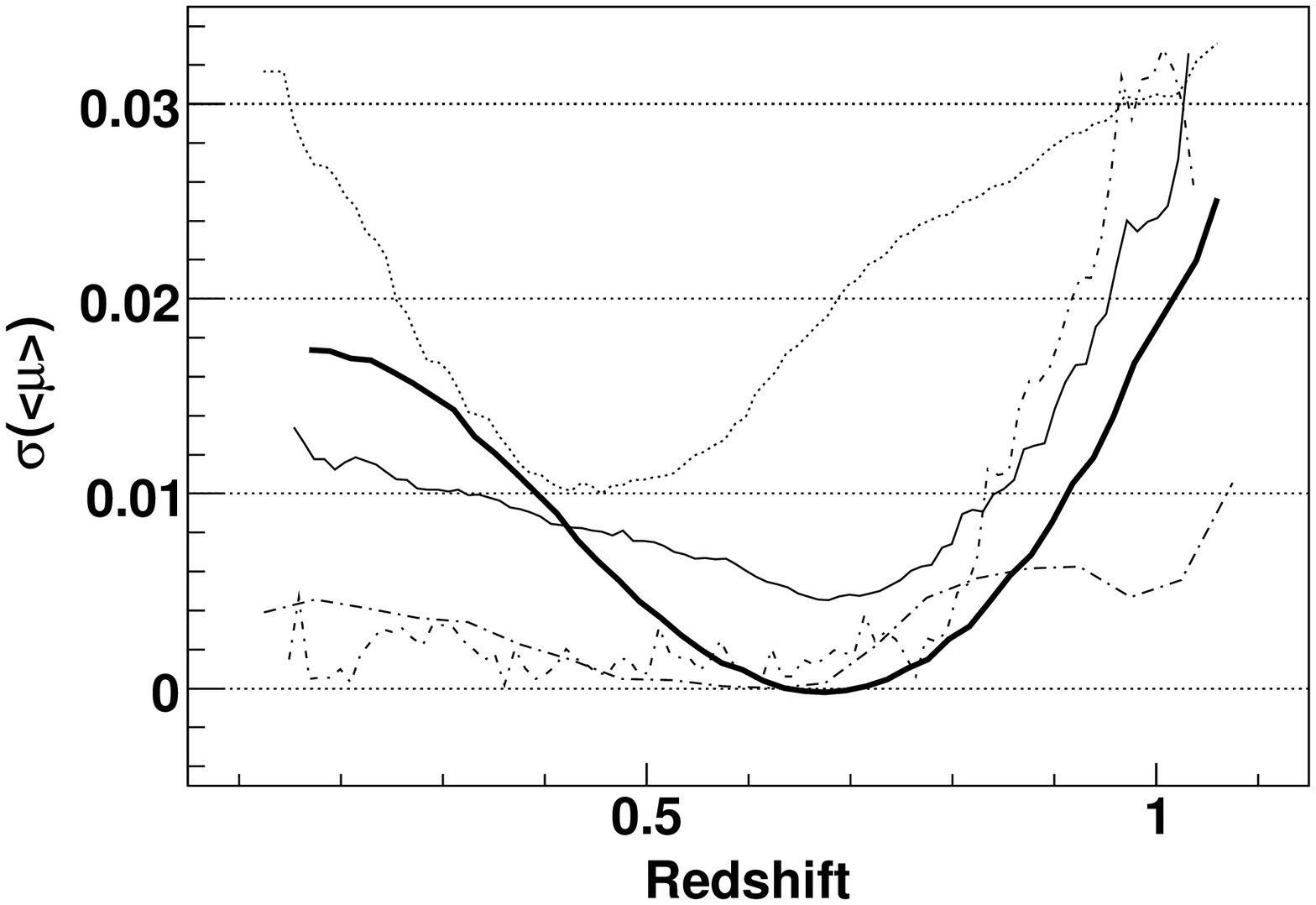}
\caption{Uncertainties on the average distance modulus $\mu$ in redshift bins of 0.2: 
 impact of the statistical uncertainty of the training (for SALT2, thin solid curve), calibration uncertainties (dotted curve), residual scatter model (dotted short dashed curve), systematic uncertainty due to SALT2 regularisation (dotted long dashed curve) and differences between results obtained with the two light curve fitters (thick solid curve). Values of $\alpha$ and $\beta$ that minimise residuals from the Hubble diagram were used.\label{fig:combined_uncertainties}
}
\end{figure}

Figure~\ref{fig:combined_uncertainties} presents the impact of light curve fitter and calibration uncertainties on distance moduli.
The uncertainty induced by the choice of light curve fitter is estimated by the differences of distance moduli obtained with SiFTO and SALT2 averaged over redshift bins of size 0.2.
The values of $M$, $\alpha$ and $\beta$ obtained in $\S$\ref{sec:hubble-diagram} have been used.
 Because of a finite supernova sample, the uncertainty induced by the differences of the light curve fitters cannot be precisely estimated. We can however see that they are of the same order as the calibration uncertainties, with $\Delta \mu \lesssim 0.03$.

The  statistical uncertainty of the light curve model and the choice of parametrisation of the residual scatter only play a significant role at the high redshift end of the survey.
This is due to the combination of Malmquist bias and a limited  wavelength coverage of the imaging survey.
Because of Malmquist bias and the colour-luminosity relation, only the bluest supernovae are measured with enough signal to noise at high redshift, which results in an average colour decrease with increasing redshift. In addition to this, the rest-frame $\bmv$ colour is not directly measured, it is rather given by measurements in rest-frame wavelength shorter than that of the $V$-band so that one has to rely on the colour relations.
As a consequence, any uncertainty on the colour relations will impact distance moduli, and especially a change of slope of the $\umb$ or $\utwomb$ colour relation due to a modification of the residual scatter model will modify the average colour and impact distance moduli.

Along with the uncertainties related to the light curve parameters, 
several other systematic uncertainties alter the precision of the
 cosmological constraints resulting from the Hubble diagram fit.
They are studied in detail in C10. Among those, we only consider 
 the most relevant one for the present analysis (based on SNLS SN sample only): a 39\% uncertainty on the Malmquist bias correction (see C10 for details).
The other systematic uncertainties presented in C10 are negligible for this analysis. As an example, C10 found that the bias on distance moduli due to a contamination of the SN~Ia sample by core collapse supernovae reaches a value of 0.005 mag at $z \simeq 1$. This is much smaller than the other uncertainties we consider in this paper (see Fig.~\ref{fig:combined_uncertainties}).

The choice of the colour cut value of 0.2 applied to the SN sample is partly arbitrary.  Choosing instead a cut of 0.3 for instance adds four SNe to the sample.
With those additional supernovae the best fit $\Omega_m$ value is shifted by $0.01$. The statistical fluctuation of this difference of $\Omega_m$ is $\pm 0.005$ (based on a jackknife analysis), so the shift we obtain is marginally consistent with a statistical fluctuation.

\subsection{Evolution of the colour-luminosity relation with redshift}
\label{sec:beta-evolution}

Recently, \citet{Kessler09} (hereafter K09) found an evolution of the $\beta$ parameter with redshift,
 based on light curve parameters resulting from SALT2 fits of the SNLS first year SNe data set~(A06).
Before presenting the final estimate of $\om$ (including all previously listed systematic uncertainties), we have to address this issue as our cosmological constraints are based on the assumption that $\beta$ is constant.
 
The measurement of the colour-luminosity relation at high redshift is
difficult because of three combined effects:\\ 
i) The fitted value of $\beta$ depends on the assumed uncertainties
of $(B-V)$ (known as a total least-squares problem, see~\citealt{TLS} for a review).\\ 
ii) Because of a limited detection/selection efficiency, we miss red
SNe at high $z$, and hence the range of colours to fit for the relation
is reduced. This makes the determination of $\beta$ both more
uncertain and more sensitive to the assumed uncertainties.\\
iii) SNe rest-frame $(B-V)$ colour is not directly measured at high $z$. We
measure a colour at shorter rest-frame wavelength, and rely on colour
relations to estimate $(B-V)$. Incorrect colour relations  
cause the apparent $\beta$ to change with redshift. Furthermore
the $(B-V)$ estimate depends on assumed uncertainties of the SN model and
intrinsic scatter of SN colours, which are difficult to evaluate (see
$\S$\ref{sec:residual-scatter}).

   Hence, the fit of $\beta$ not only
requires an unbiased estimate of $(B-V)$ but also of the {\it
uncertainties} of this estimate (including all sources of scatter): 
underestimated uncertainties on
$(B-V)$ result in a negative bias on $\beta$. The uncertainties of 
colour relations and colour scatter should be propagated.

\begin{figure}
\centering
\includegraphics[width=\linewidth]{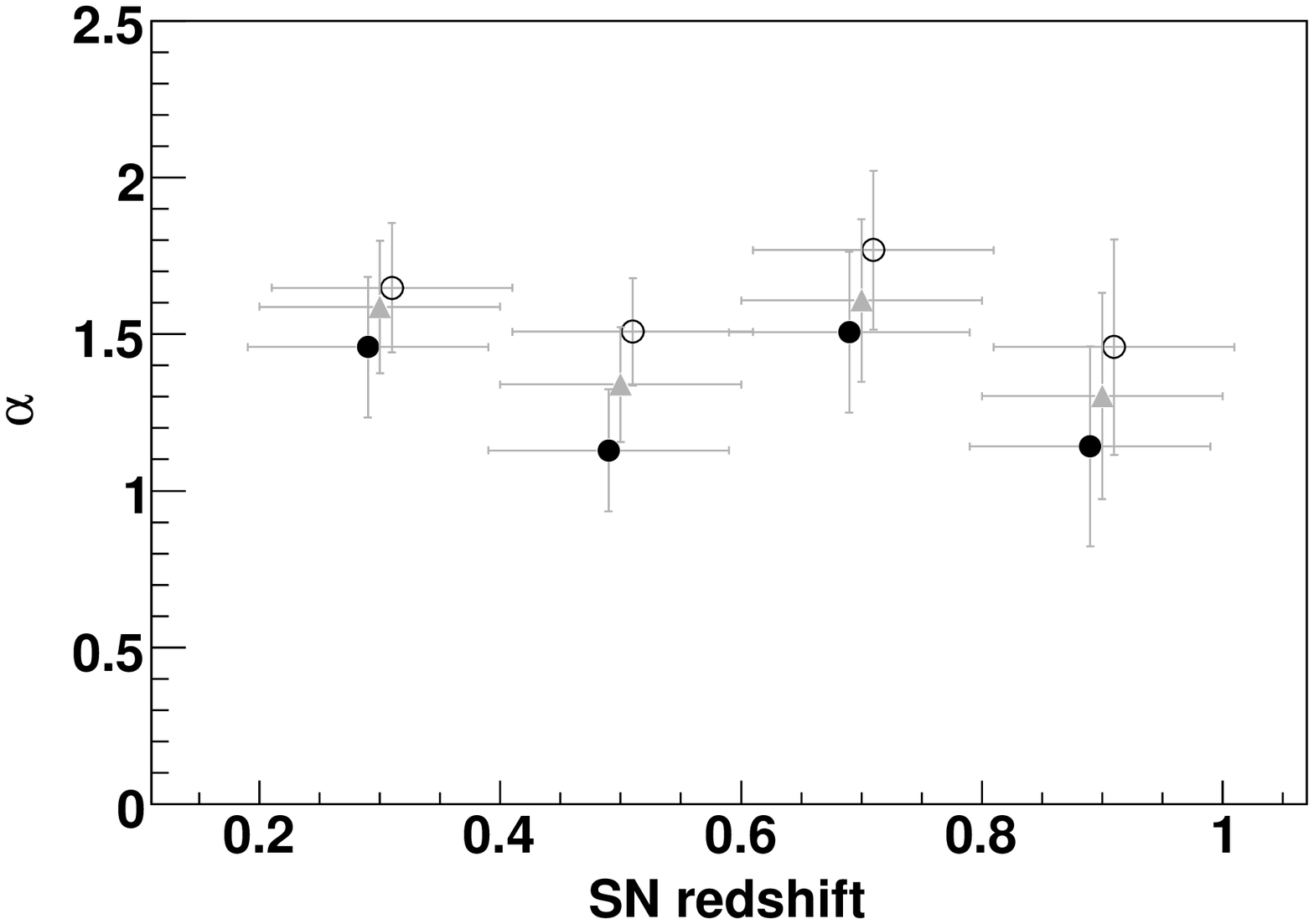}
\includegraphics[width=\linewidth]{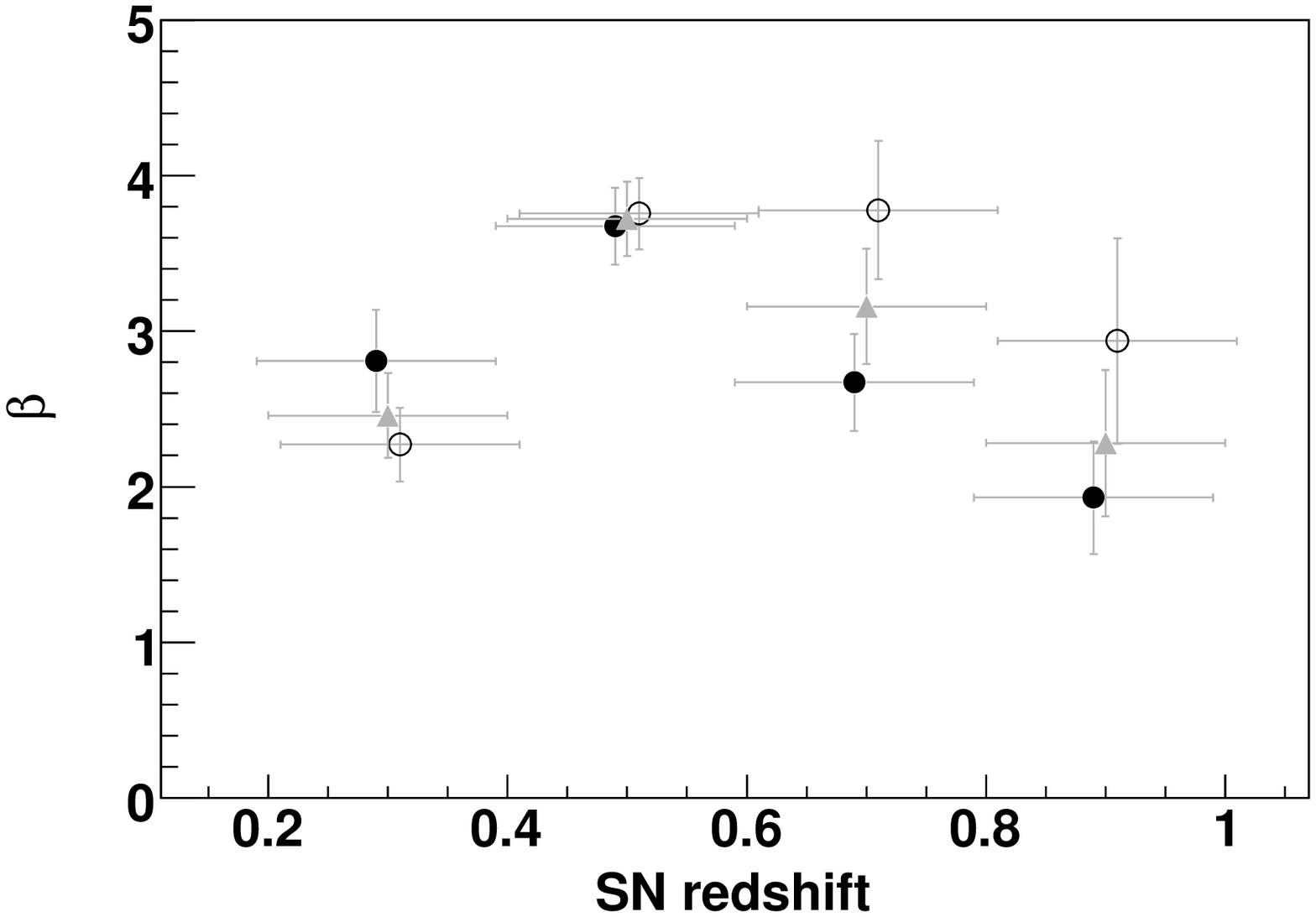}
\caption{
$\alpha$ and $\beta$ estimates in redshift bins of 0.2 using SALT2 (filled circles), SiFTO (open circles) and the combined light curve parameters (gray triangles, see $\S$\ref{sec:salt2-sifto-combination}).
\label{fig:beta-evolution}
}
\end{figure}

With the improved modeling of the residual scatter
presented in this paper with respect to the first version of SALT2
used by K09, we obtain larger uncertainties on the colour parameter and
hence larger $\beta$ at high $z$ than those presented in K09.
 We repeat their analysis which consists of fitting for $M$, $\alpha$ and
$\beta$ in redshift bins assuming a fixed cosmological model.  The
results are shown in Fig.~\ref{fig:beta-evolution} using both SALT2
and SiFTO parameters. Although our redshift--$\beta$ relation is much shallower than that obtained by
K09, we still see a $\beta$ evolution with the SALT2 parameters,
$\beta$ being smaller at higher $z$. 

This change in the estimate of  $\beta$ at high redshift between this analysis and that of K09 comes from the combination of two effects.
 
i) The SALT2 colour variation law used in this paper is steeper than that of G07 for $\lambda<3600~\AA$ mostly because of a higher number of parameters in the model training (see Fig.~\ref{fig:colorcorrection}). At $z \simeq 0.8$, a change of the observed $(r_M-i_M)$ colour translates into a change of the colour parameter $\delta \col_{this \, work} \simeq 0.66 \times \delta (r_M-i_M)$ with the version of SALT2 used in this paper whereas, using that of G07, one obtains $\delta \col_{G07} \simeq 0.81 \times \delta (r_M-i_M)$\footnote{At $z\simeq 0.8$, the $z_M$-band light curve, which have a poor signal to noise ratio, have a low weight for the colour estimate.}. Hence, $\delta \col_{this \, work} / \delta \col_{G07} (z=0.8) \simeq 0.8$ and because of this, we expect $\beta_{this \, work} / \beta_{G07} (z=0.8) \simeq 1.24$. 

ii) For supernovae at $z > 0.7$, which correspond to an average redshift $z \simeq 0.8$, we obtain an average uncertainty on SNe colours of $\sigma_\col = 0.06$ in this paper whereas one obtain an uncertainty of $0.05$ when fitting the same data with the SALT2 version presented in G07 and used in K09. If we apply the scaling factor on the colours of $0.8$ coming from the change of the colour variation law, the uncertainty on colours reduces to $\sigma_\col \simeq 0.04$. 
This difference of colour uncertainties ($0.06$ versus $0.04$) comes from differences in the model uncertainties that are shown in Fig.~\ref{fig:salt2_broadband_colour_dispersion} in this paper and  Fig. 7 in G07. 
Using a Monte Carlo simulation of a linear fit of the colour-luminosity relation for SNe at $z>0.7$, where we consider in the fit that $\sigma_\col = 0.04$ when the value is set to $0.06$ in the simulation, we find a bias $\beta_{fit}-\beta_{true} \simeq -1$\footnote{We consider in the Monte Carlo simulations a value $\beta=3$, 96 SNe, an intrinsic dispersion of distance moduli of $0.87$, and a dispersion of the true SNe colours of $0.05$ (the RMS of the observed colour distribution is of $0.08$, so that one expects an RMS of the true colours of order of $0.05$ after subtracting the contribution of uncertainties).}.

Combining the two effects (change in the estimate of colours and their uncertainties $\sigma_\col$) results in a change of $\beta$ of $\beta_{this \, work} - \beta_{G07} (z=0.8) \simeq 1.6$. At lower $z$, where we measure directly the rest-frame $(B-V)$ colour and benefit from a larger range of SNe colours to fit the slope of the correlation with luminosity, we do not expect any significant difference.

Using SiFTO parameters, we see in Fig.~\ref{fig:beta-evolution} the opposite trend than that obtained with SALT2: larger $\beta$ at higher $z$ (at a $3\sigma$ significance based on a linear fit, but again, the statistical uncertainties are not meaningful),
 which illustrates how
sensitive this measurement is to the light curve fitter modeling and
error propagation. 

As a consequence, given the current uncertainties in the empirical light curve models, we are not able to conclude on an evolution of $\beta$ with redshift.  To account for this, we have added a systematic uncertainty on the evolution of $\beta$ parametrised by 
its first derivative with redshift, for which we assume $\sigma (\partial \beta / \partial z) = 1$.

\subsection{Measurement of $\om$}
\label{sec:om-measurement}

We propagate the sources of uncertainties listed above with the technique described in C10 and summarised in $\S$\ref{sec:salt2-sifto-combination}. Except for the uncertainty associated with the difference of the two light curve fitters, the derivatives in the matrix $H$ are the average of those obtained with SALT2 and SiFTO. For instance, the two sets of parameters $\eta_{a(b)}$ result from fits with different weighting of the \allmegasnfilts~ light curves, so that they have different dependences on the calibration systematics. 

Recently \citet{Sullivan10b} found with a $4 \sigma$ evidence that the  distance moduli (rest-frame magnitude corrected for the correlation with stretch and colour) of SNe in massive host galaxies are smaller than those of SNe in faint galaxies. 
In order to account for this effect in the cosmology fits, a second  normalisation parameter $M_B$ is introduced in order to treat independently SNe in low or high mass galaxies (with a split value at a galaxy mass of $10^{10} M_{\odot}$). When we apply this procedure to the SNLS sample, the retrieved value of $\om$ is shifted by 0.0035. So this effect is negligible for the present analysis based on SNLS data only.

\begin{figure}
\centering
\includegraphics[width=\linewidth]{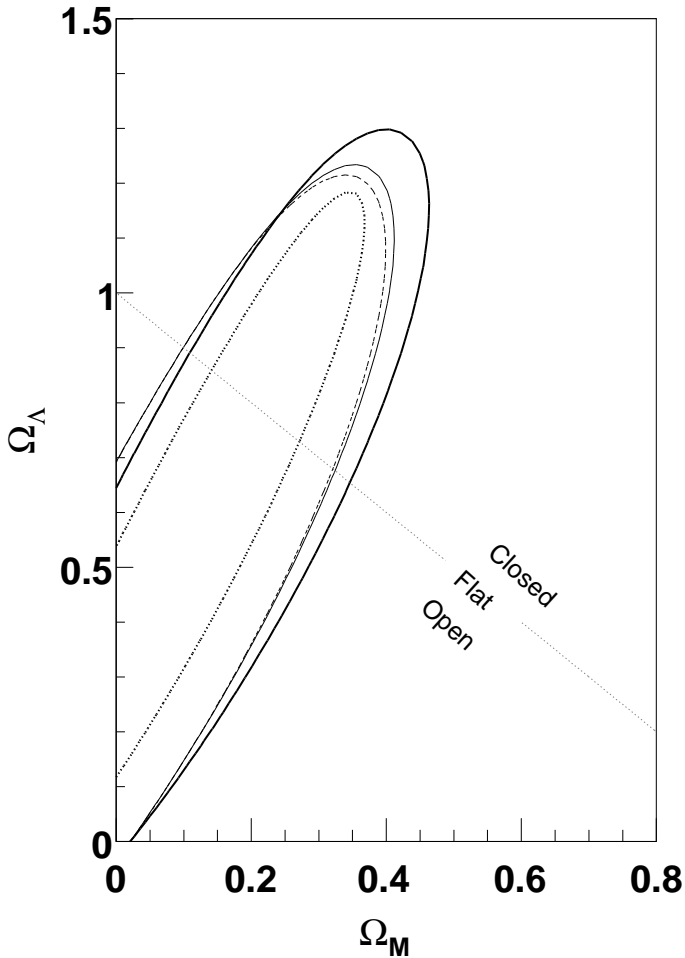}
\caption{Contours at 68.3\% confidence level for the fit to a $\Lambda$CDM cosmology from the SNLS third year sample (without additional SNe samples) accounting for: Hubble diagram statistical uncertainties and the difference between SALT2 and SiFTO results (dotted curve), photometric calibration (dashed curve), model training statistical uncertainties and Malmquist bias uncertainties (thin solid), and finally the potential $\beta$ evolution (thick solid curve).\label{fig:omol-sys}
}
\end{figure}

The contours in the $\om,\ol$ plane are presented in Fig.~\ref{fig:omol-sys}.
With a flatness prior ($\om+\ol=1$), we obtain $\om = 0.211 \pm 0.077$, the uncertainty of which can be approximately decomposed as
\begin{center}
\begin{tabular}{ll}
Hubble diagram statistical uncertainty & $\pm 0.034$ \\
Choice of empirical model (SALT2 vs SiFTO) & $\pm 0.026$ \\
Model training statistical uncertainty  & $\pm 0.034$ \\
Photometric calibration & $\pm 0.048$ \\
Uncertainty on Malmquist bias correction (40\%) & $\pm 0.004$ \\
Uncertainty on the residual scatter (see $\S$\ref{sec:residual-scatter}) & $\pm 0.010$ \\
Potential $\beta$ evolution ($\sigma (\partial \beta / \partial z) = 1$) & $\pm 0.022$  \\
SNe brightness vs host galaxy mass & $\pm 0.003$ 
\end{tabular}
\end{center}

When the potential $\beta$ evolution is accounted for, the best fit $(\om,\ol)$ is shifted. This is  due to a modification of the covariances of the distance moduli and not their values. Modifying the covariances alters the weight of SNe as a function of redshift and hence the best fit cosmological parameters.

Of course, when SNe samples at lower redshifts are added to the cosmological fit, both the statistical and systematic uncertainties are significantly reduced (though other systematics have to be included), and the relative impact of each uncertainty changes. This is studied in detail in C10.

\section{Conclusion}
\label{sec:conclusion}

We have presented photometric properties and distances measurements of SNe discovered during the first 3-year of the SNLS project. 
This paper is the first of a series of three: the two others present the cosmological constraints obtained by combining this data set with other SNe~Ia samples (C10) and taking into account constraints from other cosmological probes~\citep{Sullivan10}.
The various steps of the
analysis were cross-checked using significantly different techniques, in order to identify and minimise systematic uncertainties.
The photometric reduction was done using two independent pipelines.
The calibration transfer from tertiary stars to supernovae light curves was performed with systematic uncertainties of 0.002 mag.
Two light curve fitters were considered in this paper. Both are intended to estimate three parameters (magnitude, shape and colour) that can be subsequently linearly combined to determine the luminosity distances of the SNe~Ia.
 Despite a common goal, the two methods are significantly different in their implementation, allowing us to estimate, through the comparison of their outcome, the uncertainties associated with the choice of modeling.
Whereas SALT2 accounts for the SNe variability with a principal component estimation altered by a colour variation law, SiFTO assumes a pure time stretching of an average spectral sequence, and relies on a limited set of linear relations between rest-frame colours of SNe~Ia.
 The two techniques lead to systematically different results (up to 0.02-0.03 on distance moduli), at a level comparable to other sources of uncertainties such as calibration. 
The quantitative impact of systematic uncertainties, and their relative weight with respect to the purely statistical uncertainty on distance moduli, depends on the subsequent usage of the distance estimates to constrain cosmological parameters, as they introduce large correlations from supernova to supernova.

In contrast to~\citet{Kessler09}, we find no  convincing evidence for an evolution of the colour-luminosity relation with redshift (i.e. evolution of $\beta$). This is traced to improvements made in the empirical modeling of the SNe~Ia spectral sequence and to a more precise estimate of their remaining variability beyond that addressed by the models. Nevertheless, we account for a possible evolution of $\beta$ in the evaluation of systematic  uncertainties.

We obtain $\om = 0.211 \pm 0.034\mathrm{(stat)} \pm 0.069\mathrm{(sys)}$ when fitting a flat $\Lambda$CDM cosmology to the SNLS data set only. The uncertainty is dominated by the calibration uncertainties $(0.048)$ despite a large improvement in the precision of this calibration with respect to previous supernovae surveys. 
However, the impact of systematic uncertainties depends on the data set
considered. In C10, we add external supernova samples to this one, and
constrains further the cosmological models while accounting for
other sources of systematic uncertainties.

\begin{acknowledgements}
We thank the anonymous referee for helping clarify and improve some sections this paper.
We gratefully acknowledge the assistance of the CFHT Queued Service
Observing Team. We heavily relied on the
dedication of the CFHT staff and particularly J.-C.~Cuillandre for
continuous improvement of the instrument performance. The real-time
pipelines for supernovae detection ran on computers integrated in the
CFHT computing system, and were very efficiently installed, maintained
and monitored by K.~Withington (CFHT). We also heavily relied on the
real-time Elixir pipeline which is operated and monitored by
J-C.~Cuillandre, E. Magnier and K.~Withington. The
French collaboration members carry out the data reductions using the
CCIN2P3.  Canadian collaboration members acknowledge support from
NSERC and CIAR; French collaboration members from CNRS/IN2P3,
CNRS/INSU, PNC and CEA. This work was supported in part by the
Director, Office of Science, Office of High Energy and Nuclear
Physics, of the US Department of Energy.  The France-Berkeley Fund
provided additional collaboration support. CENTRA members were
supported by Funda\c{c}\~ao para a Ci\^encia e Tecnologia (FCT),
Portugal under POCTI/FNU/43423.  S. Fabbro and \,C. Gon\c{c}alves
acknowledge support from FCT under grants no SFRH/BPD/14682/2003 and
SFRH/BPD/11641/2002 respectively.
\end{acknowledgements}

\onecolumn
\begin{landscape}
\include{spectro_table}

\end{landscape}
\twocolumn

\include{photometry}
\onecolumn
\include{lcfit_tab}
\include{salt2-nearbysne-training-sample}

\twocolumn

\bibliographystyle{aa}
\bibliography{bibi}

\appendix

\section{Technicals details on the training of SALT2}
\label{appendix:salt2}
\subsection{Parameters and regularisation}
The SALT2 SED sequence is evaluated for phases between -20 and +50 days, in 
the wavelength range from 200 to 920~nm. The model is addressed using a basis of third order b-splines for which the phase and wavelength axis have been distorted in order to allow higher resolution in some areas of the parameter space. Those distortions have been tuned to minimise residuals from the training data set for a given number of spline nodes. For wavelength, the distortion is almost equivalent to a constant resolution in log space, except in the UV range where we lack spectroscopic constraints.  
For this paper, the number of parameters in phase $\times$ wavelength is $14 \times 100$ both for the average SED sequence and the first component that is correlated to stretch. The magnitude of the colour variation law is modelled as $\col \times P(\lambda)$ where $\col$ is the colour parameter and $P(\lambda)$ is a polynomial with linear extrapolations for wavelength smaller than 280~nm or larger than 700~nm, with four free coefficients, the two others being fixed so that $P(\lambda_B)=0$ and  $P(\lambda_V)=1$.

The choice of the number of parameters is a trade off between the statistical noise of the model, and the minimisation of biases introduced by a forced smoothness of the model. For instance, SALT2 has many more parameters than SiFTO which assumes a pure stretching of the light curves.
Also, in some areas of the parameter space, we lack data to train the model. For instance in the UV range, we have a lot of photometric information from the SNLS dataset at all phases but are missing early- and late-time spectroscopy.
In this situation, the training fit is trying to deconvolve the SN SED from broad-band integrated flux measurements, which introduces high frequency noise as for any deconvolution procedure. As a consequence, when integrated in filters other than those used for the training, the model light curves in the UV are quite noisy.
This is only a cosmetic issue when the statistical uncertainty of the model is propagated to the estimates of the light curve parameters (amplitude, shape and colour, see $\S$\ref{sec:error_model_propagation}). Still, we can apply priors on the properties of the SED sequence to cure part of the deconvolution noise in order to improve the statistical accuracy of the model. This however systematically introduces biases that have to be controlled and minimised.

A regularisation term for each of the two components (SED sequences) is added to the standard $\chi^2$ to be minimised of the following form:
\begin{eqnarray}
&& \sum_{i,j,l} w(i,j,l) \left[ m_{i,j+l} -  m_{i,j}\right]^2 \nonumber  \\
&+& \sum_{i,j,k,l} w(i,j,l) \left[ m_{i+k,j+l} \, m_{i,j} - m_{i+k,j} \, m_{i,j+l}\right]^2 \nonumber
\end{eqnarray} 
where $m_{i,j}$ is the flux value of the SED sequence on a grid of phase and wavelength indexed by $i$ and $j$ respectively (similarly $k$ and $l$ index offsets in phase and wavelength) . $w(i,j,l)$ is a weighting function  whose value is inversely proportional to the amount of spectroscopic data at the phase and wavelength position $(i,j)$; it also decreases for increasing $l$ (wavelength difference) so that the regularisation has no effect on the relative values of the SED for wavelength distant by 100~nm or more (in order to limit the biases on colours).
The first term minimises the gradient as a function of wavelength when there is no spectroscopic information, and the second is null if the model can be factorised as a dyadic product $m_{i,j}=a_{i}\times b_{j}$. Hence, the effect of the regularisation is to force the model into a simple interpolation of spectra at different phases when there is no spectroscopic information at intermediate phases, with a minimal impact on regions that are well constrained by data and a small effect on broad-band colours.
The strength of this regularisation is adjusted via the normalisation of $w(i,j,l)$ so that the value of the colour difference $(U-B)-(B-V)$ at maximum luminosity is altered by less than 0.01.
The impact of this regularisation on distance moduli obtained with SALT2 is however controlled (see $\S$\ref{sec:lcfitter-uncertainties}).

\subsection{Error modeling}
\label{appendix:salt2-errors}
For an optimal light curve fit, we need to evaluate the
 intrinsic SNe~Ia scatter about the model in order to weight data 
 accordingly. A perfect evaluation of the covariance of this remaining variability as a function of phase and wavelength would be equivalent to a complete modeling of SNe~Ia. This is of course beyond reach so some approximations have to be made. As in~G07, we evaluated the intrinsic scatter of a light curve assuming that the residuals were not correlated as a function of phase, and independently broad-band colour uncertainties defined by the scatter of light curve amplitudes as a function of the effective rest-frame wavelength of the observation pass-bands.

The intrinsic scatter of light curves as a function of phase, effective wavelength and $\x$, is evaluated using the covariance matrix of the best fit model as described in~G07~$\S$6.1.
 
The broad-band colour uncertainties ($K$-correction uncertainties is G07) can be evaluated by minimising the following restricted log likelihood (see e.g.~\citealt{Harville77}): \\
\begin{equation}
F =   R^T W R - \log \det \left( W \right) + \log \det \left(  A^T W A  \right) \label{eq:restricted_log_likelihood}
\end{equation}
where the first term is the usual $\chi^2$ ($W$ being the inverse of the covariance matrix of the data, and $R$ is the vector of residuals), and $A$ is the matrix of derivatives of the model with respect to all parameters for each measurement ($A^T W A$ is the inverse of the covariance matrix of the best fit parameters).
The second term is expected for a standard log-likelihood, 
the last one accounts for the correlation among residuals introduced by the fitted model. 

In practice, we have to assume a functional form for the  broad-band colour uncertainties as a function of wavelength since only a limited number of parameters can be fit. The following two functional forms which exhibit different asymptotic behaviour have been considered (exponential of polynomial and combination of sigmoids):
\begin{eqnarray}
k_1(\lambda) &=& \exp \left( \sum_{i=0}^4 a_i \lambda^i \right) \nonumber \\
k_2(\lambda) &=&  \frac{a_0-a_1}{1+ \exp \left[ ( \lambda - \lambda_0 )/ \Delta \lambda_0 \right]} + a_1  \nonumber \\ 
&& + \frac{a_2-a_1}{1+ \exp \left[( - \lambda + \lambda_2 )/ \Delta \lambda_2\right] } \label{eq:colorscatterfunction}
\end{eqnarray}

An additional scatter for the observed $U$-band data is evaluated at the same time, since it suffers from sizable calibration systematic uncertainties.

Instead of minimising the restricted log likelihood of Eq.~\ref{eq:restricted_log_likelihood} on the full training set of SALT2, which comprises light curves and spectra, we apply the minimisation to a compressed data set.
 For each light curve of the SALT2 training, a flux scale is fitted to improve the match to the best fit model. An effective wavelength of the corresponding rest-frame filter is also computed (accounting for the spectrum of the SN in the filter integration window). Those flux scales with their associated uncertainties and the effective wavelength provide us the data needed for an estimate of the broad-band colour uncertainties. 
In this process, we fit simultaneously corrections to the colour variation law and the colours of the average SED.

This technique has been successfully tested on Monte Carlo simulations.
 The corrections to the SED and colour variation law turn out to be negligible at the end of the minimisation: about one milli-magnitude for the colours of the SED and 4\% of the colour variation amplitude in the $U$-band (or 0.005 mag difference in $U$-band for $\col = 0.1$), with the caveat that the $U$-band dispersion $\sigma_U$ has been accounted for in the training of SALT2~\footnote{The $U$-band calibration uncertainty is accounted for with an additional term $\sigma_U^2 M M^T$ in the covariance matrices of the light curves, where $M$ is a vector of the model values for each data point.}.

The remaining scatter as a function of wavelength for the two parameterizations of Eq.~\ref{eq:colorscatterfunction} is shown in Fig.~\ref{fig:salt2_broadband_colour_dispersion}. 
We also obtain $\sigma_U = 0.11 \pm 0.02$.

\subsection{Propagation of the model statistical uncertainties}
\label{sec:error_model_propagation}
As discussed in $\S$\ref{sec:lcfitter_considerations}, the statistical uncertainties of the SED sequence model have to be propagated to the estimates of luminosity, shape and colour of the SNe. Indeed, they introduce redshift-dependent uncertainties on distances that do not cancel with an infinite number of SNe.

For SALT2, we use the covariance matrix of the training, including the residual scatter evaluated in the previous section. This latter is accounted for by adding to the covariance matrix of each light curve a term $k(\lambda) M M^T$, where $M$ is a vector containing the model values corresponding to each photometric point, and $\lambda$ is the effective rest-frame wavelength of the filter used for the observations.
 Let 
$\theta$ be the parameters of the model, $C_{\theta}$ their covariance matrix,  
and $\eta_i$ the parameters $(\msb, \x, \col)$ of a given SN,
we may write the $\chi^2$  as:

\begin{equation}
\chi^2 = \sum_{i=1,N} \left[ f_i - m_i(\theta, \eta_i) \right]^T W_i  \left[ f_i - m_i(\theta, \eta_i) \right]
\label{eq:annexe:chi2_glob}
\end{equation}
where
\begin{itemize}
\item $f_i$ are the flux measurements of SN $i$
\item $m_i$ is the model.
\item $W_i$ incorporates both the measurement uncertainties and the error model  discussed previously. 
\end{itemize}
For small deviations $(\delta \theta, \delta \eta_i)$ to a set of parameters $(\theta_0, \eta_{i,0})$, the normal equations provide us with the following relation between $\delta \eta_i$ and $\delta \theta$:
\begin{equation}
\label{eq:eta-theta}
\eta_i = (B_i W_i B_i^T)^{-1} \left[ B_i W_i (r_i  -  A_i^T \delta \theta) \right]
\end{equation}
where $r_i = f_i-m_i(\theta_0,\eta_{i,0})$, $A_i = d m_i / d \theta$ and $B_i = d m_i / d \eta_i$.
This can be inserted into the ``global'' normal equation:
\begin{equation}
\label{eq:theta-sol}
\left[ \sum_{i=1,N}  A_i W_i A_i^T - D_i^T E_i^{-1} D_i \right] \delta \theta = \sum_{i=1,N}  \left[ A_iW_i -D_i^TE_i^{-1}D_i B_i W_i \right] r_i
\end{equation}
with:
\begin{eqnarray}
D_i &= & B_i W_i A_i^T \nonumber \\
E_i &= & B_i W_i B_i^T \nonumber 
\end{eqnarray}
This defines $\delta \theta$, and the parameters $\delta \eta_i$ are obtained 
from Eq. \ref{eq:eta-theta}.
The covariance matrix blocks can be computed:
\begin{eqnarray}
Cov(\theta,\theta)^{-1} & \equiv  & C_{\theta}^{-1} = \sum_{i=1,N} A_i W_i A_i^T - D_i^T E_i^{-1} D_i \nonumber \\
Cov(\theta,\eta_i) & = & - C_{\theta} D_i^T E_i^{-1} \nonumber \\
Cov(\eta_i,\eta_j) & = & \delta_{i,j} E_i^{-1} + E_i^{-1} D_i C_{\theta} D_j^T E_j^{-1} \label{eq:cov-eta-eta} 
\end{eqnarray}
In the expression (\ref{eq:cov-eta-eta}) for $Cov(\eta_i,\eta_j)$, the first term accounts
for the photometric noise, the second accounts for the model noise. 
Note that when considering $Cov(\eta_i,\eta_i)$, the second term does not vanish
when the photometric data of this SN becomes infinitely precise, i.e. 
$W_i \rightarrow \infty$.

These expressions are valid for SNe in the training sample. We can
as well compute the contribution of model uncertainty to the fit
of a single SN which is not in the training sample. The $\chi^2$
is the same as above, but $\theta$ is fixed (and $\delta \theta = 0$).
Using the same notations as above, the normal
equation for $\delta \eta_i$ now reads:
$$
B_i W_i B_i^T \delta \eta_i = B_i W_i r_i
$$
and the uncertainty of the model affects the $\eta_i$ estimator
$\hat{\eta_i}$ by offsetting $r_i$:
$$
\frac{d \hat{\eta_i}}{d \theta} = \frac {d \hat\eta_i}{ d r_i} \frac{d r_i}{d \theta} = \left [ (B_i W_i B_i^T)^{-1} B_i W_i \right] \left [ A_i^T \right ] = E_i^{-1} D_i
$$
So the covariance of SN parameters due to model uncertainties reads:
$$
Cov_{\theta}(\eta_i, \eta_j) = E_i^{-1} D_i C_\theta D_j^T E_j^{-1}
$$
which is identical to the second term in expression (\ref{eq:cov-eta-eta})
above. So the expression of the covariance of SN parameters' estimators
does not depend on their belonging to the training sample.
Note however that when the training sample increases, this
covariance decreases, as we expect.

We can now compute a covariance matrix of distance moduli $(\mu_i)$ of the whole
sample due to light curve model statistical uncertainties. 
This covariance matrix is added to other noise sources and used 
in the cosmological fit. It reads:
$$
C_{\mu}(i,j) \equiv Cov(\mu_i, \mu_j) = V^T E_i^{-1} D_i C_\theta D_j^t E_j V
$$
where $V$ is defined by $\mu_i = V^T \eta_i$ ($V=\{1,\alpha,-\beta\}$ from Eq.~\ref{eq:distance-modulus}).

The uncertainty of the difference between the average distance modulus in a redshift bin $[z_1,z_2]$ and the average on the whole sample (see Fig.~\ref{fig:combined_uncertainties}) is given by 
$$
\sigma_{z_1 < z < z_2} = \left( V_{z_1 z_2}^T C_{\mu} V_{z_1 z_2} \right)^{1/2}
$$
where  $V_{z_1 z_2}(i) = (1/N_{z_1 z_2}-1/N) $ if $z_1< z_i < z_2$, and $-1/N$ if not, $z_i$ being the redshift of the SN number $i$, $N_{z_1 z_2}$ the number of SNe in the bin, and $N$ the total number of supernovae in the sample. 

\subsection{Public release of the light curve fitter}
\label{sec:salt2-release}
The SALT2 model parameters used in this paper, the code to fit
light curves, the calibration files and models of the instrument
responses are public. They can be found at {\tt
http://supernovae.in2p3.fr/~guy/salt/} along with some
documentation. Note however that the usage of this package requires
some care, in particular the rest-frame wavelength validity range of
the model is limited. Also, the photometric calibration of the SNe
light curves of the training sample is imprinted in the spectral model
broadband colours. Hence it is better suited to retrain the model when
more precise calibrations of those samples become available, or when
new data sets that may improve the model can be used. SALT2 is indeed
more a method than a final spectral model.  Finally, as described in
detail in this paper, the uncertainties of the model parameters have
to be taken into account (training sample statistics, calibration
uncertainties, residual scatter model), and the codes to propagate
those are not easy to distribute. For all these reasons, users are
welcome to discuss with the author of their usage of the light curve
fitter.

\section{Flux bias on PSF photometry}
\label{section:psf-bias}

We consider the standard PSF photometry when flux and position are fitted
simultaneously, using a PSF $P(x,y)$, normalised to 1. We call the
flux and position parameters $(f,\delta_x,\delta_y)$. Assuming the noise is stationary and the object faint, 
the weight matrix of their estimators reads:
\begin{equation}
W = w
\begin{pmatrix}
\int P ^2  &  f\int  P \, \partial_x P    & f \int P \, \partial_y P  \\
 & f^2 \int (\partial_x P )^2  & f^2 \int \partial_x P \, \partial_y P \\
 &  & f^2 \int ( \partial_y P  )^2  
\nonumber
\end{pmatrix}
\end{equation}
where $w$ is the inverse of the noise variance per unit area.

Assuming that the PSF is symmetric in x and y, the non-diagonal terms vanish,
and the flux and position estimates are thus independent. The sums over pixels 
have been replaced by integrals which is adequate for our well sampled images.
The flux estimator reads:
$$ 
\widehat{f} = \frac {\int I(x,y)P(x-\widehat{\delta_x},y-\widehat{\delta_y})dx dy }{\int P^2(x,y)dxdy}
$$
where both sums run over pixels, the true position is assumed to be (0,0),
the fitted position is $(\widehat{\delta_x},\widehat{\delta_y})$, and I is the sky-subtracted 
image. Assuming that the image reads:
$$
 I(x,y) = f \, P(x,y)+\epsilon(x,y)
$$
(with $\epsilon$ representing the measurement noise: $E[\epsilon]=0$, and its variance reads $w^{-1}$ per unit area), the flux 
estimator expectation value can be written as:
\begin{eqnarray}
E[\widehat{f}] &= & f \frac {E[\int P(x,y)P(x-\widehat{\delta_x},y-\widehat{\delta_y})dxdy] }{\int P^2(x,y)dxdy} \nonumber \\ 
           & = &  f \left\{ 1 + \frac{1}{2}\frac{E[\widehat{\delta_x}^2]\int P\partial^2_xP+E[\widehat{\delta_y}^2]\int P\partial^2_yP  }{\int P^2(x,y)dxdy} +{\mathcal O}(\delta^4) \right\} \nonumber
\end{eqnarray}
The  flux and position variances read:
\begin{align*}
Var[\widehat{f}] &= \left( w \int P^2 \right) ^{-1}\nonumber \\  
Var[\widehat{\delta_x}] &= E[\widehat{\delta_x^2}] = \left( w f^2 \int (\partial_x P)^2 \right) ^{-1}=  \left(- w f^2 \int P \partial^2_x P \right) ^{-1} \nonumber  
\end{align*}
Hence,
\begin{equation}
E[\widehat{f}] \simeq  f \left\{ 1 - \frac{Var[\widehat{f}]}{f^2} \right\} \label{eq:bias_var_flux}
\end{equation}
So, the PSF flux is underestimated at low S/N, due to the inaccuracy of the
position, because position shifts yield on average smaller PSF fluxes. 
Note that expression \ref{eq:bias_var_flux}
assumes that flux and position are measured from the same data, and also 
assumes a stationary noise and a faint object. The neglected
higher order terms may become important below S/N of a few. For S/N above 1,
the bias is smaller than the uncertainty, but should anyway be studied
if averages of faint fluxes are used in the analysis, such as at the
high redshift end of a Hubble diagram.

For a Gaussian PSF, we have: 
\begin{align*}
Var[\widehat{f}] &= w^{-1} 4 \pi \sigma^2 \nonumber \\
Var[\widehat{\delta_x}] &= w^{-1} 8 \pi \sigma^4 / f^2 \nonumber
\end{align*}
where $\sigma$ stands for the Gaussian RMS
This, combined with Eq.~\ref{eq:bias_var_flux}, gives
\begin{equation}
E[\widehat{f}] = f \left\{ 1 - \frac{E[\widehat{\delta_x}^2+\widehat{\delta_y}^2]}{4 \sigma^2} \right\} \label{eq:bias_var_pos}
\end{equation}
  
Schematically, the measurement of a supernova light curve consists of
a fit of a common position and varying fluxes to an image series.
As for a single image, the flux estimators are biased:
$$
E[\widehat{f_i}] = f_i \left\{ 1 - \frac{E[\widehat{\delta_x}^2+\widehat{\delta_y}^2]}{4 \sigma_i^2} \right\} \label{eq:bias_var_pos2}
$$
where $f_i$ is the supernovae flux at epoch $i$ and $\sigma_i$ is the 
RMS of the PSF (assumed Gaussian) at the same epoch. One may note that
the images with the poorest seeing are the less affected.  
We assume the position to be measured from the images series. Its variance reads:
$$
E[\widehat{\delta_x}^2+\widehat{\delta_y}^2] = 2 Var[\widehat{\delta_x}] = \left( \sum_i \frac {f_i^2}{4 \sigma_i^2 Var[\widehat{f_i}]} \right)^{-1}
$$
where the sum runs over images.
The amplitude of the light curve will be fitted to the individual image 
measurements $\widehat{f_i}$, and is a linear combination of those. 
The least squares estimator of the amplitude A of the light curve reads:
$$
\widehat{A} = k \sum_i a_i \widehat{f_i}, {\rm with} \,\,\, a_i = \frac{f_i}{Var[\widehat{f_i}]}  
$$
where $k$ is a global constant depending on the chosen definition of A. 
We now propagate the flux bias at individual epochs into the 
lightcurve amplitude:
\begin{align*}
\widehat{A} &= k \sum_i \frac{f_i^2}{Var[\widehat{f_i}]} \left\{ 1 - \frac{E[\widehat{\delta}^2]}{4\sigma_i^2}\right\} \nonumber \\
   &= k \left\{ \sum_i \frac{f_i^2}{Var[\widehat{f_i}]} - {E[\widehat{\delta}^2]} \left(\sum_i \frac{f_i^2}{4 Var[\widehat{f_i}] \sigma_i^2} \right) \right \} \nonumber \\
   &= k \left\{ \sum_i \frac{f_i^2}{Var[\widehat{f_i}]} - 1 \right\} \nonumber
\end{align*}
Using that:
$$
\frac{Var[\widehat{A}]}{A^2} = \left( \sum_i \frac{f_i^2}{Var[\widehat{f_i}]} \right)^{-1}
$$
We finally have:
$$
E[\widehat{A}] = A \left( 1 - \frac{Var[\widehat{A}]}{A^2} \right )
$$
and we can hence use the relative uncertainty of the lightcurve amplitude
as an indicator of its own bias, when position and amplitude are measured
from the same data.

\end{document}

%% file: authors.tex
\author{
J.~Guy\inst{1}
\and
M.~Sullivan\inst{2}
\and
A.~Conley\inst{3,4}
\and
N.~Regnault\inst{1}
\and
P.~Astier\inst{1}
\and
C.~Balland\inst{1,5}
\and
S.~Basa\inst{6}
\and
R.G.~Carlberg\inst{4}
\and
D.~Fouchez\inst{7}
\and
D.~Hardin\inst{1}
\and
I.M.~Hook\inst{2,8}
\and
D.A.~Howell\inst{9,10}
\and
R.~Pain\inst{1}
\and
N.~Palanque-Delabrouille\inst{11}
\and
K.M.~Perrett\inst{4,12}
\and
C.J.~Pritchet\inst{13}
\and
J.~Rich\inst{11}
\and
V.~Ruhlmann-Kleider\inst{11}
\and
D.~Balam\inst{13}
\and
S.~Baumont\inst{14}
\and
R.S.~Ellis\inst{2,15}
\and
S.~Fabbro\inst{13,16}
\and
H.K.~Fakhouri\inst{17}
\and
N.~Fourmanoit\inst{1}
\and
S.~Gonz\'alez-Gait\'an\inst{}
\and
M.L.~Graham\inst{13}
\and
E.~Hsiao\inst{17}
\and
T.~Kronborg\inst{1}
\and
C.~Lidman\inst{18}
\and
A.M.~Mourao\inst{16}
\and
S.~Perlmutter\inst{17}
\and
P.~Ripoche\inst{17,1}
\and
N.~Suzuki\inst{17}
\and
E.S.~Walker\inst{2,19}
}
\institute{
LPNHE, CNRS/IN2P3, Universit\'e Pierre et Marie Curie Paris 6, Universit\'e Denis Diderot Paris 7, 4 place Jussieu, 75252 Paris Cedex 05, France
\and
Department of Physics (Astrophysics), University of Oxford, DWB, Keble Road, Oxford OX1 3RH, UK
\and
Department of Astrophysical and Planetary Sciences, University of Colorado, Boulder, CO 80309-0391, USA
\and
Department of Astronomy and Astrophysics, University of Toronto, 50 St. George Street, Toronto ON M5S 3H4, Canada
\and
Universit\'e Paris 11, Orsay, F-91405, France
\and
LAM, CNRS, BP8, Traverse du Siphon, 13376 Marseille Cedex 12, France
\and
CPPM, CNRS-IN2P3 and Universit\'e Aix-Marseille II, Case 907, 13288 Marseille Cedex 9, France
\and
INAF - Osservatorio Astronomico di Roma, via Frascati 33, 00040 Monteporzio (RM), Italy.
\and
Las Cumbres Observatory Global Telescope Network, 6740 Cortona Dr., Suite 102, Goleta, CA 93117
\and
Department of Physics, University of California, Santa Barbara, Broida Hall, Mail Code 9530, Santa Barbara, CA 93106-9530
\and
CEA, Centre de Saclay, Irfu/SPP, F-91191 Gif-sur-Yvette, France
\and
Network Information Operations, DRDC-Ottawa, 3701 Carling Avenue, Ottawa, ON, K1A 0Z4, Canada
\and
Department of Physics and Astronomy, University of Victoria, PO Box 3055 STN CSC, Victoria BC V8T 1M8, Canada
\and
LPSC, CNRS-IN2P3, 53 rue des Martyrs, 38026 Grenoble Cedex, France
\and
Department of Astrophysics, California Institute of Technology, MS 105-24, Pasadena, CA 91125, USA
\and
CENTRA-Centro M. de Astrofisica and Department of Physics, IST, Lisbon, Portugal
\and
LBNL, 1 Cyclotron Rd, Berkeley, CA 94720, USA
\and
Anglo-Australian Observatory, P.O. Box 296, Epping, NSW 1710, Australia
\and
Scuola Normale Superiore, Piazza dei Cavalieri 7, 56126 Pisa, Italy
}

%% file: spectro_table.tex
\begin{longtable}{lcccccccccc}
\caption{\label{table:spectro} List of SNLS 3 years type Ia supernova sample.}\\
\hline\hline
Name & RA (J2000) & DEC (J2000) & Spectrum Date(s) & Telescope(s)  & Type & $z$  & Ref. & $E(B-V)$ & Phot. \\
     &            &             & JD-2450000       \\
\hline
\endfirsthead
\caption{continued.}\\
\hline\hline
Name & RA (J2000) & DEC (J2000) & Spectrum Date(s) & Telescope(s)  & Type & $z$  & Ref. & $E(B-V)$ & Phot. \\
     &            &             & JD-2450000       \\
\hline
\endhead
\hline
\endfoot
03D1ar & 2:27:14.669 & -4:19:4.75 & 2909 & VLT & SNIa$\star$ & $0.4080 \pm 0.0010 $~(H) & Ba09 & 0.025 & griz \\
03D1au & 2:24:10.370 & -4:2:14.89 & 2906 & Keck & SNIa & $0.5043 \pm 0.0005 $~(H) & E08 & 0.030 & griz \\
03D1aw & 2:24:14.792 & -4:31:1.77 & 2906 & Keck & SNIa & $0.5820 \pm 0.0050 $~(H) & E08 & 0.025 & griz \\
03D1ax & 2:24:23.349 & -4:43:14.79 & 2912 & Gem-N & SNIa & $0.4960 \pm 0.0010 $~(H) & H05 & 0.026 & griz \\
03D1bf & 2:24:2.391 & -4:55:57.11 & 2908 & VLT & SNIa$\star$ & $0.7030 \pm 0.0010 $~(H) & Ba09 & 0.027 & none (c) \\
03D1bk & 2:26:27.414 & -4:32:12.12 & 2911 & Gem-N & SNIa & $0.8650 \pm 0.0010 $~(H) & H05 & 0.028 & griz \\
03D1bm & 2:24:36.072 & -4:12:44.51 & 2911 & VLT & SNIa$\star$ & $0.5750 \pm 0.0010 $~(H) & Ba09 & 0.028 & griz \\
03D1bp & 2:26:37.736 & -4:50:19.61 & 2909 & VLT & SNIa$\star$ & $0.3470 \pm 0.0010 $~(H) & Ba09 & 0.024 & griz \\
\end{longtable}
\noindent
{\bf Notes.} Photometric status. (a) : no photometry because of variable host or second SN in host, (b) close to a bright star, (c) edge of field; (s): missing epochs for light curve fit.
\tablebib{H05: \citet{Howell05}, H06: \citet{Howell06}, Ba09: \citet{Balland09}, Br08: \citet{Bronder08}
E08: \citet{Ellis08}, F10: \citet{Fakhouri10}, W10: \citet{Walker10}.}

%% file: photometry.tex
\begin{table*}
\caption{Photometry of the SNLS 3-year SNe~Ia.
\label{table::photometry}}
\centering
\begin{tabular}{ccccc}
\hline \hline 
Name & Filter\tablefootmark{a} & MJD & Flux\tablefootmark{b} & Flux uncertainty\tablefootmark{b} \\ 
\hline 
03D1ar & g & 52880.58 & 1.3035E+01 & 5.4246E+00 \\
03D1ar & i & 52881.50 & 1.6157E+01 & 1.8629E+01 \\
03D1ar & r & 52881.54 & 7.0282E+00 & 9.9105E+00 \\
03D1ar & z & 52881.56 & 2.2086E+01 & 7.1951E+01 \\
03D1ar & i & 52886.60 & 1.8944E+02 & 1.9855E+01 \\
03D1ar & g & 52900.50 & 1.4280E+02 & 1.4970E+01 \\
03D1ar & i & 52900.53 & 6.5402E+02 & 2.3202E+01 \\
03D1ar & r & 52900.56 & 4.5705E+02 & 1.6919E+01 \\
03D1ar & z & 52900.59 & 5.7204E+02 & 8.9594E+01 \\
03D1ar & i & 52904.53 & 6.9411E+02 & 1.7710E+01 \\
03D1ar & r & 52904.57 & 3.9937E+02 & 1.2907E+01 \\
03D1ar & g & 52904.60 & 1.1473E+02 & 7.5546E+00 \\
03D1ar & g & 52908.53 & 8.4923E+01 & 7.5163E+00 \\
03D1ar & r & 52908.56 & 3.8457E+02 & 1.2797E+01 \\
03D1ar & i & 52908.60 & 6.0466E+02 & 2.0464E+01 \\
03D1ar & z & 52909.61 & 6.6974E+02 & 8.0965E+01 \\
03D1ar & i & 52912.49 & 5.7083E+02 & 1.9573E+01 \\
03D1ar & r & 52914.57 & 2.0934E+02 & 2.1674E+01 \\
03D1ar & i & 52915.62 & 5.0601E+02 & 2.5140E+01 \\
03D1ar & r & 52916.46 & 2.7066E+02 & 1.2909E+01 \\
... \\ 
\end{tabular}
\tablefoot{An electronic version of this table is available at the {\it Centre de Donn\'ees astronomiques de Strasbourg} (CDS). Light curves can also be downloaded at the University of Toronto's Research Repository {\tt https://tspace.library.utoronto.ca/handle/1807/24512}.\\ 
\tablefoottext{a}{The average focal plane coordinates of SNe are available at CDS. Those are needed to estimate the filter transmission function, see $\S$~\ref{sec:calibration}.} 
\tablefoottext{b}{Fluxes and uncertainties are given for a fiducial zero point of 30, a magnitude is $\mathrm{mag} = -2.5 \log_{10}{(\mathrm{Flux})} + 30$. The interpretation of those magnitudes for supernovae study is detailed in $\S$~\ref{sec:using-calibration}.} 
}
\end{table*}

%% file: lcfit_tab.tex
\begin{longtable}{l|ccc|ccc}
\caption{\label{table:lcfit} SNLS Type Ia supernovae parameters.}\\
\hline\hline
Name & $\msb$(SALT2) & $\x$(SALT2) & $\col$(SALT2) & $\msb$(SIFTO) & $s$(SIFTO) & $\col$(SIFTO) \\
\hline
\endfirsthead
\caption{continued.}\\
\hline\hline
Name & $\msb$(SALT2) & $\x$(SALT2) & $\col$(SALT2) & $\msb$(SIFTO) & $s$(SIFTO) & $\col$(SIFTO) \\
\hline
\endhead
\hline
\endfoot
03D1ar & $23.941 \pm 0.033$ & $-0.945 \pm 0.209$ & $0.266 \pm 0.035$ & $23.969 \pm 0.036$ & $0.878 \pm 0.026$ & $0.289 \pm 0.024$ \\
03D1au & $23.009 \pm 0.026$ & $1.273 \pm 0.198$ & $-0.015 \pm 0.030$ & $23.004 \pm 0.031$ & $1.144 \pm 0.019$ & $0.015 \pm 0.025$ \\
03D1aw & $23.584 \pm 0.030$ & $1.161 \pm 0.342$ & $-0.021 \pm 0.037$ & $23.562 \pm 0.040$ & $1.065 \pm 0.030$ & $-0.015 \pm 0.042$ \\
03D1ax & $22.961 \pm 0.026$ & $-0.574 \pm 0.114$ & $-0.112 \pm 0.031$ & $23.011 \pm 0.032$ & $0.916 \pm 0.013$ & $-0.048 \pm 0.025$ \\
03D1bk & $24.296 \pm 0.047$ & $0.211 \pm 0.274$ & $-0.249 \pm 0.054$ & $24.363 \pm 0.067$ & $1.016 \pm 0.030$ & $-0.155 \pm 0.044$ \\
03D1bm & $24.335 \pm 0.039$ & $0.567 \pm 0.347$ & $0.358 \pm 0.056$ & $24.330 \pm 0.047$ & $1.061 \pm 0.042$ & $0.350 \pm 0.052$ \\
03D1bp & $22.387 \pm 0.032$ & $-1.145 \pm 0.135$ & $-0.051 \pm 0.028$ & $22.453 \pm 0.030$ & $0.893 \pm 0.013$ & $0.030 \pm 0.019$ \\
03D1co & $24.076 \pm 0.046$ & $0.833 \pm 0.445$ & $-0.047 \pm 0.064$ & $24.075 \pm 0.049$ & $1.082 \pm 0.031$ & $-0.017 \pm 0.041$ \\
\end{longtable}
\noindent
{\bf Notes.} An electronic version of this table including covariances is available at the {\it Centre de Donn\'ees astronomiques de Strasbourg} (CDS). The SALT2 and SiFTO fit parameters can also be downloaded at the University of Toronto's Research Repository {\tt https://tspace.library.utoronto.ca/handle/1807/24512}.\\ 

%% file: salt2-nearbysne-training-sample.tex
\begin{longtable}{cccc}
\caption{The SALT2 low-$z$ training sample of Type Ia supernovae light curves and spectra
\label{table:salt2_nearby_training_sample}}\\
%\begin{tabular}{cccc}
\hline\hline
SN name & $z^{a}$ &  Bands$^{b}$  & Number of spectra$^{b}$ \\
\hline
\endfirsthead
\caption{continued.}\\
\hline\hline
SN name & $z^{a}$ &  Bands$^{b}$ & Number of spectra$^{b}$ \\
\hline
\endhead
\hline
\endfoot

sn1981b & 0.0060 & UBVR (B83) & 0 \\
sn1981d & 0.0059 & UBV (H91) & 0 \\
sn1989b & 0.0024 & UBVRI (W94) & 3 (AC) \\
sn1990af & 0.0503 & BV (H96) & 0 \\
sn1990n & 0.0034 & UBVRI (L91,L98) & 12 (AC,IUE,CFA) \\
sn1991t & 0.0060 & UBVRI (F92,L98,A04,K04b) & 32 (AC,CFA) \\
sn1992a & 0.0063 & UBVRI (S92,A04) & 19 (CFA,IUE) \\
sn1992al & 0.0146 & BVRI (H96) & 0 \\
\end{longtable}
\noindent
$^{a}$ Heliocentric redshift.\\
$^{b}$ {\bf References. } B83:~\citet{Buta83},
Br83:~\citet{Branch83},
P87:~\citet{Phillips87},
H91:~\citet{Hamuy91},
L91:~\citet{Leibundgut91},
F92:~\citet{Filippenko92},
S92:~\citet{Suntzeff92},
W94:~\citet{Wells94},
R95:~\citet{Richmond95},
H96:~\citet{Hamuy96b},
M96:~\citet{Meikle96},
P96:~\citet{Patat96},
L98:~\citet{Lira98},
R99:~\citet{riess99a},
S99:~\citet{Suntzeff99},
K00:~\citet{Krisciunas00},
K01:~\citet{Krisciunas01},
H02:~\citet{Hamuy02},
S02:~\citet{Stritzinger02},
St02:~\citet{Strolger02},
K03:~\citet{Krisciunas03},
V03:~\citet{Valentini03},
Vi03:~\citet{Vinko03},
W03:~\citet{Wang03sn01el},
Z03:~\citet{Zapata03},
A04:~\citet{Altavilla04},
B04:~\citet{Benetti04},
G04:~\citet{Garavini04},
K04a:~\citet{Krisciunas04a},
K04b:~\citet{Krisciunas04b},
P04:~\citet{Pignata04},
A05:~\citet{Anupama05},
J05:~\citet{Jha05},
R05:~\citet{Riess05},
K06:~\citet{Krisciunas06},
T06:~\citet{Tsvetkov06b},
P07:~\citet{Pastorello07},
K08:~\citet{Kowalski08},
M08:~\citet{Matheson08},
H09:~\citet{Hicken09},
AC:~\citet{AsiagoCatalogue},
IUE:~\citet{IUE},
CFA: {\tt www.cfa.harvard.edu/oir/Research/supernova/}.